%% file: main.tex
\documentclass[twocolumn]{aastex631}
\renewcommand{\email}[1]{\relax}

\usepackage{bmpsize}

\newcommand{\mic}{\ensuremath{\mu\rm m}}

\newcommand{\um}{\ensuremath{\mu{\rm m}}}

\begin{document}

\title{Overview of the JWST Advanced Deep Extragalactic Survey (JADES)}

\author[0000-0002-2929-3121]{Daniel J.\ Eisenstein} \affiliation{Center for Astrophysics $|$ Harvard \& Smithsonian, 60 Garden St., Cambridge MA 02138 USA}\email{deisenstein@cfa.harvard.edu}
\author[0000-0002-4201-7367]{Chris Willott} \affiliation{NRC Herzberg, 5071 West Saanich Rd, Victoria, BC V9E 2E7, Canada}\email{chris.willott@nrc.ca}
\author[0000-0002-8909-8782]{Stacey Alberts} \affiliation{Steward Observatory, University of Arizona, 933 N. Cherry Avenue, Tucson AZ 85721 USA}\email{salberts@stsci.edu}
\author[0000-0001-7997-1640]{Santiago Arribas} \affiliation{Centro de Astrobiolog\'ia (CAB), CSIC–INTA, Cra. de Ajalvir Km.~4, 28850- Torrej\'on de Ardoz, Madrid, Spain}\email{arribas@cab.inta-csic.es}
\author[0000-0001-8470-7094]{Nina Bonaventura} \affiliation{Cosmic Dawn Center (DAWN), Copenhagen, Denmark} \affiliation{Niels Bohr Institute, University of Copenhagen, Jagtvej 128, DK-2200, Copenhagen, Denmark} \affiliation{Steward Observatory, University of Arizona, 933 N. Cherry Avenue, Tucson AZ 85721 USA}\email{nbonaventura@arizona.edu}
\author[0000-0002-8651-9879]{Andrew J.\ Bunker} \affiliation{Department of Physics, University of Oxford, Denys Wilkinson Building, Keble Road, Oxford OX1 3RH, UK}\email{andy.bunker@physics.ox.ac.uk}
\author[0000-0002-0450-7306]{Alex J.\ Cameron} \affiliation{Department of Physics, University of Oxford, Denys Wilkinson Building, Keble Road, Oxford OX1 3RH, UK}\email{alex.cameron@physics.ox.ac.uk}
\author[0000-0002-6719-380X]{Stefano Carniani} \affiliation{Scuola Normale Superiore, Piazza dei Cavalieri 7, I-56126 Pisa, Italy}\email{stefano.carniani@sns.it}
\author[0000-0003-3458-2275]{Stephane Charlot} \affiliation{Sorbonne Universit\'e, CNRS, UMR 7095, Institut d'Astrophysique de Paris, 98 bis bd Arago, 75014 Paris, France}\email{charlot@iap.fr}
\author[0000-0002-9551-0534]{Emma Curtis-Lake} \affiliation{Centre for Astrophysics Research, Department of Physics, Astronomy and Mathematics, University of Hertfordshire, Hatfield AL10 9AB, UK }\email{e.curtis-lake@herts.ac.uk}
\author[0000-0003-2388-8172]{Francesco D'Eugenio} \affiliation{Kavli Institute for Cosmology, University of Cambridge, Madingley Road, Cambridge CB3 0HA, UK} \affiliation{Cavendish Laboratory, University of Cambridge, 19 JJ Thomson Avenue, Cambridge CB3 0HE, UK}\email{fd391@cam.ac.uk}
\author[0000-0001-8895-0606]{Pierre Ferruit} \affiliation{European Space Agency, European Space Astronomy Centre, Camino Bajo del Castillo s/n, 28692 Villafranca del Castillo, Madrid, Spain}\email{pierre.ferruit@esa.int}
\author[0000-0002-9262-7155]{Giovanna Giardino} \affiliation{ATG Europe for the European Space Agency, ESTEC, Noordwijk, The Netherlands}\email{ggiardin@cosmos.esa.int}
\author[0000-0003-4565-8239]{Kevin Hainline} \affiliation{Steward Observatory, University of Arizona, 933 N. Cherry Avenue, Tucson AZ 85721 USA}\email{kevinhainline@arizona.edu}
\author[0000-0002-8543-761X]{Ryan Hausen} \affiliation{Department of Physics and Astronomy, The Johns Hopkins University, 3400 N. Charles St., Baltimore, MD 21218}\email{rhausen@jhu.edu}
\author[0000-0002-6780-2441]{Peter Jakobsen} \affiliation{Cosmic Dawn Center (DAWN), Copenhagen, Denmark} \affiliation{Niels Bohr Institute, University of Copenhagen, Jagtvej 128, DK-2200, Copenhagen, Denmark}\email{pjakobsen@nbi.ku.dk}
\author[0000-0002-9280-7594]{Benjamin D.\ Johnson} \affiliation{Center for Astrophysics $|$ Harvard \& Smithsonian, 60 Garden St., Cambridge MA 02138 USA}\email{benjamin.johnson@cfa.harvard.edu}
\author[0000-0002-4985-3819]{Roberto Maiolino} \affiliation{Kavli Institute for Cosmology, University of Cambridge, Madingley Road, Cambridge CB3 0HA, UK} \affiliation{Cavendish Laboratory, University of Cambridge, 19 JJ Thomson Avenue, Cambridge CB3 0HE, UK} \affiliation{Department of Physics and Astronomy, University College London, Gower Street, London WC1E 6BT, UK}\email{rm665@cam.ac.uk}
\author[0000-0003-2662-6821]{Bernard J. Rauscher}\affiliation{Observational Cosmology Laboratory, NASA Goddard Space Flight Center, 8800 Greenbelt Road, Greenbelt MD 20771 USA}\email{Bernard.J.Rauscher@nasa.gov}
\author[0000-0002-7893-6170]{Marcia Rieke} \affiliation{Steward Observatory, University of Arizona, 933 N. Cherry Avenue, Tucson AZ 85721 USA}\email{mrieke@gmail.com}
\author[0000-0003-2303-6519]{George Rieke} \affiliation{Steward Observatory and Dept of Planetary Sciences, University of Arizona 933 N. Cherry Avenue Tucson AZ 85721 USA}\email{ghrieke@gmail.com}
\author[0000-0003-4996-9069]{Hans-Walter Rix}
\affiliation{Max-Planck-Institut f\"ur Astronomie, K\"onigstuhl 17, D-69117, Heidelberg, Germany}\email{rix@mpia.de}
\author[0000-0002-4271-0364]{Brant Robertson} \affiliation{Department of Astronomy and Astrophysics University of California, Santa Cruz, 1156 High Street, Santa Cruz CA 96054 USA}\email{brant@ucsc.edu}
\author[0000-0001-6106-5172]{Daniel P.\ Stark} \affiliation{Steward Observatory, University of Arizona, 933 N. Cherry Avenue, Tucson AZ 85721 USA}\email{dpstark@berkeley.edu}
\author[0000-0002-8224-4505]{Sandro Tacchella} \affiliation{Kavli Institute for Cosmology, University of Cambridge, Madingley Road, Cambridge CB3 0HA, UK} \affiliation{Cavendish Laboratory, University of Cambridge, 19 JJ Thomson Avenue, Cambridge CB3 0HE, UK}\email{st578@cam.ac.uk}
\author[0000-0003-2919-7495]{Christina C.\ Williams} \affiliation{NSF’s National Optical-Infrared Astronomy Research Laboratory, 950 North Cherry Avenue, Tucson, AZ 85719 USA}\email{christina.williams@noirlab.edu}
\author[0000-0001-9262-9997]{Christopher N.\ A.\ Willmer} \affiliation{Steward Observatory, University of Arizona, 933 N. Cherry Avenue, Tucson AZ 85721 USA}\email{cnaw@as.arizona.edu}
\author[0000-0003-0215-1104]{William M.\ Baker} \affiliation{Kavli Institute for Cosmology, University of Cambridge, Madingley Road, Cambridge CB3 0HA, UK} \affiliation{Cavendish Laboratory, University of Cambridge, 19 JJ Thomson Avenue, Cambridge CB3 0HE, UK}\email{william.baker@nbi.ku.dk}
\author[0000-0002-4735-8224]{Stefi Baum} \affiliation{Department of Physics and Astronomy, University of Manitoba, Winnipeg, MB R3T 2N2, Canada}\email{stefi.baum@umanitoba.ca}
\author[0000-0003-0883-2226]{Rachana Bhatawdekar} \affiliation{ European Space Agency (ESA), European Space Astronomy Centre (ESAC), Camino Bajo del Castillo s/n, 28692 Villanueva de la Ca\~nada, Madrid, Spain} \affiliation{European Space Agency, ESA/ESTEC, Keplerlaan 1, 2201 AZ Noordwijk, NL}\email{rachanab@gmail.com}
\author[0000-0003-4109-304X]{Kristan Boyett} \affiliation{School of Physics, University of Melbourne, Parkville 3010, VIC, Australia} \affiliation{ARC Centre of Excellence for All Sky Astrophysics in 3 Dimensions (ASTRO 3D), Australia} \email{kit.boyett@physics.ox.ac.uk}
\author[0000-0002-2178-5471]{Zuyi Chen} \affiliation{Steward Observatory, University of Arizona, 933 N. Cherry Avenue, Tucson AZ 85721 USA}\email{zychen@arizona.edu}
\author[0000-0002-7636-0534]{Jacopo Chevallard} \affiliation{Department of Physics, University of Oxford, Denys Wilkinson Building, Keble Road, Oxford OX1 3RH, UK}\email{chevalla@iap.fr}
\author[0000-0001-8522-9434]{Chiara Circosta} \affiliation{European Space Agency (ESA), European Space Astronomy Centre (ESAC), Camino Bajo del Castillo s/n, 28692 Villanueva de la Ca\~nada, Madrid, Spain}\email{chiara.circosta@esa.int}
\author[0000-0002-2678-2560]{Mirko Curti} \affiliation{European Southern Observatory, Karl-Schwarzschild-Strasse 2, 85748 Garching, Germany} \affiliation{Kavli Institute for Cosmology, University of Cambridge, Madingley Road, Cambridge CB3 0HA, UK} \affiliation{Cavendish Laboratory, University of Cambridge, 19 JJ Thomson Avenue, Cambridge CB3 0HE, UK}\email{mirko.curti@eso.org}
\author[0000-0002-9708-9958]{A.\ Lola Danhaive} \affiliation{Kavli Institute for Cosmology, University of Cambridge, Madingley Road, Cambridge CB3 0HA, UK}\email{ald66@cam.ac.uk}
\author[0000-0002-4781-9078]{Christa DeCoursey} \affiliation{Steward Observatory, University of Arizona, 933 N. Cherry Avenue, Tucson AZ 85721 USA}\email{cndecoursey@arizona.edu}
\author[0000-0003-4564-2771]{Ryan Endsley} \affiliation{Department of Astronomy, University of Texas, Austin, TX 78712 USA}\email{ryan.endsley@austin.utexas.edu}
\author[0000-0002-2380-9801]{Anna de Graaff} \affiliation{Max-Planck-Institut f\"ur Astronomie, K\"onigstuhl 17, D-69117, Heidelberg, Germany}\email{degraaff@mpia.de}
\author[0000-0002-6317-0037]{Alan Dressler} \affiliation{The Observatories of the Carnegie Institution for Science, 813 Santa Barbara St., Pasadena, CA 91101}\email{dressler@carnegiescience.edu}
\author[0000-0003-1344-9475]{Eiichi Egami} \affiliation{Steward Observatory, University of Arizona, 933 N. Cherry Avenue, Tucson AZ 85721 USA}\email{egami@arizona.edu}
\author[0000-0003-4337-6211]{Jakob M.\ Helton} \affiliation{Steward Observatory, University of Arizona, 933 N. Cherry Avenue, Tucson AZ 85721 USA}\email{jakobhelton@psu.edu}
\author[0000-0002-4684-9005]{Raphael E.\ Hviding} \affiliation{Steward Observatory, University of Arizona, 933 N. Cherry Avenue, Tucson AZ 85721 USA}\email{rehviding@arizona.edu}
\author[0000-0001-7673-2257]{Zhiyuan Ji} \affiliation{Steward Observatory, University of Arizona, 933 N. Cherry Avenue, Tucson AZ 85721 USA}\email{zhiyuanji@arizona.edu}
\author[0000-0002-0267-9024]{Gareth C.\ Jones} \affiliation{Department of Physics, University of Oxford, Denys Wilkinson Building, Keble Road, Oxford OX1 3RH, UK}\email{gj283@cam.ac.uk}
\author[0000-0002-5320-2568]{Nimisha Kumari} \affiliation{AURA for European Space Agency, Space Telescope Science Institute, 3700 San Martin Drive. Baltimore, MD, 21210}\email{kumari@stsci.edu}
\author[0000-0002-4034-0080]{Nora L\"utzgendorf} \affiliation{European Space Agency, Space Telescope Science Institute, Baltimore, Maryland, US}\email{nora.luetzgendorf@esa.int}
\author[0000-0003-4323-0597]{Isaac Laseter} \affiliation{Department of Astronomy, University of Wisconsin-Madison, 475 N. Charter St., Madison, WI 53706 USA}\email{laseter@wisc.edu}
\author[0000-0002-3642-2446]{Tobias J.\ Looser} \affiliation{Kavli Institute for Cosmology, University of Cambridge, Madingley Road, Cambridge CB3 0HA, UK}\email{tobias.looser@cfa.harvard.edu}
\author[0000-0002-6221-1829]{Jianwei Lyu} \affiliation{Steward Observatory, University of Arizona, 933 N. Cherry Avenue, Tucson AZ 85721 USA}\email{jianwei@arizona.edu}
\author[0000-0003-0695-4414]{Michael V.\ Maseda} \affiliation{Department of Astronomy, University of Wisconsin-Madison, 475 N. Charter St., Madison, WI 53706 USA}\email{maseda@astro.wisc.edu}
\author[0000-0002-7524-374X]{Erica Nelson} \affiliation{Department for Astrophysical and Planetary Science, University of Colorado, Boulder, CO 80309 USA}\email{Erica.June.Nelson@Colorado.edu}
\author[0000-0002-7392-7814]{Eleonora Parlanti} \affiliation{Scuola Normale Superiore, Piazza dei Cavalieri 7, I-56126 Pisa, Italy}\email{eleonora.parlanti@sns.it}
\author[0000-0002-0362-5941]{Michele Perna} \affiliation{Centro de Astrobiolog\'ia (CAB), CSIC–INTA, Cra. de Ajalvir Km.~4, 28850- Torrej\'on de Ardoz, Madrid, Spain}\email{michele.perna@cab.inta-csic.es}
\author[0000-0001-8630-2031]{D\'avid Pusk\'as} \affiliation{Kavli Institute for Cosmology, University of Cambridge, Madingley Road, Cambridge CB3 0HA, UK} \affiliation{Cavendish Laboratory, University of Cambridge, 19 JJ Thomson Avenue, Cambridge CB3 0HE, UK}\email{dp670@cam.ac.uk}
\author[0000-0002-7028-5588]{Tim Rawle} \affiliation{European Space Agency (ESA), European Space Astronomy Centre (ESAC), Camino Bajo del Castillo s/n, 28692 Villafranca del Castillo, Madrid, Spain}\email{tim.rawle@esa.int}
\author[0000-0001-5171-3930]{Bruno Rodríguez Del Pino} \affiliation{Centro de Astrobiolog\'ia (CAB), CSIC–INTA, Cra. de Ajalvir Km.~4, 28850- Torrej\'on de Ardoz, Madrid, Spain}\email{brunorodriguez85@gmail.com}
\author[0000-0002-0303-499X]{Wiphu Rujopakarn}\affiliation{National Astronomical Research Institute of Thailand, Don Kaeo, Mae Rim, Chiang Mai 50180, Thailand}\affiliation{Department of Physics, Faculty of Science, Chulalongkorn University, 254 Phayathai Road, Pathumwan, Bangkok 10330, Thailand}\email{wiphu@narit.or.th}
\author[0000-0001-9276-7062]{Lester Sandles} \affiliation{Kavli Institute for Cosmology, University of Cambridge, Madingley Road, Cambridge CB3 0HA, UK} \affiliation{Cavendish Laboratory, University of Cambridge, 19 JJ Thomson Avenue, Cambridge CB3 0HE, UK}\email{ls861@cam.ac.uk}
\author[0000-0001-5333-9970]{Aayush Saxena} \affiliation{Department of Physics, University of Oxford, Denys Wilkinson Building, Keble Road, Oxford OX1 3RH, UK} \affiliation{Department of Physics and Astronomy, University College London, Gower Street, London WC1E 6BT, UK}\email{aayush.saxena@physics.ox.ac.uk}
\author{Jan Scholtz} \affiliation{Kavli Institute for Cosmology, University of Cambridge, Madingley Road, Cambridge CB3 0HA, UK} \affiliation{Cavendish Laboratory, University of Cambridge, 19 JJ Thomson Avenue, Cambridge CB3 0HE, UK}\email{honzascholtz@gmail.com}
\author[0000-0001-8225-8969]{Katherine Sharpe} \affiliation{Center for Astrophysics $|$ Harvard \& Smithsonian, 60 Garden St., Cambridge MA 02138 USA}\email{kesharpe@berkeley.edu}
\author[0000-0003-4702-7561]{Irene Shivaei} \affiliation{Steward Observatory, University of Arizona, 933 N. Cherry Avenue, Tucson AZ 85721 USA}\email{ishivaei@cab.inta-csic.es}
\author[0009-0002-0651-5761]{Maddie S.\ Silcock} \affiliation{Centre for Astrophysics Research, Department of Physics, Astronomy and Mathematics, University of Hertfordshire, Hatfield AL10 9AB, UK}\email{m.s.silcock@herts.ac.uk}
\author[0000-0003-4770-7516]{Charlotte Simmonds} \affiliation{Kavli Institute for Cosmology, University of Cambridge, Madingley Road, Cambridge CB3 0HA, UK} \affiliation{Cavendish Laboratory, University of Cambridge, 19 JJ Thomson Avenue, Cambridge CB3 0HE, UK}\email{cs2210@cam.ac.uk}
\author[0009-0004-0844-0657]{Maya Skarbinski} \affiliation{Center for Astrophysics $|$ Harvard \& Smithsonian, 60 Garden St., Cambridge MA 02138 USA}\affiliation{William H. Miller III Department of Physics and Astronomy, Johns Hopkins University,
Baltimore, MD 21218, USA}\email{mskarbi1@jh.edu}
\author[0000-0001-8034-7802]{Renske Smit} \affiliation{Astrophysics Research Institute, Liverpool John Moores University, 146 Brownlow Hill, Liverpool L3 5RF, UK}\email{R.Smit@ljmu.ac.uk}
\author[0000-0002-9720-3255]{Meredith Stone} \affiliation{Steward Observatory, University of Arizona, 933 N. Cherry Avenue, Tucson AZ 85721 USA}\email{meredithstone@arizona.edu}
\author[0000-0002-1714-1905]{Katherine A.\ Suess} \affiliation{Department of Astronomy and Astrophysics University of California, Santa Cruz, 1156 High Street, Santa Cruz CA 96054 USA} \affiliation{Kavli Institute for Particle Astrophysics and Cosmology and Department of Physics, Stanford University, Stanford, CA 94305 USA}\email{suess@ucsc.edu}
\author[0000-0002-4622-6617]{Fengwu Sun} \affiliation{Steward Observatory, University of Arizona, 933 N. Cherry Avenue, Tucson AZ 85721 USA}\email{fengwu.sun@cfa.harvard.edu}
\author[0000-0001-5940-338X]{Mengtao Tang} \affiliation{Steward Observatory, University of Arizona, 933 N. Cherry Avenue, Tucson AZ 85721 USA}\email{tangmtasua@arizona.edu}
\author[0000-0001-8426-1141]{Michael W.\ Topping} \affiliation{Steward Observatory, University of Arizona, 933 N. Cherry Avenue, Tucson AZ 85721 USA}\email{michaeltopping@arizona.edu}
\author[0000-0003-4891-0794]{Hannah \"Ubler} \affiliation{Kavli Institute for Cosmology, University of Cambridge, Madingley Road, Cambridge CB3 0HA, UK} \affiliation{Cavendish Laboratory, University of Cambridge, 19 JJ Thomson Avenue, Cambridge CB3 0HE, UK}\email{hannah@mpe.mpg.de}
\author[0000-0001-6917-4656]{Natalia C.\ Villanueva} \affiliation{Center for Astrophysics $|$ Harvard \& Smithsonian, 60 Garden St., Cambridge MA 02138 USA}\email{ncv375@my.utexas.edu}
\author[0000-0002-0695-8485]{Imaan E.\ B.\ Wallace} \affiliation{Department of Physics, University of Oxford, Denys Wilkinson Building, Keble Road, Oxford OX1 3RH, UK}\email{imaan.wallace@physics.ox.ac.uk}
\author[0000-0003-1432-7744]{Lily Whitler} \affiliation{Steward Observatory, University of Arizona, 933 N. Cherry Avenue, Tucson AZ 85721 USA}\email{lwhitler@arizona.edu}
\author[0000-0002-7595-121X]{Joris Witstok} \affiliation{Kavli Institute for Cosmology, University of Cambridge, Madingley Road, Cambridge CB3 0HA, UK} \affiliation{Cavendish Laboratory, University of Cambridge, 19 JJ Thomson Avenue, Cambridge CB3 0HE, UK}\email{joris.witstok@nbi.ku.dk}
\author[0000-0001-5962-7260]{Charity Woodrum} \affiliation{Steward Observatory, University of Arizona, 933 N. Cherry Avenue, Tucson AZ 85721 USA}\email{cwoodrum@arizona.edu}

\begin{abstract}
We present an overview of the James Webb Space Telescope (JWST) Advanced Deep Extragalactic Survey (JADES), an ambitious program of infrared imaging and spectroscopy in the GOODS-S and GOODS-N deep fields, designed to study galaxy evolution from high redshift to cosmic noon.  JADES uses about 770 hours of Cycle 1 guaranteed time largely from the Near-Infrared Camera (NIRCam) and Near-Infrared Spectrograph (NIRSpec) instrument teams.  
In GOODS-S, in and around the Hubble Ultra Deep Field and Chandra
Deep Field South, JADES produces a deep imaging region of $\sim$42
arcmin$^2$ with over 100 hrs of exposure time spread
over 9 NIRCam filters, including two medium-band filters.  This is extended at medium depth in GOODS-S
and GOODS-N with NIRCam imaging of $\sim$167 arcmin$^2$, averaging 25 hrs of exposure over 8--10 filters.  In both
fields, we conduct extensive NIRSpec multi-object spectroscopy,
including 2 deep pointings of 55 hrs exposure time, 14 medium
pointings of $\sim$12 hrs, and 15 shallower pointings of $\sim$4 hrs, targeting
over 5000 HST and JWST-detected faint sources with 5 low, medium, and high-resolution dispersers covering 0.6--5.3~$\mu$m.  Finally, JADES
extends redward via coordinated parallels with the JWST Mid-Infrared
Instrument (MIRI), featuring $\sim$10 arcmin$^2$ with 43
hours of exposure at 7.7~\mic\ and thrice that area with 1.4--6.8
hours of exposure at 12.8~\mic\ and 15~\mic.  For nearly 30 years, the GOODS-S
and GOODS-N fields have been developed as the premier deep fields
on the sky; JADES is now providing a compelling start on the JWST
legacy in these fields.  
\end{abstract}


\section{Introduction}\label{sec:intro}

\input{introduction}


\section{JADES Science Goals}\label{sec:science}

\input{motivation}

\section{JADES Survey Design}\label{sec:survey}
\input{survey_design}

\input{footprint}

\section{NIRCam Observations}\label{sec:nircam}
\input{nircam}

\section{NIRSpec Observations}\label{sec:nirspec}
\input{nirspec}

\section{MIRI Observations}\label{sec:miri}
\input{miri}

\section{Preparing for JADES}\label{sec:prep}
\input{preparations}

\section{Conclusions}\label{sec:conclusions}
\input{conclusion}

\begin{acknowledgements}
The JADES Collaboration thanks the Instrument Development Teams and the instrument teams at the European Space Agency and the Space Telescope Science Institute for the support that made this program possible. We also thank our program coordinators at STScI for their help in planning complicated parallel observations.

This work is based in part on observations made with the NASA/ESA/CSA James Webb Space Telescope. The data were obtained from the Mikulski Archive for Space Telescopes at the Space Telescope Science Institute, which is operated by the Association of Universities for Research in Astronomy, Inc., under NASA contract NAS 5-03127 for JWST. These observations are associated with programs 1180, 1181, 1210, 1286, and 1287.

Processing for the JADES NIRCam data release was performed on the \emph{lux} cluster
at the University of California, Santa Cruz, funded by NSF MRI grant AST 1828315.
This research makes use of ESA Datalabs (datalabs.esa.int), an initiative by ESA’s Data Science and Archives Division in the Science and Operations Department, Directorate of Science.
This work was performed using resources provided by the Cambridge Service for Data Driven Discovery (CSD3) operated by the University of Cambridge Research Computing Service (www.csd3.cam.ac.uk), provided by Dell EMC and Intel using Tier-2 funding from the Engineering and Physical Sciences Research Council (capital grant EP/T022159/1), and DiRAC funding from the Science and Technology Facilities Council (www.dirac.ac.uk).
\end{acknowledgements}

\begin{acknowledgements}
MR, AD, EE, DJE, BDJ, BR, GR, FS, and CNAW acknowledge support from the NIRCam Science Team contract to the University of Arizona, NAS5-02015.  DJE is further supported as a Simons Investigator.  SAr acknowledges support from Grant PID2021-127718NB-I00 funded by the Spanish Ministry of Science and Innovation/State Agency of Research (MICIN/AEI/ 10.13039/501100011033).  SAl acknowledges support from the JWST Mid-Infrared Instrument (MIRI) Science Team Lead, grant 80NSSC18K0555, from NASA Goddard Space Flight Center to the University of Arizona.  AJB, AJC, JC, IEBW, AS \& GCJ acknowledge funding from the ``FirstGalaxies" Advanced Grant from the European Research Council (ERC) under the European Union’s Horizon 2020 research and innovation programme (Grant agreement No. 789056).  AJC acknowledges funding from the "FirstGalaxies" Advanced Grant from the European Research Council (ERC) under the European Union’s Horizon 2020 research and innovation programme (Grant agreement No. 789056).  ECL acknowledges support of an STFC Webb Fellowship (ST/W001438/1).  
Funding for this research was provided by the Johns Hopkins University, Institute for Data Intensive Engineering and Science (IDIES).  The Cosmic Dawn Center (DAWN) is funded by the Danish National Research Foundation under grant no.140.  RM, WB, FDE, TJL, JS, LS, and JW acknowledge support by the Science and Technology Facilities Council (STFC) and by the ERC through Advanced Grant 695671 ``QUENCH". RM also acknowledges funding from a research professorship from the Royal Society.  JW further acknowledges support from the Fondation MERAC.  The research of CCW is supported by NOIRLab, which is managed by the Association of Universities for Research in Astronomy (AURA) under a cooperative agreement with the National Science Foundation.  
ALD thanks the University of Cambridge Harding Distinguished Postgraduate Scholars Programme and Technology Facilities Council (STFC) Center for Doctoral Training (CDT) in Data intensive science at the University of Cambridge (STFC grant number 2742605) for a PhD studentship.  BRP acknowledges support from the research project PID2021-127718NB-I00 of the Spanish Ministry of Science and Innovation/State Agency of Research (MICIN/AEI/ 10.13039/501100011033) RS acknowledges support from a STFC Ernest Rutherford Fellowship (ST/S004831/1).  CWo is supported by the National Science Foundation through the Graduate Research Fellowship Program funded by Grant Award No. DGE-1746060.  DP acknowledges support by the Huo Family Foundation through a P.C.\ Ho PhD Studentship.  
\end{acknowledgements}

\begin{acknowledgements}
H{\"U} gratefully acknowledges support by the Isaac Newton Trust and by the Kavli Foundation through a Newton-Kavli Junior Fellowship.  LW acknowledges support from the National Science Foundation Graduate Research Fellowship under Grant No. DGE-2137419.  
MP acknowledges support from the research project PID2021-127718NB-I00 of the Spanish Ministry of Science and Innovation/State Agency of Research (MICIN/AEI/ 10.13039/501100011033), and the Programa Atracci\'on de Talento de la Comunidad de Madrid via grant 2018-T2/TIC-11715 MSS acknowledges support by the Science and Technology Facilities Council (STFC) grant ST/V506709/1.  REH acknowledges acknowledges support from the National Science Foundation Graduate Research Fellowship Program under Grant No. DGE-1746060.  SC acknowledges support by European Union’s HE ERC Starting Grant No. 101040227 -- WINGS.  The research of KB is supported in part by the Australian Research Council Centre of Excellence for All Sky Astrophysics in 3 Dimensions (ASTRO 3D), through project number CE170100013. Support for program JWST-GO-1963 was provided in part by
NASA through a grant from the Space Telescope Science Institute,
which is operated by the Associations of Universities for Research
in Astronomy, Incorporated, under NASA contract NAS 5-26555.
\end{acknowledgements}

\software{
Aladin Sky Atlas \url{http://aladin.u-strasbg.fr/};
Astronomer’s Proposal Tools \url{https://www.stsci.edu/scientific-community/software/astronomers-proposal-tool-apt};
BEAGLE \protect\citep{chevallard16};
eMPT code \protect\citep{bonaventura23a};
FitsMap \protect\citep{hausen22};
Guitarra \url{https://github.com/cnaw/guitarra};
grizli \url{https://doi.org/10.5281/zenodo.7963066};
JADESview \url{https://github.com/kevinhainline/JADESView};
Montage \url{http://www.ascl.net/1010.036};
NCDhas (Misselt, private communication);
NIRSpec Instrument Performance Simulator \protect\citep{dorner16}
}

\bibliography{library,library2}

\end{document}

%% file: introduction.tex
The James Webb Space Telescope (JWST) is revolutionizing the study of 
galaxy evolution by giving us unprecedented access to deep, sharp, and 
nuanced infrared imaging and spectroscopy.  Designed to push the redshift
frontier and bring the early growth of galaxies into clear focus, the
telescope is performing at, or even better than, expectations \citep{rigby23a}.  JWST takes marvelous 
advantage of the faintness of the zodiacal foregrounds at 2--10~$\mu$m
and state-of-the-art infrared detectors to unlock the rest-frame
optical at redshifts $z>4$, combining large collecting area and diffraction-limited
imaging.  Exploiting this telescope are ambitious multi-purpose instruments  \citep{gardner23a}
that give us dozens of selectable filters, 
exquisite slitless, integral-field or multi-object spectroscopy (MOS) and multiple coronagraphs.

Unraveling the physics of high-redshift galaxies will require the 
combination of many different kinds of observations.  While there are important observations of rare, extreme phenomena, many goals in studying the general population
are well served by deep-sky general 
surveys, as each image contains an unbiased superposition of all 
epochs of galaxy evolution.  Of course, low redshifts are best served
by wider, shallower data, but great depth is required to identify and characterize
high-redshift galaxies.

The Hubble Deep Field was a dramatic advance in this regard.  It boldly unveiled the high-redshift Universe
in a single multi-color blank-field image \citep{williams96,ferguson00}.  Since then, such surveys
have been vigorously pursued, utilizing virtually every high-sensitivity
narrow-field telescope.  This field and that of the Chandra Deep
Field South \citep{giacconi02} were broadened out to form the Great Observatories
Origins Deep Survey \citep[GOODS][]{giavalisco04}, utilizing the new opportunities of the Hubble Space Telescope
(HST) Advanced Camera for Surveys (ACS) and Spitzer infrared telescope
to partner with deep Chandra X-ray imaging \citep{luo08}.  Soon after, the Hubble
Ultra Deep Field (HUDF) was sited in the heart of GOODS-S \citep{beckwith06}.  It has
since become the standard bearer of deep fields, pushing into the epoch
of reionization by leveraging exceptional
optical and infrared Hubble Space Telescope imaging, e.g., the UDF09 \citep{bouwens10c} and UDF12 programs \citep{ellis13a}, with tremendous
investments across the electromagnetic spectrum from many other imaging and spectroscopic facilities.

JWST is designed to pursue such surveys; the telescope provides exquisite
image quality and depth, but it moves slowly enough that deep fields are
operationally favored.  Every pointing of practical depth reveals tens of thousands
of galaxies, including many at $z>6$ where the intergalactic medium (IGM) 
completely blocks optical light.  Further, JWST provides a novel opportunity
to conduct detailed faint multi-object spectroscopy beyond 2~$\mu$m, where
critical rest-optical lines are shifted at high redshift.  Numerous projects
have already started this work, which we expect will be
one of the enduring legacies of the telescope \citep{robertson22a}.

Here, we describe the JWST Advanced Deep Extragalactic Survey (JADES),
a collaboration of the Near-Infrared Camera (NIRCam) and Near-Infrared
Spectrograph (NIRSpec) Instrument Development Teams.  The plans to conduct deep-field
imaging and spectroscopy were featured in the original instrument proposals, 
with the intent to devote a substantial amount of guaranteed time to this
topic.  In 2015, the teams joined to form a larger and more coordinated
project, now called JADES, to focus on the exceptional opportunities of JWST 
onto the GOODS-S and GOODS-N fields.  In doing so, it became possible
to carry out a project with fewer compromises: deep and wide enough
to support the geometry of the instrument footprints and utilize
efficient parallel observations, with robust well-dithered imaging
and spectroscopy in many filters and several dispersion modes.

At about 770 hours of observing time plus coordinated parallels, JADES is the largest program
operating in JWST Cycle 1 and is a very large investment of instrument
team guaranteed time.  The time was roughly evenly split between
the NIRCam and NIRSpec GTO budgets, with a supplementary contribution from the 
MIRI-US team.  By applying the experience of the teams
that designed and commissioned the instruments, we aim to provide
an exquisite legacy data set for these deep fields.  

In this overview paper, we describe the scientific motivations and resulting
survey design of JADES, providing an overview of how we developed our
strategy to maximize performance in both imaging and spectroscopy in the targeted
fields.  Descriptions of the data reduction and
spectroscopic target selection are being presented in the data release papers \citep[][Curtis-Lake et al, in prep., Scholtz et al., in prep.]{bunker23r,rieke23r,eisenstein23jof,deugenio2025}.

%% file: motivation.tex
\subsection{Motivations}

The sensitivity and instrumentation of JWST provide a singular opportunity
to study the evolution of galaxies from the earliest epochs $\lesssim 300$~Myr
after the Big Bang, through the Epoch of Reionization during the first
billion years of cosmic history, and on to Cosmic High Noon where the stellar
mass and black hole mass densities of the universe were well-established. 
Unlike all other previous studies of high-redshift (i.e., $z>3$) galaxy populations where only
rest-frame ultraviolet spectral properties have been accessible, JWST 
enables for the first time, via its instruments NIRCam \citep{rieke23a},
NIRSpec \citep{jakobsen22,ferruit22}, NIRISS \citep{doyon23} and MIRI \citep{wright23}, photometry and spectroscopy extending from blueward
of the Lyman break to redward of the Balmer/$4000$\AA\ break region.
In this section we describe how our view of the Universe before the launch of JWST shaped the design of JADES and how early JWST observations support our decisions.

At the earliest times in cosmic history (e.g., $<500$~Myr), the first
abundant population of star-forming galaxies developed. Galaxy
formation is a self-regulated process, and the ways in which early
galaxies respond to the rapid accretion and cooling of gas greatly
affects their bulk properties like luminosity and size. Through
perseverance in HST imaging surveys, a handful of galaxies at
$z\sim10-11$ were discovered \citep{coe13,oesch16}, enabling a first
glance at the primitive galaxy formation process.

JWST has the sensitivity and the required array of infrared filters to
identify galaxies selected in rest-frame ultraviolet (UV) at $z>12$, farther than any tentative
HST detections. Hundreds of hours of NIRCam multi-filter imaging 
yields sufficient source counts to measure the UV luminosity function
evolution out to $z\sim10$ and beyond. 
JADES includes medium-band filters in the NIRCam long wave channel that can help identify and distinguish high-z candidates from dusty, strong emission line sources at lower redshifts (e.g. \citealt{zavala23, fujimoto23, arrabal23}). 
The most distant galaxies are inevitably very faint and
these NIRCam discoveries require very long integrations with the
NIRSpec low-resolution prism for redshift confirmation, especially at
redshifts $z>10$ where the strongest optical lines are redshifted beyond
the NIRSpec range \citep{robertson23,curtis-lake23}.

Extending redder than Hubble and reaching many magnitudes deeper than
ever achieved with Spitzer, JWST greatly improves our census of the early Universe,
tracing the growth of galaxies at this early epoch.  Combining the SFR and stellar masses derived
from spectral-energy distributions (SED) spanning the rest-UV and optical, NIRCam and MIRI imaging allow estimates of the stellar
birthrate of galaxies out to $z\sim10$ to deliver our earliest
constraints on the efficiency of galaxy formation
\citep[e.g.][]{labbe22,endsley22,tacchella23gz}. Stellar masses measured at
$z\sim10$ allow us to infer a bulk star formation rate to $z\gtrsim12$
(given the minimum $\sim100$~Myr timescale typically required for the
development of strong rest-frame optical breaks;
\citealt{whitler23,dressler23r}). Rest-UV emission lines, such as [CIV](1549\AA),
HeII(1640\AA), OIII](1663\AA) and CIII](1909\AA), are accessible to
NIRSpec to the highest redshifts to measure the physical properties of
nebular gas and infer their sources of ionization \citep{bunker23gz,hsiao23,tang23}.  Our window into
early galaxies with JWST becomes dramatically richer at just slightly
later times ($z\sim8-9$) where NIRSpec can measure the rest-frame
optical strong lines (e.g., [OII], [NeIII], H$\beta$, [OIII], and even
H$\alpha$ at $z<7$) and both the Lyman and
Balmer/$4000$\AA\ breaks \citep{cameron23,sanders23,reddy23a,tang23}. 
NIRCam imaging in medium and wide filters
can measure these breaks and strong lines, and given enough filters can differentiate between the two \citep{endsley22,williams23a,withers23}.

\begin{figure*}[ht]
\hspace*{\fill}
\includegraphics[width=\textwidth]{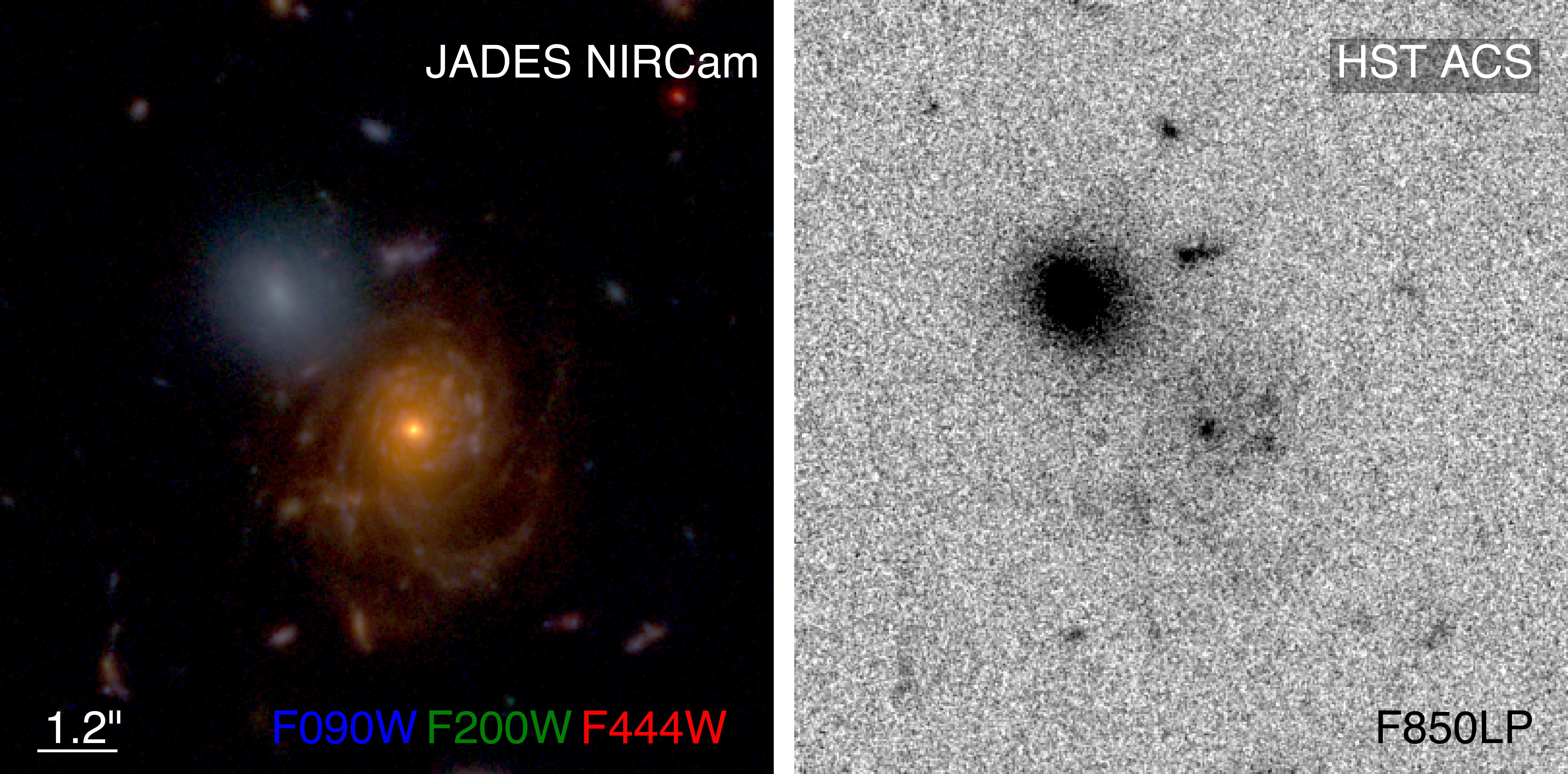}
\hspace*{\fill}
\caption{A grand design spiral at redshift 2 revealed in JADES imaging in GOODS-S.  ({\it left}) The JWST NIRCam image combining F090W, F200W, and F444W filters in the Deep Prime region of program 1180.  
({\it right}) The HST ACS F850LP image \protect\citep{illingworth16a,whitaker19}, where this star-forming galaxy is all but invisible.  While JADES F090W imaging is deeper than the HST F850LP data here, this image shows the essential value of observing at redder wavelengths. 
\label{fig:grand_design}}
\end{figure*} 

From previous surveys of the cosmic microwave background, quasars and
galaxies, we know that these epochs experience the first important
contributions of galaxies to the cosmic reionization process \citep{planck-collaboration20,fan22,robertson22}. Here, JWST simultaneously
constrains the evolving rest-frame UV galaxy luminosity density and
provides information on the hardness of the ionizing continuum and the
escape fraction of Lyman continuum photons \citep{simmonds23,mascia23,endsley22,donnan23,bouwens23}. These
measurements result in a more accurate ``balancing of the
budget'' for cosmic reionization, where we weigh the cosmic ionization
rate against the recombination of the intergalactic hydrogen in
determining the evolving bulk IGM neutrality.  Mapping with spectra
the ``Lyman-$\alpha$ disappearance'', measuring the evolving fraction
of UV-dropout selected galaxies that show (or not) Lyman-$\alpha$
emission, can track how the increased IGM neutrality at earlier times
extinguishes observed line emission in progressively more of the
sources \citep{stark10,fontana10,pentericci14,mason18,ouchi20,jones23}. Correlating the transmission of
Lyman-$\alpha$ with photometric or spectroscopic measurements of
environment provides further insight into the topology of ionized
bubbles during this epoch \citep{tang23,witstok23r,endsley23r,Jung23,lu23,whitler23b}. 

Rest-frame optical line spectroscopy at $z\sim6$--9 dramatically
extends our knowledge of the chemical enrichment of galaxies and
reveals the physical conditions in the warm interstellar medium of
early star-forming galaxies. Line excitation diagrams provide insights
into the ionization state of the star-forming ISM in these systems;
NIRSpec allows to apply these diagnostics in an entirely new redshift
regime \citep[e.g.,][]{cameron23,sanders23,reddy23a}, connecting them
with the properties of the exciting stellar populations.  Via the combined measures of
star formation rate in the rest-UV, stellar mass in the rest-optical,
and metallicity from nebular lines, we can explore whether the
fundamental metallicity and mass-metallicity relations are already in
place after only $\sim1$ billion years of cosmic history \citep{curti23, nakajima23}.

JWST is revealing the emergence of morphological structures at $z>2$
through superb infrared imaging \citep{Robertson23b, kartaltepe23, ferreira22, ferreira23, jacobs23, Huertas-company23, magnelli23, baker23}. JWST can resolve
these galaxies from the rest-UV to the rest-optical, providing
spatially-resolved measures of color gradients and stellar population
properties, vastly outperforming HST (e.g. Figure~ \ref{fig:grand_design}). We
can distinguish the clumpy UV-bright morphology from the
rest-optical light on a galaxy-by-galaxy basis, and thereby constrain
the role of large-scale gravitational instability in setting galaxy
structures at $z\sim2-3$. NIRSpec spectroscopy with the medium- or
high-resolution gratings connects these morphological measures to
the dynamics of the galaxies, and through measuring outflows further
constrain the role of feedback in shaping these maturing
galaxies.

HST has found compact red galaxies at $z=2$, but JWST's angular
resolution, sensitivity and redder bands are proving revolutionary to
explore old stellar populations and their morphology at z$>$3
\citep{suess22,Carnall23,ji23}.  With NIRCam imaging, including medium
filters, Balmer and D$_{4000}$ breaks can be cleanly picked out.
Spectroscopy at $R=100$ with the NIRSpec prism provides precise
redshifts and break strengths, but higher resolution spectroscopy
enables more detailed constraints on SFH and abundances at $z>3-4$, an era that prior to JWST was prohibitive or impossible to study, but critical to our understanding of why and how galaxies stop forming stars
\citep{Carnall23, nanayakkara23}. The importance of burstiness in the
evolution of galaxies is now becoming clear with several examples of
$z>5$ galaxies undergoing `mini-quenching' episodes
\citep{Looser23,strait23}. Additional evidence for burstiness is now becoming apparent in statistical samples of NIRCam SEDs \citep{endsley23r,dressler23r}.

\begin{figure*}[t]
\includegraphics[scale=0.6]{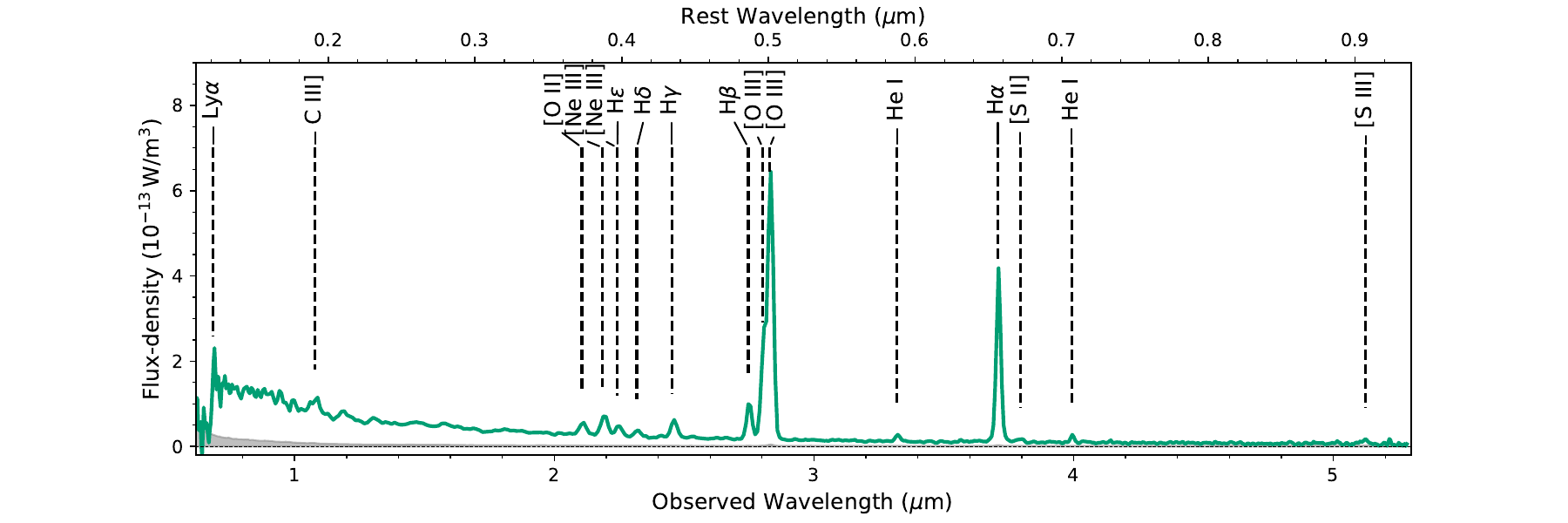} 
\hspace{1.0in}
\includegraphics[scale=0.53]{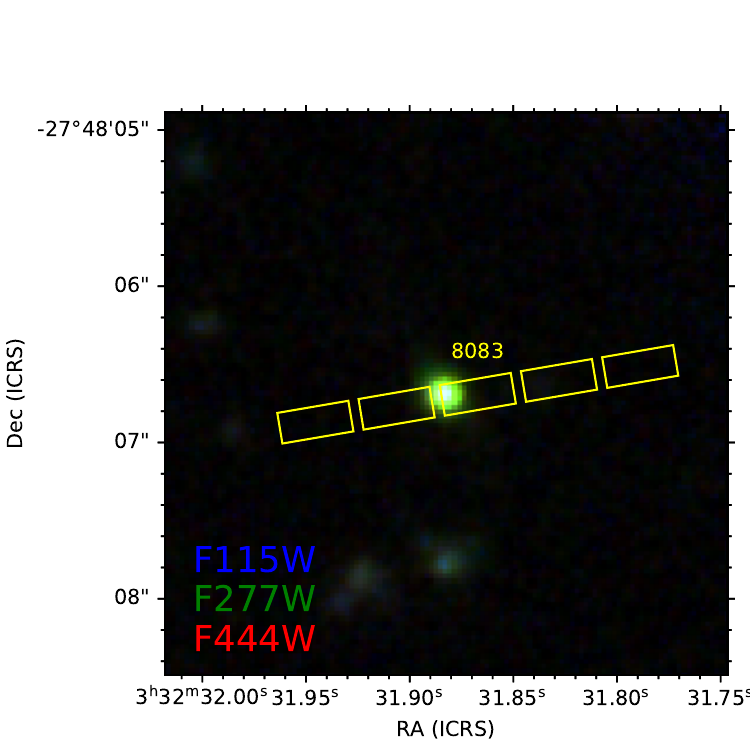}
\includegraphics[scale=0.62]{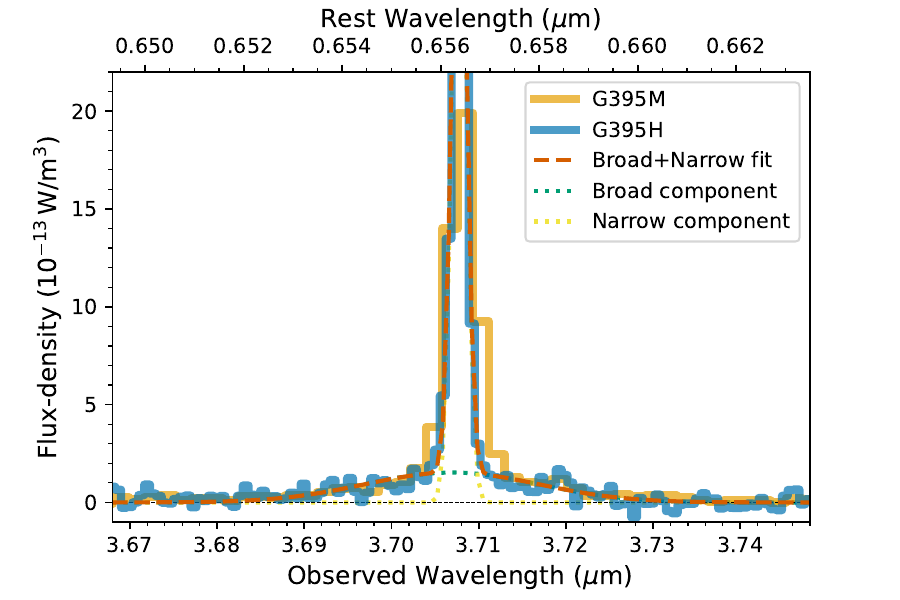} 

\caption{Example spectra for the $z=4.65$ galaxy JADES-GS+53.13284-27.80185 (ID=00008083) from the JADES Deep/HST observations. The upper panel shows the low-resolution prism spectrum (total integration time $10^5$ seconds). This spectrum reveals many emission lines and high S/N continuum. However, some emission lines are blended or have low equivalent width, motivating the acquisition of grating spectra for most of the galaxies with prism spectra in JADES. The rms uncertainty is shown in the gray shaded spectrum at the bottom of the plot. The lower-left panel shows the JADES Deep imaging data of this galaxy with an overlay of the position of the microshutters during the three nods. The green color of the galaxy indicates [OIII] and H$\beta$ line emission dominating in the F277W filter.  For the angular scale, we remind that the individual microshutters are 0.2$''$ by 0.46$''$. The lower-right panel shows the H$\alpha$ spectra obtained with the medium G395M and high G395H resolution gratings. Both these spectra reveal a broad ($\sigma=800$\,km\,s$^{-1}$) emission line from a low-luminosity AGN broad line region \citep{2024A&A...691A.145M}. However, the G395H grating is required to spectrally resolve the narrow line emission from the galaxy itself (observed $\sigma=65$\,km\,s$^{-1}$, compared to instrumental line spread function $\sigma=30$\,km\,s$^{-1}$ \citep{2024A&A...684A..87D}.
}
\label{fig:spectralexamples}
\end{figure*}

Dust attenuation and reddening in the rest-frame UV and optical
spectra and SEDs of galaxies strongly affect the inferred physical
quantities, so it is important to study the underlying dust
properties. 
JWST opens a new window to identify and characterize obscured populations that were completely missed by even the deepest Spitzer and HST surveys. ALMA has revealed that such hidden galaxies likely contribute significantly to the cosmic star formation rate density at $3<z<8$ 
\citep[e.g.][]{williams19, fudamoto21, algera23}, and early JWST data is supporting this finding \citep{barrufet23}. The stellar SEDs and morphologies of dust obscured galaxies at $z>3$ can now be characterized in detail for the first time \citep{gomez-guijarro23, nelson22, perez-gonzalez23ceers}. JWST is indicating that massive galaxies have non-negligible dust content at high-redshift, pointing to efficient production mechanisms even out to $z\sim8$ \citep{mckinney23, akins23}.
The combination of multi-band NIRCam photometry and
multi-line NIRSpec spectroscopy offers the opportunity to tackle these
issues through SED-fitting and through line diagnostics such as the
Paschen and Balmer decrements and 2200\AA\ bump
\citep[e.g.,][]{witstok23,shapley23,reddy23b,sandles23r}.

The intimate connection between the growth of galaxies and their
supermassive black holes can be traced back to the earliest epochs
with deep JWST imaging and spectroscopy. Active galactic nuclei (AGN) are being discovered via
their pointlike morphology \citep[particularly in the redder bands, e.g.][]{labbe22,furtak22}, broad
wings of Balmer emission lines (e.g., Figure
\ref{fig:spectralexamples}, but also from NIRCam grism spectroscopy \citep[such as][]{matthee24}), and highly-ionized narrow lines
\citep{kocevski23,harikane23b,maiolino23a,scholtz23r,larson23}. In most cases
these AGN are not detectable even by very deep Chandra or JVLA
imaging, putting JWST at the forefront of the quest for the earliest
supermassive black holes. With JADES we obtain deep NIRCam and MIRI
imaging plus deep NIRSpec spectrocopy to discover these previously
hidden AGN and investigate the evolving relationship between
supermassive black holes and their host galaxies.

Throughout these epochs, the development of the galaxy populations
remains tightly connected with the structure formation process
in our $\Lambda$CDM cosmology universe. The rates of star formation,
stellar population aging, merging, and dynamical
and morphological transformation are ultimately 
manifestations of the growth of dark matter halos. JWST is providing
a new context for understanding the connection between galaxy and
dark matter structure formation by aiming to discover the earliest
galaxies that form in rare peaks of the density field, establishing
both the SFR-halo mass and stellar mass-halo mass relations out to
$z\sim10$, watching the emergence of dynamically cold galactic structures,
and by observing the assembly of the first massive galaxies that
form primarily through dissipationless mergers. The new spectroscopic
capabilities allow us to identify physically-associated galaxies in the
early universe rather than just projected overdensities \citep{kashino23,helton23,morishita23,sun24hdf850} and enable 
us to distinguish between how central and satellite galaxies evolve
further back in time than has previously been possible. The combination
of area and depth allows for clustering analyses down to very faint
magnitudes on spectroscopically-informed samples with well-constrained
redshift selection functions.  This combination can also address critical gaps in our knowledge of environmentally-driven galaxy evolution.  The key epochs of stellar growth and the subsequent quenching in groups and (proto-)clusters likely often occur in a dust-obscured phase \citep[see][for a review]{alberts2022}, necessitating rest-frame near- and mid-infrared observations that are robust against extinction and directly probe obscured activity.  In all, JWST allows a more physically
complete view of galaxy formation that builds directly from the underlying
$\Lambda$CDM framework.

We stress that most if not all of these science drivers require a
substantial survey volume, not just depth.  We aim to slice the galaxy
samples in a variety of parameters for inter-comparison.  Going deep
may (slowly) reveal the less luminous galaxies, but we need to gather
sufficient samples of the $L^*$ and brighter ones as well.  Rare
phases, such as AGN and extreme starbursts, can be important for the
evolutionary story.  Large-scale structure is prominent even at high
redshifts because galaxies are extremely biased tracers of the
underlying density field.  On the scale of one NIRCam or NIRSpec MOS
pointing, this can cause the fluctuations in the number of objects,
particularly those from the most massive halos, to vary substantially \citep{steinhardt21}.
Larger surveys allow one to measure more accurate luminosity
functions, but also to potentially measure the clustering amplitude
itself, which bears on the mass of the host halos as well as on
possible Mpc-scale environmental drivers in galaxy evolution.

The above science cases can all be addressed efficiently through a deep extragalactic
survey.  Typical high-redshift galaxies are common on the sky but very faint.
The scientific exploitation of images is necessarily broad simply because of
projection of the line of sight, but this is also true for efficient use of
multi-object spectroscopy.  Since the advent of the Hubble Deep Field, the
community has been focusing its resources onto a small number of deep fields,
so that the synergies between different types of data can be best exploited.
Our survey follows this same logic.

\subsection{Opportunities of a Combined Imaging \& Spectroscopy Program}

As the combination of imaging and spectroscopy is a key driver of our
coordinated parallel strategy, we want to stress that
JWST imaging and spectroscopy reinforce each other in numerous critical ways.

First, one has the obvious aspect that imaging and spectroscopy constrain
different physical properties of the galaxies, which we seek to combine.

Second, having accurate redshifts is important for the interpretation
of imaging in terms of luminosities, rest-frame colors, and proper sizes.
For example, the conversion of SEDs to stellar masses and star formation
histories can easily be degenerate with redshift uncertainty.
Spectroscopy is the gold standard for redshifts, and JWST has 
sufficient sensitivity and multiplex to provide spectroscopic redshifts
for thousands of galaxies all the way to z$>$10.
Moreover, our program is providing large training samples for the
photometric redshift methods that supplies redshifts for the rest
of the imaging sample.

Third, there are technical synergies.  NIRCam broad-band filters in
the rest-frame optical can have substantial contribution from very
strong emission lines, as illustrated by the recent JWST results
\citep{cameron23,matthee23}. NIRSpec spectroscopy is providing the
location and fluxes of these lines, enabling us to subtract them to
accurately measure the continuum SED.  This is important for the
estimation of stellar population age distributions.  We can do this
subtraction directly in thousands of objects but also measure the
trends and variations needed to model the purely photometric samples.

NIRCam, in turn, is important for NIRSpec MOS to understand its slit losses,
background subtraction, and to aid in the interpretation of emission line kinematics.
Unlike ground-based slit masks, the NIRSpec
micro-shutter array (MSA) provides a fixed grid of slits.  Galaxies 
fall at various registrations relative to those slits and with a
range of sizes.  Achieving accurate line flux calibration requires
the imaging to provide a model of this.  Further, NIRSpec MOS background 
subtraction requires the subtraction of neighboring shutters; only 
NIRCam can provide a deep 2--5 $\mu$m probe of contaminating objects
in these shutters.  Eventually, we expect that the sharp NIRCam images
will provide morphological templates for more ambitious extraction of
undersampled NIRSpec spectra, going beyond just summing along the
spatial direction of a slit.

Fourth, JWST imaging can allow more efficient target selection for
NIRSpec MOS spectroscopy of rare populations.  While HST can provide
Lyman-dropout selection for UV bright targets, the longer wavelength
coverage of JWST yields much improved photometric redshifts for
redder objects.  NIRCam medium-band imaging can isolate objects with
strong rest-frame optical line emission that can then be targeted 
for line profile studies with the NIRSpec gratings.

Finally, there are more subtle astrophysical synergies.  For strong
line emitters, the high S/N and the excellent angular resolution of
NIRCam provides, via the comparison of different filters,
measurements of size and morphology of the line emission relative
to the stellar light.  NIRCam imaging can reveal color gradients 
to be correlated with spectral properties.  For spatially extended galaxies, one can even connect these to resolved spectral
variations along the slits.

We note that while the combination of imaging and spectroscopy is
critically important, it is not the case that one requires
spectroscopy for every imaging object.  Rather one intends to use the
spectroscopy to build models of the trends, so that one can perform
statistical work on the non-spectroscopic sample.

%% file: survey_design.tex
\subsection{Field Selection}\label{sec:fields}
JADES seeks to combine deep multi-band imaging and spectroscopy 
in pursuit of the science goals described in \S~\ref{sec:science}.  It is designed
to bring NIRCam and NIRSpec MOS together on a common region of the 
sky, while covering substantial areas in two different fields
in order to increase the statistical reach for rare objects and
sample large-scale structure.

We observe two fields in JADES, to avoid concern that the large-scale structure
of highly biased tracers and the radiation transport of reionization
could make any one field peculiar and limit confidence in any unusual results.  Further, we wanted to spread the
observing around the year to ease the constraints on scheduling such a large program. Of course, even more fields would better mitigate 
the concerns about cosmic variance, but this would limit the depth
and area of each.  

The choice of location was driven by the availability of 
deep pan-chromatic imaging and spectroscopy, as the study of 
galaxy evolution draws on a wide range of such input.
This led us clearly to the GOODS-South and GOODS-North fields,
which have received huge investments of telescope time over
the past 25 years from essentially every facility that bears
on the high-redshift Universe.  

GOODS-South, home of the Chandra Deep Field South and the Hubble
Ultra Deep Field as well as very deep ALMA \citep{walter16,dunlop17,franco18,hatsukade18} and JVLA data \citep{rujopakarn16,alberts20}, is the preeminent deep field on the sky.  We 
chose this as the primary field for JADES and focused the majority
of the observing time there.  GOODS-North, home of the Hubble Deep Field and 
exceptionally deep Chandra data, was chosen as the second field.

By placing JADES in the legacy GOODS fields, we seek to augment the rich HST community
data products with a comprehensive set of JWST imaging and spectroscopy. The GOODS \citep{giavalisco04},
CANDELS \citep{grogin11a,koekemoer11a}, and UDF \citep{beckwith06,ellis13a,illingworth13a}
data have been reduced and released as components of the
Hubble Legacy Fields \citep[HLF,][]{illingworth16a,whitaker19a}.
The HLF reductions provide an excellent matched set of HST images in multiple ACS and WFC3 bands for use
with the JADES JWST NIRCam imaging.
We also utilize astrometric registration to Gaia performed by G.\ Brammer (private communication) using the methods of \citet{kokorev22} and {\it grizli}\footnote{https://doi.org/10.5281/zenodo.7963066}.

\subsection{Tiers and Geometrical Constraints}\label{sec:tiers}

JADES is built as a two-layer wedding cake, with Deep portions of 
both imaging and spectroscopy, flanked by larger Medium depth regions.
Bringing imaging and spectroscopy to bear on the same targets, while
making efficient use of coordinated parallel observations, are driving
goals of the survey design.  Here we begin to describe these considerations.

The differing on-sky geometries of the NIRCam and NIRSpec instruments
mean that these two cannot be efficiently overlapped with a single
pointing of NIRCam.  It takes at least a 2x2 mosaic of NIRCam to produce a filled
area large enough to cover one NIRSpec MOS pointing.  Further, the
ability to use two instruments at once is an important opportunity
to increase the science return, but the angular separation between
the instantaneous fields of the instruments drives one to a large
field.  We note that, as a consequence of the visibility constraints, neither of the JADES fields allows JWST to
return at a 180$^\circ$ position angle, so as to swap the instrument
locations.  Instead, we have to construct an adequately sized mosaic,
and choose the parallels to maximize the science return.  Most of the MIRI parallel data and the NIRSpec MOS parallel data falls on NIRCam imaging, and nearly all of the NIRSpec parallel data falls on the GOODS/CANDELS HST imaging.

We placed the Deep portion of JADES in GOODS-S, while the Medium data are
in both fields.  We considered placing Deep pointings in GOODS-N, but
as full support of the NIRSpec MOS footprint requires 4 NIRCam pointings, this would have become overly expensive.

The NIRCam data in JADES fall into 4 categories.  There are contiguous
portions in regular mosaics, of both deep and medium variety; we
call these ``Prime''.  Other portions occur as
parallel exposures to NIRSpec MOS pointings, whose positions are therefore
dictated by the location and position angle of the spectroscopy;
we call these ``Parallel''.  Again, these come in deep and medium
variety.

In detail, ten of the Medium-depth Prime NIRCam exposures were
taken with NIRSpec in parallel, but the structure of JWST
coordinated parallel observations is that NIRSpec is formally prime
in the planning tool.  We refer to NIRCam as prime (and NIRSpec as
parallel) despite this, because the exact pointing and exposure
times were being dictated by the NIRCam science goals.

Other NIRCam pointings were taken with MIRI in parallel; these yield
Deep and Medium MIRI imaging.

The JADES NIRSpec MOS data fall into 3 tiers.  One tier comprises
two deep pointings, one scheduled early in the program and targeted
without JWST imaging, the other scheduled at the end and targeted
from the JADES imaging.  These are called Deep/HST and Deep/JWST, respectively.
Then there are two tiers of medium
depth, called Medium/HST and Medium/JWST for the same reason.  Unlike the Deep pointings, Medium/HST is somewhat shallower than Medium/JWST.
All of the NIRSpec MOS data is taken with NIRCam in parallel.

The Medium-depth designs with both NIRCam and NIRSpec MOS are shaped heavily by a 
desire to take well-dithered data, with at least 6 pixel locations to provide robustness to bad pixels and the undersampled point-spread function, 
and to use long enough exposures to keep observatory overheads low (these concerns are easily satisfied with
the Deep data).  Hence, even this flanking data is quite deep in comparison to pre-JWST opportunities.
Because of the full use of coordinated 
parallel observations, we paid particular attention to minimizing the data rate for telemetry, typically utilizing the DEEP8 readout pattern for NIRCam and SLOW
readouts for MIRI.  This kept the program to about 2 GB/hr of data volume.

Although JADES was designed to be observed in a single year, over-scheduling
in Cycle 1 resulted in it being scheduled over 18 months.  In particular,
the large investment in GOODS-S was spread over two observing seasons.
This led to some reoptimization relative to the original design, induced by
exact position angles, instrument problems for some observations, and 
further on-orbit appreciation of science opportunities.  It also created a richer data set for deep-sky infrared variability than originally planned \citep{decoursey25}.  We will focus
in this paper on the observed program, making only passing mention of the
original layout.

%% file: footprint.tex
\subsection{JADES Footprint}

\begin{figure*}[p]
\includegraphics[scale=0.35]{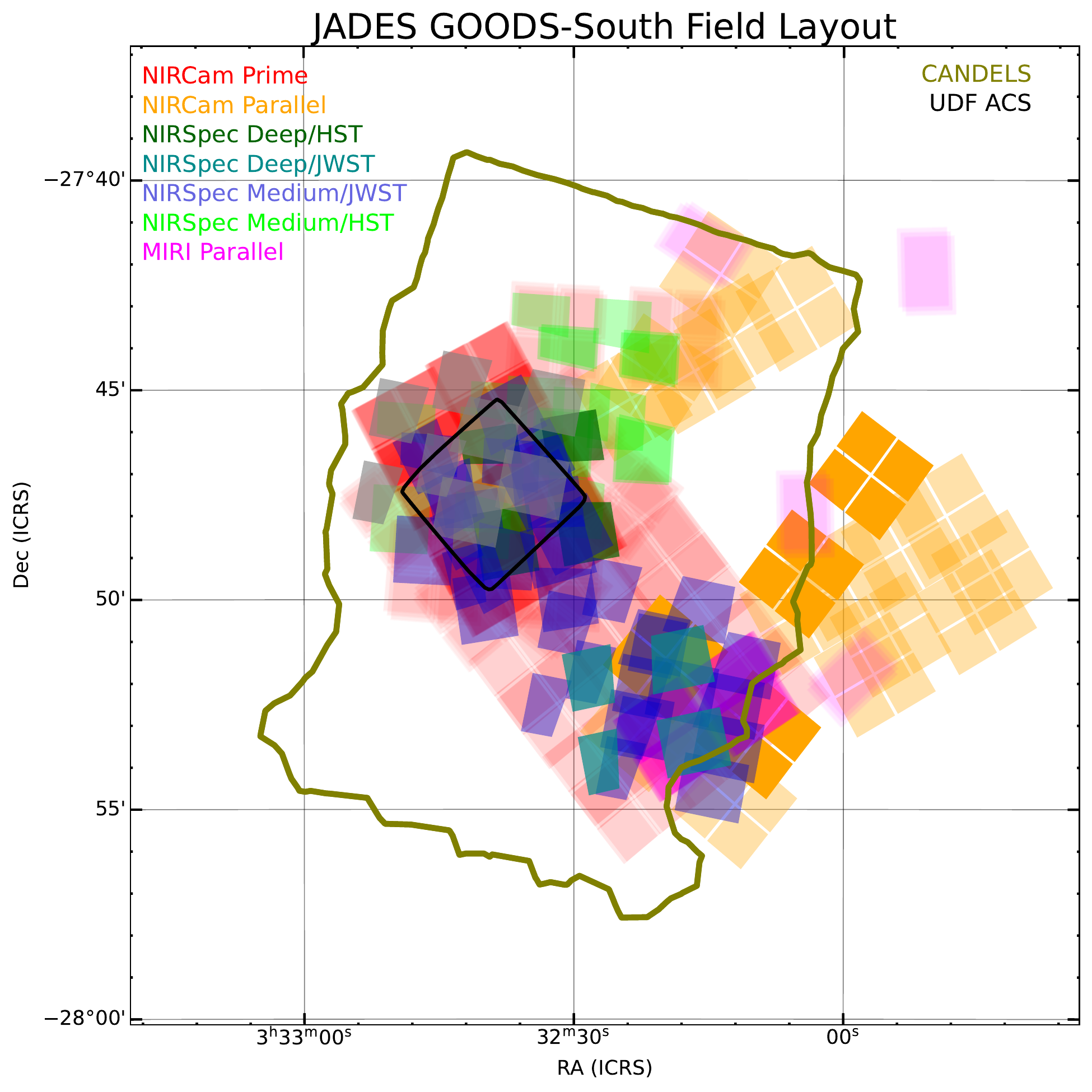} 
\begin{minipage}[b]{2.0in}
\includegraphics[scale=0.18]{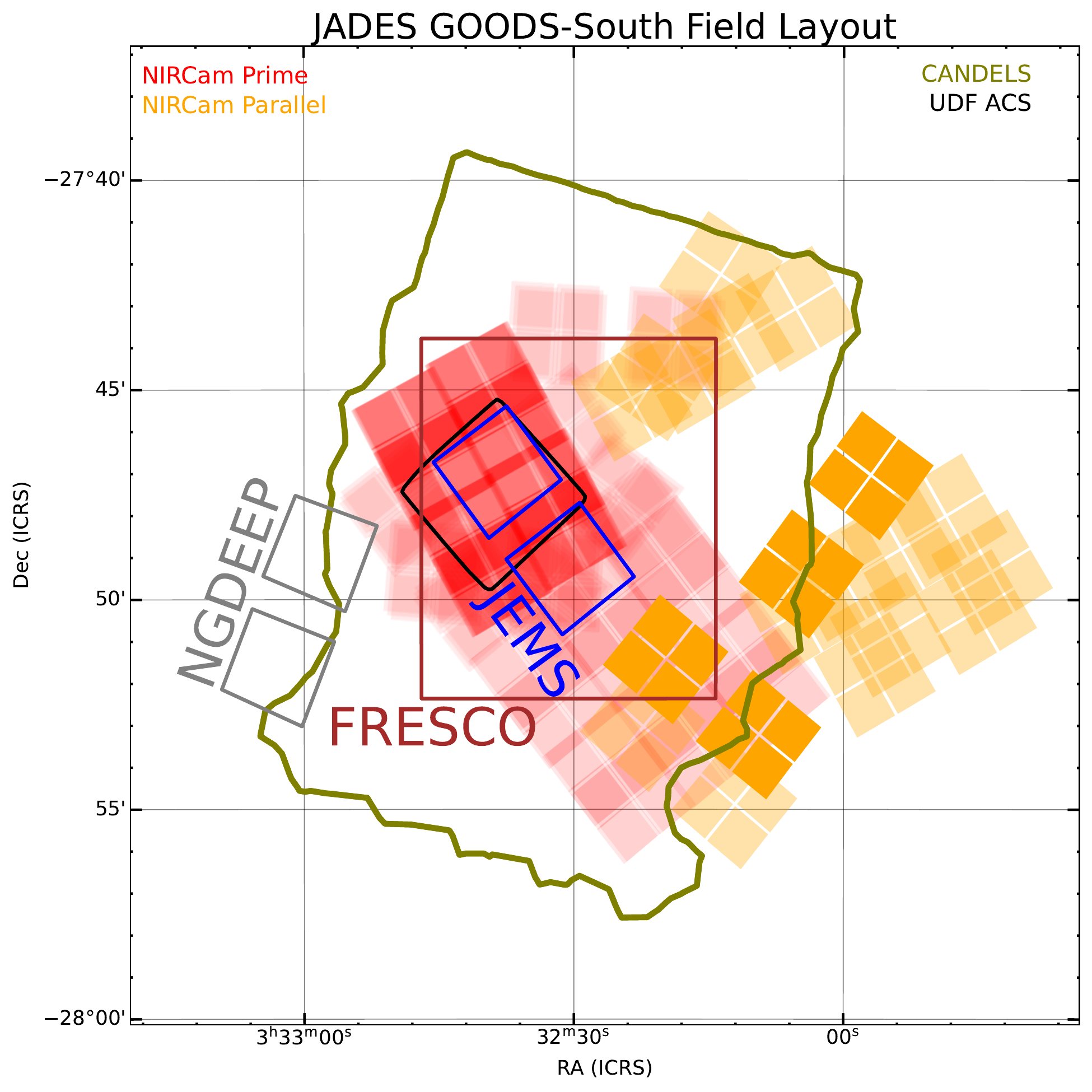}
\includegraphics[scale=0.18]{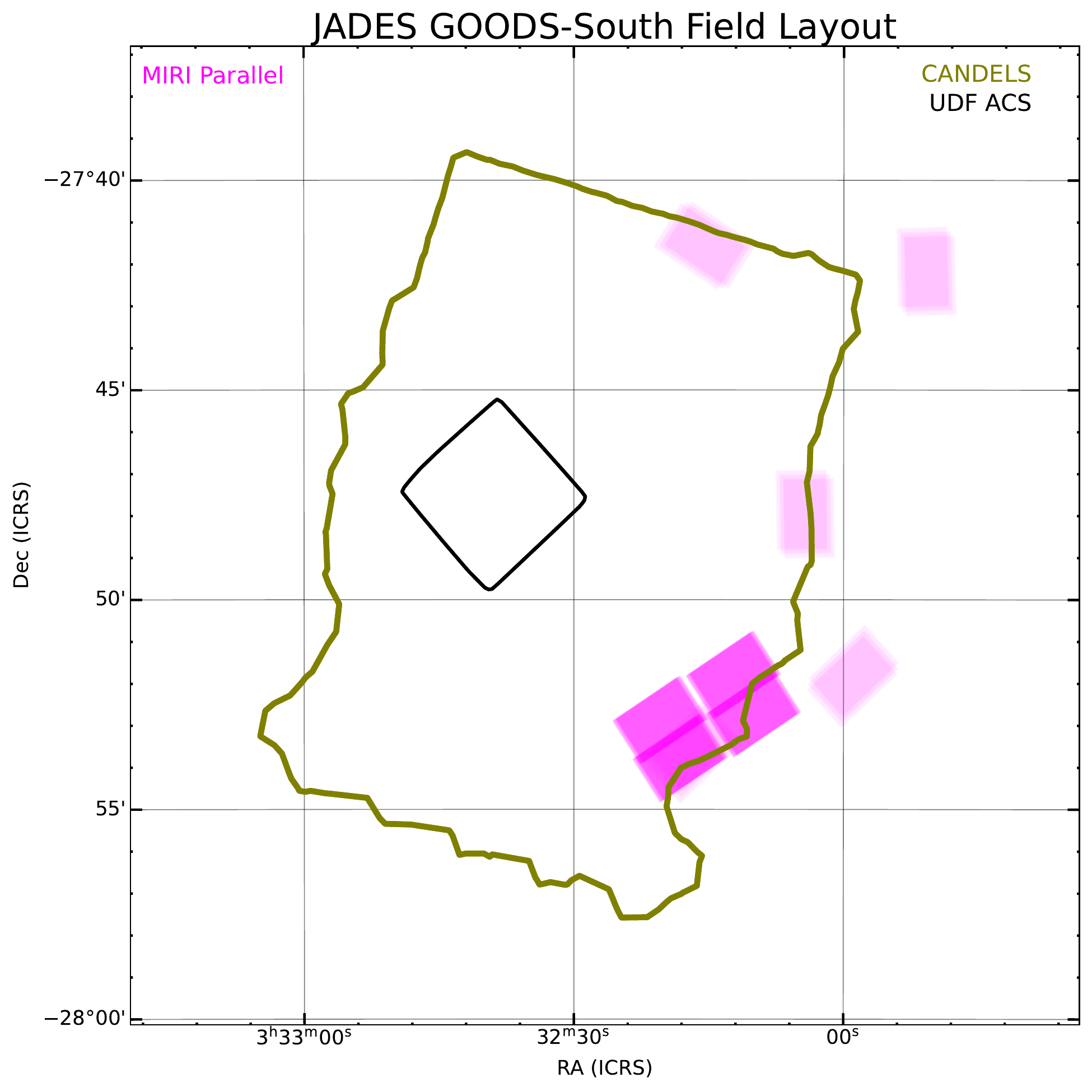} 
\end{minipage}
\\
\includegraphics[scale=0.18]{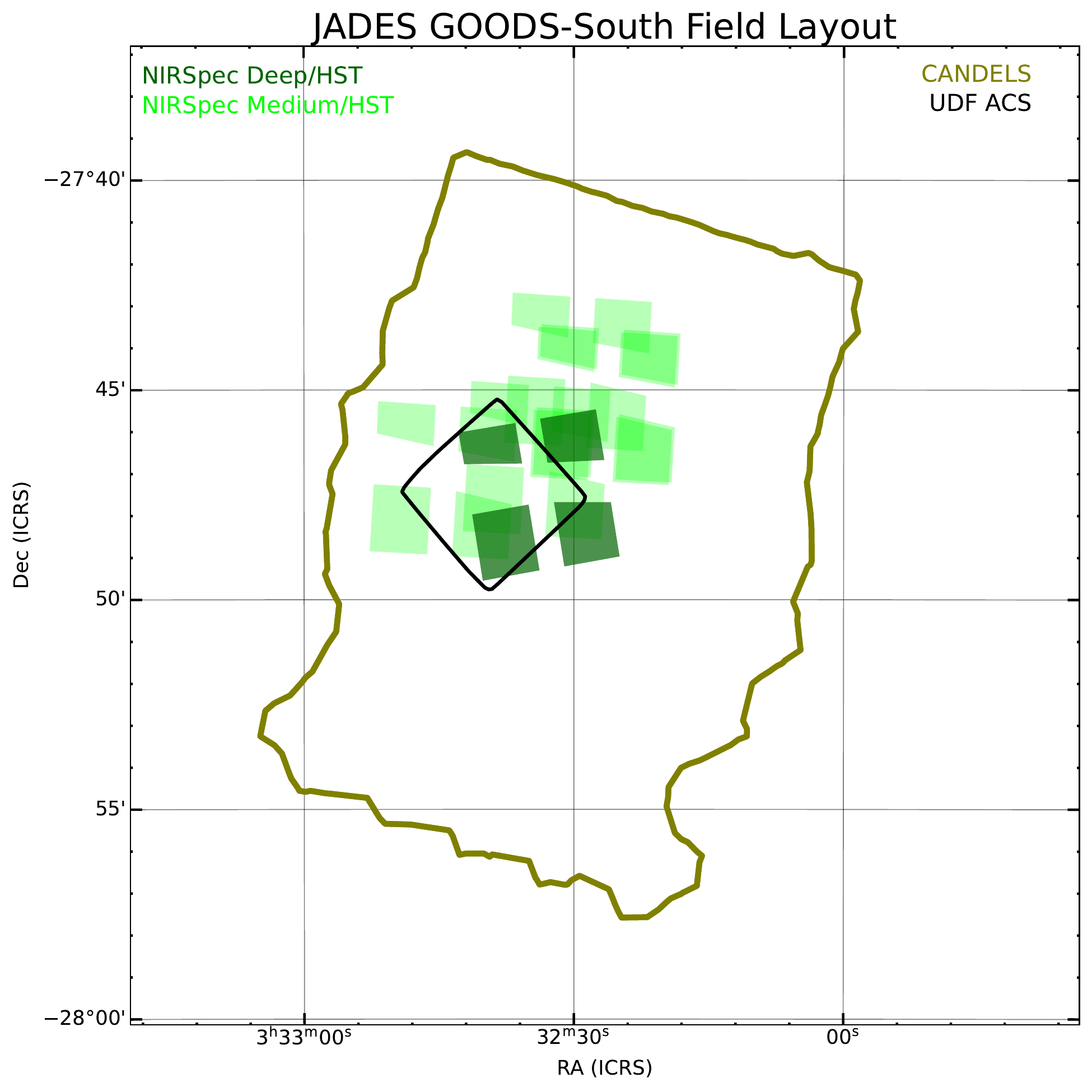} 
\includegraphics[scale=0.18]{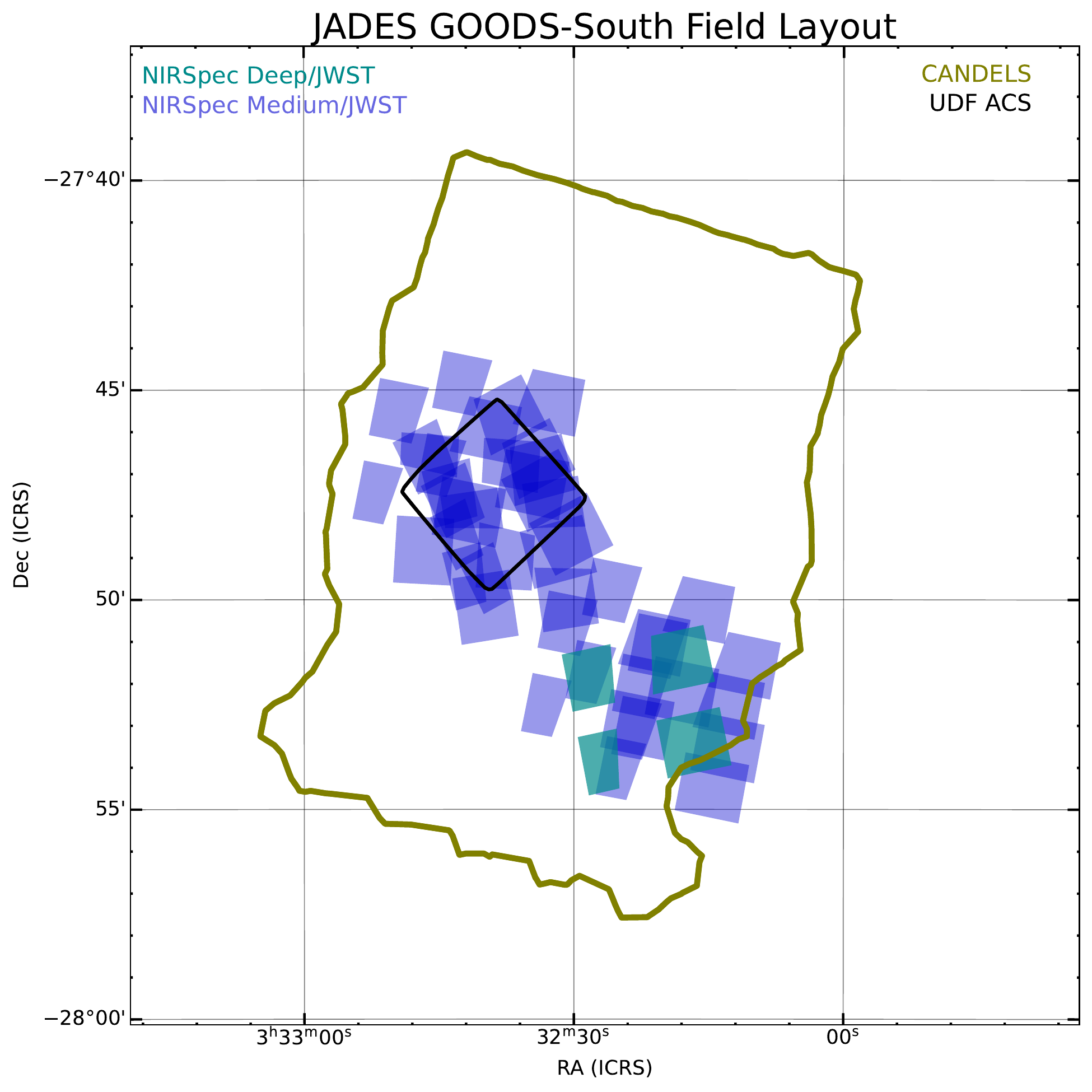} 
\hspace*{\fill} \\
\caption{Layout of the JADES observations executed in the GOODS-South field. JADES observations with NIRCam, NIRSpec MOS, and MIRI are shown as colored shaded regions. Higher opacity indicates higher exposure time for NIRCam and MIRI or overlapping MSA pointings for NIRSpec. Dithers and nods smaller than 2 arcseconds are mostly not plotted. For NIRCam, only the SW quadrants are shown for clarity. For NIRSpec, only the active area of the MSA that was used for target placement, excluding regions that lead to truncated prism spectra, is shown.  Two MIRI parallels with short F1500W exposures are not visualized. Outlines of other surveys, including the HST/ACS UDF and CANDELS, are shown with black and olive green curves, respectively. The smaller sub-panels show the same information split by instrument for clarity because it can be difficult to see the details when all observations are plotted together. The NIRCam sub-plot in upper-right additionally includes field outlines for the public JWST Cycle 1 NIRCam imaging from the JEMS, FRESCO and NGDEEP programs.  
\label{fig:south}}
\end{figure*}

\begin{figure*}[p]
\includegraphics[scale=0.35]{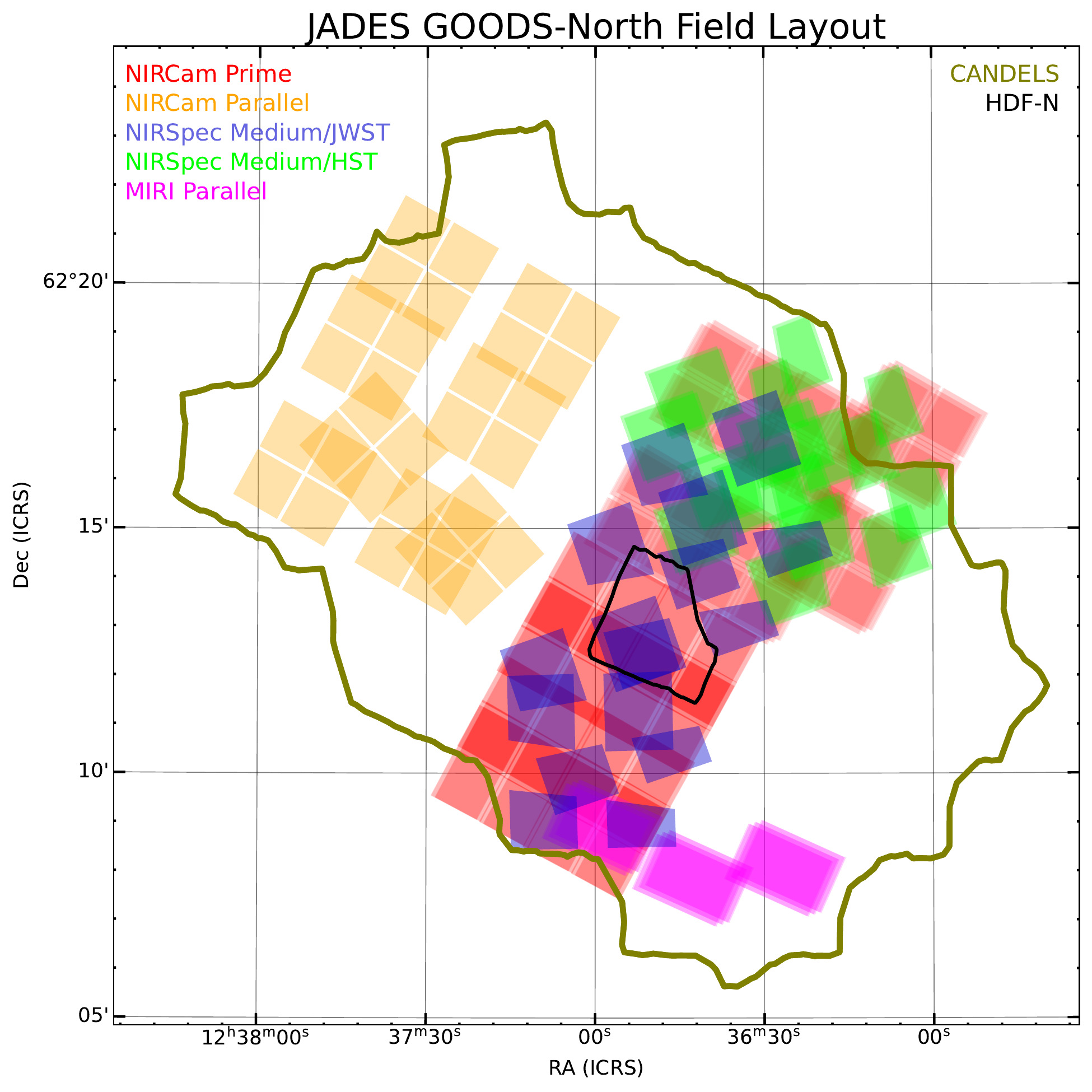} 
\begin{minipage}[b]{2.0in}
\includegraphics[scale=0.18]{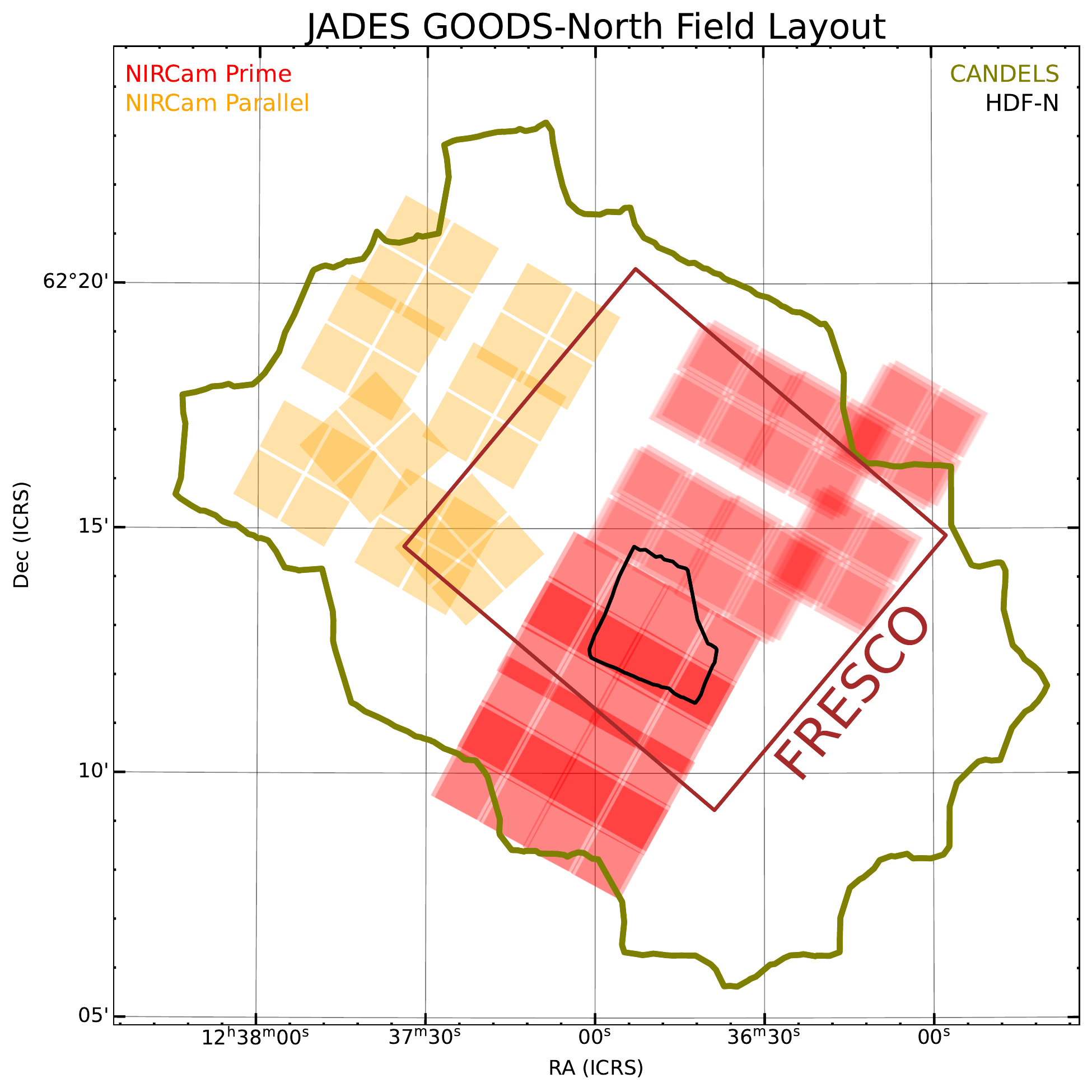}
\includegraphics[scale=0.18]{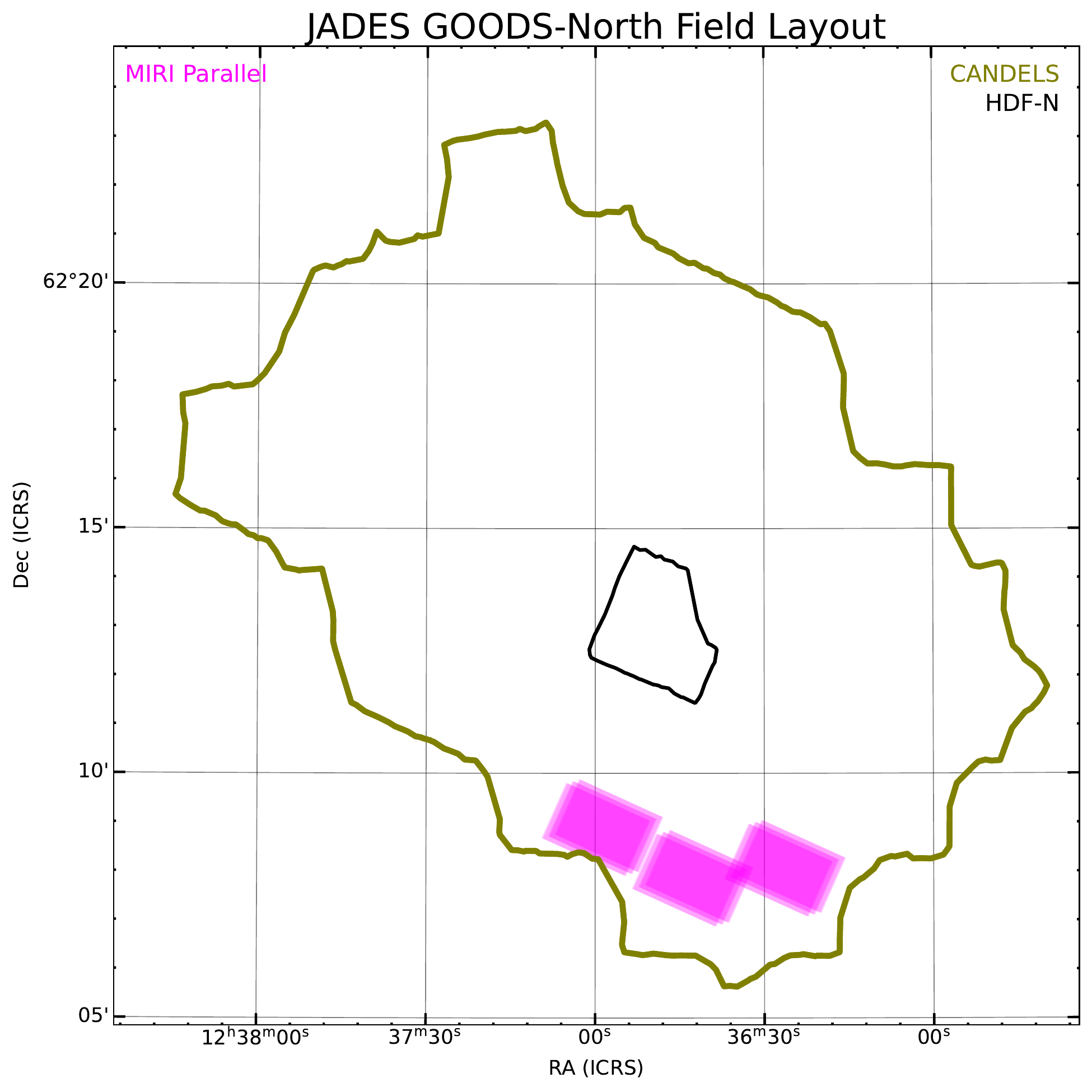} 
\end{minipage}
\\
\includegraphics[scale=0.18]{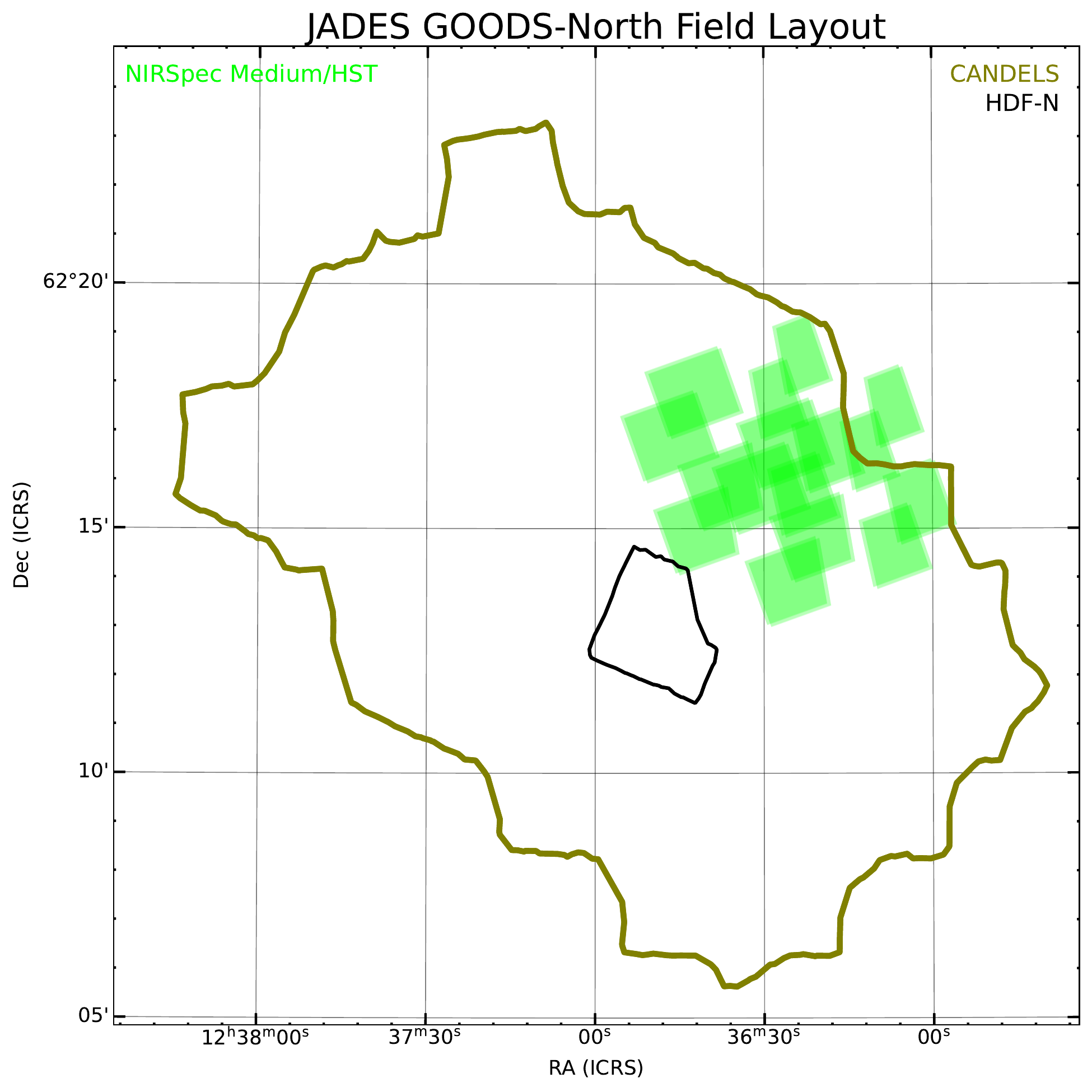} 
\includegraphics[scale=0.18]{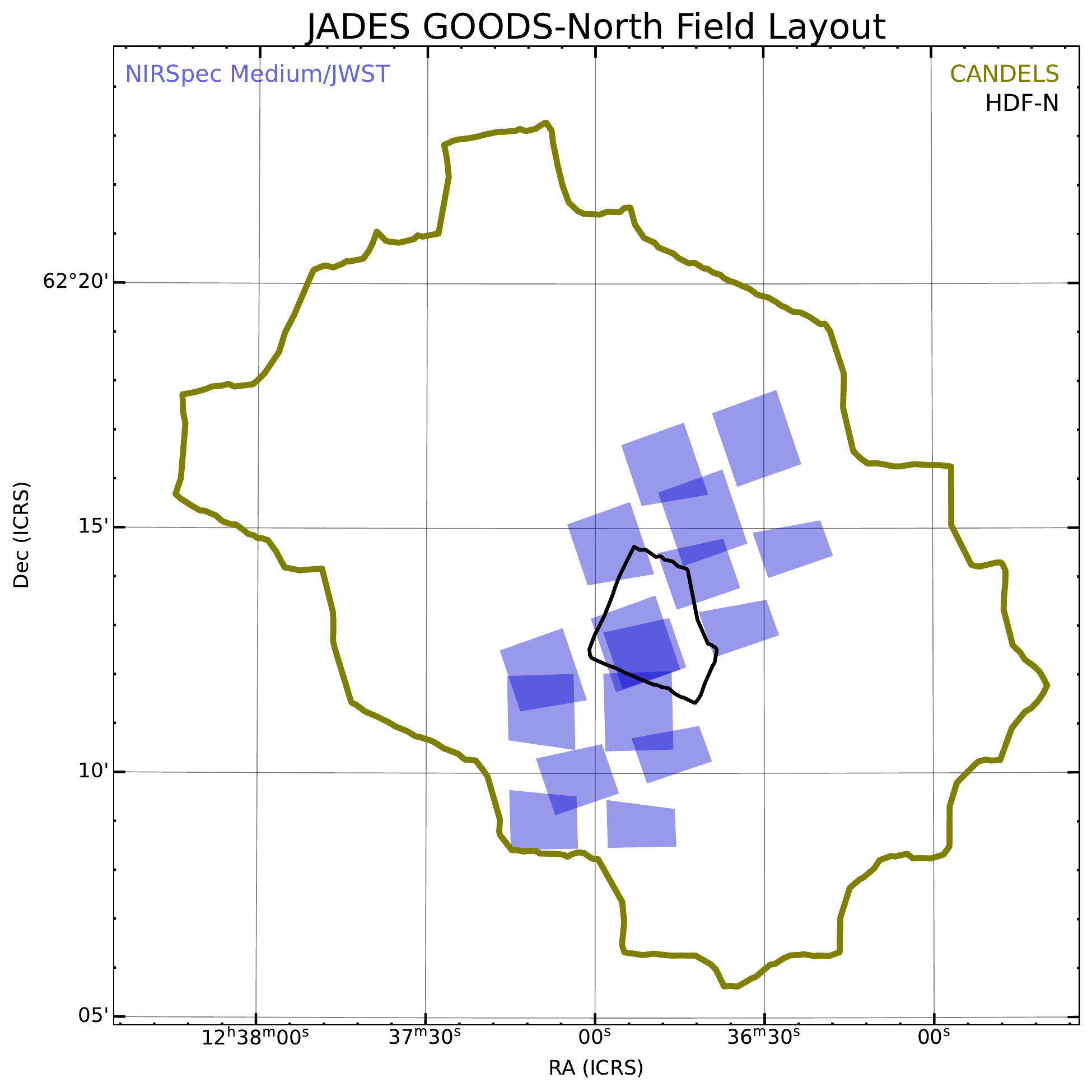} 
\hspace*{\fill} \\
\caption{Layout of the JADES observations executed in the GOODS-North field. Details as in Figure \ref{fig:south}. 
The NIRCam sub-plot in the upper-right additionally includes the outlines of the public JWST Cycle 1 NIRCam imaging from the FRESCO program.
\label{fig:north}}
\end{figure*}

The footprint of the JADES survey is shown in Figures \ref{fig:south} for GOODS-S and \ref{fig:north} for GOODS-N.  As the geometry is complicated with the various overlapping footprints, we include separate images of each major portion of the survey.  

In GOODS-S, one can see how the primary Deep portion of the survey,
centered on the UDF, and the Medium portion to the west support
each other with coordinated parallel observations.  For instance,
the MIRI parallels are largely partnered with NIRCam imaging, including 
the first NIRCam Deep Parallel.  
We also generate NIRSpec Medium/HST
parallels that fall back on the UDF, providing high-multiplex spectroscopy
for our Deep imaging.  Finally, the NIRSpec Medium/JWST and Deep/JWST
pointing create NIRCam parallels that fan out to create more
imaging area.

In GOODS-N, the area of the prime survey is a little smaller, which
makes it harder to fully utilize the parallels.  We do utilize the
NIRCam Medium Prime fields to cover the HDF and provide imaging for
the initial NIRSpec Medium/HST spectroscopy as well as targets for
the Medium/JWST followup.  The parallels from that later spectroscopy
produce NIRCam images that cover additional GOODS/CANDELS imaging.
Two of the NIRCam prime pointings did not have MIRI or NIRSpec
parallels that would fall on NIRCam imaging or GOODS; we opted to use
MIRI, to be partnered with HST CANDELS imaging and perhaps future NIRCam
data.

A major constraint on the layout of the survey comes from the intention to be able to conduct NIRSpec follow-up of NIRCam-selected targets within a single observing window. 
We judged that 60 days would be the minimum separation of these visits; this implies performing the NIRCam imaging early in the window and the NIRSpec follow-up late in the window.  In GOODS-N, this plan was scheduled in this manner.  
In GOODS-S, because some of the imaging was delayed until year 2, we still chose
to conduct the prime imaging early in the window and the spectroscopy late in the window.

In GOODS-S, the year 1 NIRCam Deep Prime imaging was taken at V3 PA = 298.56$^\circ$, in early October 2022.  The year 1 Medium Prime imaging was taken a week later at V3 PA = 308$^\circ$.  The first NIRSpec Deep/HST pointing followed at V3 PA = 321$^\circ$.  
In year 2, we sought to match the Deep Prime and Medium Prime position angles to those of year 1.  This was successful for Deep, but observatory interruptions disturbed the Medium plans.  The locations of the spectroscopic fields was chosen based on the exact availability of high-priority high-redshift candidates given the complex MSA constraints.

In GOODS-N, we observed the NIRCam Medium Prime data at PA = 241$^\circ$ in early February 2023.  The NIRSpec Medium/JWST data then followed in early May, at PA = 150.48$^\circ$.  Due to a fault with acquiring guide stars, one observation was skipped and had to be redesigned and observed at PA = $132.93^\circ$.

\subsection{Other Overlapping JWST Data}
\label{sec:external}

The GOODS fields, UDF, and HDF have of course been observed by other
programs in JWST Cycle 1 and Cycle 2.  Here we briefly describe programs whose
data overlap JADES.  
Figures \ref{fig:south} and \ref{fig:north} show the JADES NIRCam footprint with overlays
with some of these and other programs.
We surely expect this list to grow in future cycles.

The First Reionization Epoch Spectroscopic Complete program \citep[FRESCO, PI: Oesch, Program 1895,][]{oesch23a} conducted NIRCam F444W
slitless spectroscopy (2 hr depth), paired with F182M and F210M medium-band
imaging (1 hr each).  These 8-pointing mosaics in GOODS-S and GOODS-N
overlap heavily with JADES.  FRESCO has proven to be highly complementary
to JADES, as the strong emission lines are imaged as excesses in JADES
photometry and then redshifts are obtained in FRESCO.  The additional
medium-band imaging, while shallower than JADES, provides more spectral
resolution on mid-redshift galaxies.

The JWST Extragalactic Medium-band Survey \citep[JEMS; PIs: Williams, Tacchella, \& Maseda; Program 1963,][]{williams23a} observed one pointing on the UDF in F182M, F210M,
F430M, F460M, and F480M, with 4--8 hrs/filter.  This dataset heavily 
overlaps with JADES, providing additional filters with compelling
depth.  
In addition, JEMS conducted a NIRISS parallel in F430M and F480M, about
half of which overlaps JADES NIRCam imaging. 

Between these two, we note that F182M and F210M are available 
for a notable portion of JADES.  We have co-reduced JEMS and FRESCO 
NIRCam imaging with JADES, and include these data
in JADES photometric catalogs.  

In addition to FRESCO's wider medium-depth slitless spectroscopy,
the Next Generation Deep Extragalactic Exploratory Public survey (NGDEEP, PI: Finkelstein, Program 2079) is producing 
very deep NIRISS slitless spectroscopy
on the UDF \citep{bagley23a}.  We expect that these programs will complement the 
deeper but targeted NIRSpec MOS spectroscopy.  The NGDEEP NIRCam imaging
parallel falls off the JADES footprint, to the south-east.

There are also other deep MIRI GTO imaging programs in GOODS-S.  
The MIRI Deep Imaging Survey \citep[MIDIS, Program 1283, PI: Oestlin,][]{ostlin25,2024ApJ...969L..10P} is observing one extremely deep pointing in the UDF,
reaching 41 hrs in F560W and 8.5 in F1000W.  The NIRISS parallel from this pointing
falls at the southern edge of JADES NIRCam imaging. 
For wider MIRI coverage, program 1207 (SMILES, PI: G.~Rieke)
is observing 15 pointings in 8 filters, all overlapping the JADES
NIRCam imaging \citep{Alberts24MIRI}.  

The push for deep imaging and spectroscopy in GOODS-S continued
in Cycle 2 with an approved GO program 3215 (PI: Eisenstein) that
 returned to the footprint of the JADES 1210 program, now dubbed the JADES Origins Field, to add very
deep NIRCam imaging in six medium bands---F162M, F182M, F210M,
F250M, F300M, and F335M---to refine the search for $z>15$ Lyman
dropouts \citep{eisenstein23jof}.  At the same time, ultra-deep NIRSpec MOS spectroscopy
was obtained on top of the JADES Deep Prime imaging.

The scheduling of JADES over two years created new opportunities to study deep 
infrared time variability.  This was pursued by a Director's Discretionary program 6541 (PI: Egami) to conduct NIRSpec MOS spectroscopy of transients and host galaxies in the Deep Prime region, as well as two additional epochs of multi-band imaging.

Finally, there are several other NIRSpec programs in the fields.  
Two examples are program 2674 (PI: Arrabal Haro), which
conducted NIRSpec MOS and coordinated NIRCam imaging
in GOODS-N that complement the JADES footprint, and program 2198 \cite[PI: Barrufet,][]{barrufet25}, which conducted NIRSpec MOS and NIRCam pre-imaging on two fields in GOODS-S.
Other spectroscopic programs are the GTO NIRSpec WIDE MOS survey from programs 1211 and 1212 (PI: Luetzgendorf) and IFU program 1216 (PI: Luetzgendorf).
Additional NIRCam pure-parallel imaging overlapping both fields is being obtained by the PANORAMIC program 
\citep[PI: Williams, Program 2514,][]{williams25pan} and program 3990 \citep[PI: Morishita,][]{morishita25}.

%% file: nircam.tex
In this section, we detail the JADES NIRCam imaging.  We begin by 
describing the details of the four categories of NIRCam imaging:
Deep Prime, Medium Prime, Deep Parallel, and Medium Parallel.
We then discuss cross-cutting aspects: filter selection,
data quality, and a summary of depths.

\begin{table}[t]
\begin{center}
\begin{tabular}{|c|cc|cc|cc|}
\hline
& \multicolumn{2}{|c|}{Deep Prime}
& \multicolumn{2}{|c|}{Parallel 1210}
& \multicolumn{2}{|c|}{Parallel 1287} \\
Filter  
& $N_{\rm exp}$ & $t_{\rm exp}$ 
& $N_{\rm exp}$ & $t_{\rm exp}$ 
& $N_{\rm exp}$ & $t_{\rm exp}$  \\
\hline
F090W & 26 & 35.7 & 36 & 49.5 & 24 & 33.0 \\
F115W & 44 & 60.5 & 48 & 66.0 & 42 & 57.8 \\
F150W & 26 & 35.7 & 36 & 49.5 & 24 &  33.0 \\
F162M & \nodata & \nodata & \nodata & \nodata & 18 & 24.7 \\
F182M & \nodata & \nodata & \nodata & \nodata & 18 & 24.7 \\
F200W & 18 & 24.7 & 24 & 33.0 & 18 & 24.7  \\
\hline
F277W & 26 & 35.7 & 30 & 41.2 & 18 & 24.7  \\
F300M & \nodata & \nodata & \nodata & \nodata & 18 & 24.7 \\
F335M & 18 & 24.7 & 18 & 24.7 & 18 & 24.7  \\
F356W & 18 & 24.7 & 24 & 33.0 & 24 & 33.0 \\
F410M & 26 & 35.7 & 36 & 49.5 & 30 & 41.3 \\
F444W & 26 & 35.7 & 36 & 49.5 & 36 & 49.5  \\
\hline
Total & 114 & 156.7 & 144 & 198.0 & 144 & 198.0 \\
\hline
\end{tabular}
\end{center}
\caption{\label{tab:ncexp}Overview of the NIRCam Deep imaging in GOODS-S in different filters, 
listing the number of separate exposures and the total exposure time per
pointing, in ksec.  We note that some Deep prime pointings overlap, doubling the depth,
and that many Deep pointings also overlap with Medium, further increasing the depth.
}\end{table}

\begin{figure*}[ht]
\hspace*{\fill}
\includegraphics[width=\textwidth]{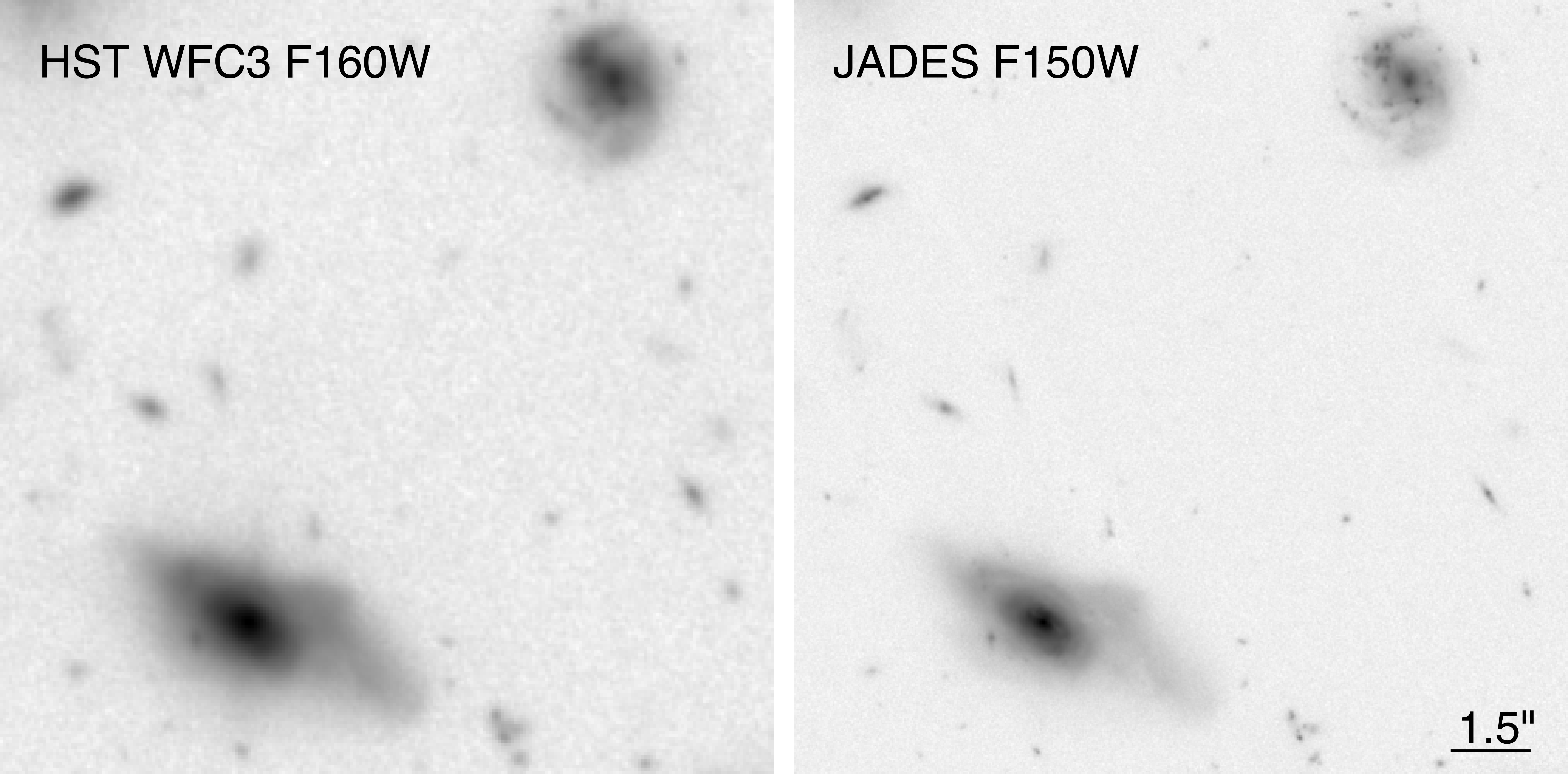}
\hspace*{\fill}
\caption{A small portion of the Hubble Ultra Deep Field, comparing HST and JWST imaging.  ({\it left}) The F160W image from HST WFC3, from the Hubble Legacy Field reduction  \protect\citep{illingworth16a,whitaker19}, with an exposure time of about 65 hrs.  This is the single deepest H-band image taken with HST.
({\it right}) The F150W JWST NIRCam image from the first year of JADES with an exposure time of about 10 hours.  The superior depth and image quality of JWST is clear.
}
\label{fig:HUDF_comparison}
\end{figure*}

\subsection{NIRCam Deep Prime}

In the heart of GOODS-S centered on the UDF, we observe 4 NIRCam
fields to image a $4.4'$ by $6.1'$ rectangular field.  A pair
of NIRCam pointings is offset by $\sim$$61.5''$ in V2 so as to
cover the NIRCam inter-module gap, and each pointing contains many
exposures including offsets in V3 to cover the shortwave chip gap.
This pattern is then repeated in the second pair of pointings,
offset in V3 leaving only a small overlap with the first pair.
We size this offset to the height of the long-wave (LW) footprint, as this is 
slightly smaller than the short-wave (SW) footprint.  We note that the mosaic
was laid out for V3~PA of 300$^\circ$ but observed at 298.56$^\circ$,
causing a small cosmetic deviation from a rectangular layout.

We use the DEEP8 readout pattern with 7 groups, yielding individual
exposure times of 1375 seconds.  DEEP8 was used to reduce the data
volume, but does yield somewhat worse recovery from cosmic rays
than a MEDIUM choice would have.  In total, each pointing utilizes
114 such exposures, each with a SW and LW filter choice.  Because
each pointing includes more than 2 days of exposure time, we must split the observations
into 3 visits.  The first two use 9-point subpixel dithers (and no
primary dither), each with 5 filter pairs, yielding 45 exposures.
These two visits are offset in V3 by $\sim$5$''$ to step over the SW chip
gap.  The third visit uses a mosaic of two pointings, each with
4-point subpixel dithers and 3 filter pairs, so 24 total exposures.
The mosaic is designed with a row overlap chosen to result in the
same V3 step and with the pointing chosen to result in the same
footprint as the first two visits.  Due to a mistake in correcting for 
a small PA shift, the third visit of pointing 2 is offset mildly, $3''$,
from the exact overlap.
All filters are included in
both 9-point subpixel dither visits and therefore most points in
the image are observed by at least 18 different NIRCam pixels (9
if the location falls in the SW chip gap in one of the visits).

These observations are all taken with MIRI in coordinated parallel,
the results of which will be described in \S\ref{sec:miri}.  Because
of this, in the first season, we selected the subpixel dither based on
F770W stepsizes.  After inspected these results, we concluded that a 
mildly larger stepsize would be better for background subtraction; 
we therefore changed to the F1500W dither pattern in year 2.

We utilize 9 filters in the Deep Prime pointings: F090W, F115W, F150W,
and F200W on the SW arm, and F277W, F335M, F356W, F410M, and F444W
on the LW arm.  The exposure times per filter are listed in Table \ref{tab:ncexp}.
It is important to note that about over a third of the total field is 
covered by two of the pointings, doubling the total exposure times.
Because of the importance of high-redshift dropouts, we slant the
SW exposure times toward the F115W and F150W filters, while on the LW
side we favor the longer filters where the zodiacal background is 
reducing the sensitivity.  A comparison of the imaging we obtained in the NIRCam Deep Prime with that from the HUDF HST imaging is shown in Fig. \ref{fig:HUDF_comparison}.

Although JADES was designed to be observed in one year, Cycle 1
scheduling constraints caused the program to be split.  We opted
to observe all 4 pointings in each year.  In year 1, all pointings
were observed with one of the 9-point dither visits, and two were
observed with the 4-point dithers.  This was repeated in year
2, completing the observing.  Unfortunately, this segmentation also
caused a delay in the spectroscopic followup of the Deep imaging.
However, it did create an opportunity to consider year-scale time
variation in this nearly 25 arcmin$^2$ deep field.

In total, this part of the JADES program was an investment of 229 hours
and resulted in 174 open-shutter hours of data, a utilization of 76\%.

\subsection{NIRCam Medium Prime}

\newcommand{\nod}{\nodata & \nodata}

\begin{table*}[t]
\begin{center}
\hspace*{-60pt}
\begin{tabular}{|c|cc|cc|cc|cc|}
\hline
& \multicolumn{2}{|c|}{w/NS-HST}
& \multicolumn{2}{|c|}{w/MIRI}
& \multicolumn{2}{|c|}{w/MIRI Replan$^a$}
& \multicolumn{2}{|c|}{w/NS-JWST} \\
Filter  
& $N_{\rm exp}$ & $t_{\rm exp}$ 
& $N_{\rm exp}$ & $t_{\rm exp}$ 
& $N_{\rm exp}$ & $t_{\rm exp}$ 
& $N_{\rm exp}$ & $t_{\rm exp}$ \\
\hline
 F070W &  \nod      &  \nod       &  \nod       &  6 &  5.03 \\ 
 F090W &  6 &  5.67 &  6 &  6.96  &  6 &  5.67  &  9 &  8.50 \\ 
 F115W &  6 &  6.96 & 12 & 11.34  & 12 & 11.34  & 12 & 10.05 \\ 
 F150W &  6 &  5.67 &  6 &  6.96  &  6 &  5.67  &  9 &  8.50 \\ 
 F200W &  6 &  5.67 &  6 &  5.67  &  6 &  5.03  &  9 &  7.54 \\ 
\hline
F277W &  6 &  5.67 &  6 &  5.67  &  6 &  5.03  &  9 &  8.50 \\ 
 F335M &  \nod      &  6 &  5.67  &  6 &  5.67  &  6 &  5.03 \\ 
 F356W &  6 &  5.67 &  6 &  5.67  &  6 &  5.67  &  9 &  7.54 \\ 
 F410M &  6 &  5.67 &  6 &  6.96  &  6 &  5.67  &  9 &  8.50 \\ 
 F444W &  6 &  6.96 &  6 &  6.96  &  6 &  5.67  & 12 & 10.05 \\ 
\hline
Total & 24 & 23.96 & 30 & 30.92  & 30 & 27.70  & 45 & 39.62 \\ 
\hline
\end{tabular}
\end{center}
\caption{\label{tab:ncexpGSmed}Overview of the NIRCam GOODS-S Medium imaging in different filters, 
listing the number of separate exposures and the total exposure time per
pointing, in ksec.  We note that some pointings overlap, increasing depth.
Pointing 25 of Medium/HST set lost half of the LW exposure time to illumination from a short circuit in NIRSpec.
$^a$Three pointings (observations 22, 219, and 223) with MIRI parallels were replanned and shortened to balance the time within the allocation.  In addition, observation 20 missed its F200W and F277W filters, which were replaced with the shorter exposure time as observations 220 and 222.
}
\end{table*}

\begin{table*}[t]
\begin{center}
\hspace*{-60pt}
\begin{tabular}{|c|cc|cc|cc|cc|}
\hline
& \multicolumn{2}{|c|}{w/NS-HST}
& \multicolumn{2}{|c|}{w/MIRI P1\&2}
& \multicolumn{2}{|c|}{w/MIRI P3}
& \multicolumn{2}{|c|}{w/NS-JWST} \\
Filter  
& $N_{\rm exp}$ & $t_{\rm exp}$ 
& $N_{\rm exp}$ & $t_{\rm exp}$ 
& $N_{\rm exp}$ & $t_{\rm exp}$ 
& $N_{\rm exp}$ & $t_{\rm exp}$ \\
\hline
 F070W &  \nod      &  \nod       &  \nod       &  6 &  5.67 \\ 
 F090W &  6 &  5.67 &  6 &  3.09  &  6 &  3.09  & 12 & 11.34 \\ 
 F115W & 12 & 11.34 & 12 &  6.18  &  6 &  3.74  & 12 & 11.34 \\ 
 F150W &  6 &  5.67 &  6 &  3.09  &  6 &  3.09  &  9 &  8.50 \\ 
 F200W &  6 &  5.67 &  6 &  3.09  &  6 &  3.09  &  6 &  5.67 \\ 
 \hline
 F277W &  6 &  5.67 &  6 &  3.09  &  6 &  3.09  &  9 &  8.50 \\ 
 F335M &  6 &  5.67 &  6 &  3.09  &  \nod       &  6 &  5.67 \\ 
 F356W &  6 &  5.67 &  6 &  3.09  &  6 &  3.09  &  6 &  5.67 \\ 
 F410M &  6 &  5.67 &  6 &  3.09  &  6 &  3.74  & 12 & 11.34 \\ 
 F444W &  6 &  5.67 &  6 &  3.09  &  6 &  3.09  & 12 & 11.34 \\ 
 \hline
 Total & 30 & 28.34 & 30 & 15.46  & 24 & 13.00  & 45 & 42.52 \\ 
\hline
\end{tabular}
\end{center}
\caption{\label{tab:ncexpGNmed}Overview of the NIRCam GOODS-N Medium imaging in different filters, 
listing the number of separate exposures and the total exposure time per
pointing, in ksec.  We note that some pointings overlap, increasing depth.
Pointing 3 (observation 3) with MIRI parallels is shorter than the other two to balance the time within the allocation.
Pointings 4-7 are in concert with NIRSpec Medium/HST, while pointings 8-11 are with Medium/JWST.
}
\end{table*}

To provide shallower flanking coverage and increase the area available
for NIRSpec MOS targeting, JADES includes 18 pointings in (mostly) regular mosaics
yielding contiguous coverage to medium depth.  7 of these are in
GOODS-N and 11 in GOODS-S.  Because of the footprint, 8 of the pointings
have MIRI in parallel, while 10 have NIRSpec MOS in parallel.
In most cases, the pointings are paired with a $62''$ 
step in V2
so as to fill a long rectangle covering the inter-module gap, with
some double coverage.  In detail, the step is chosen to match the
width of one SW chip, so that the pointings cover the V3-parallel
chip gap of their partner.

Cycle 1 scheduling constraints caused half of the Medium Prime time
in GOODS-S to be delayed until fall 2023.  We opted to observe in
2022 the six pointings of the mosaic that did not overlap the NIRCam
Deep Prime field, so that most of the prime area could be observed
in fall 2022, including the footprint of the deep MIRI parallels.
These pointings also provided NIRSpec MOS parallels on and around the UDF.
The GOODS-S mosaic was designed for a V3 PA of 308$^\circ$.
The small PA difference from the Deep Prime mosaic was included to 
make the program more easily schedulable.

For these six pointings, the filters are as in Deep Prime save for
omission of F335M.  Each of the 8 filters received 6 exposures,
falling on 6 different pixels.  The filter pair of F115W and F444W
received 1159 second exposures using DEEP8 and 6 groups.  The other
three filter pairings receive 945 second exposures; these use DEEP8
with 5 groups to reduce data volume.  Unfortunately, one pointing failed
due to shorts in the NIRSpec MSA; this one was repeated in fall 2023.
Another pointing had half of the LW imaging impacted by the shorts; we opted 
to accept this one and adjust a future pointing to compensate.

After this start, and in view of the ongoing scheduling of the NIRCam
parallel observations, we opted to make some minor adjustments.
We had originally planned for 19 Medium pointings, but we decided to 
remove one pointing in GOODS-S and compress certain exposure times in GOODS-N 
in order to add F335M back into 
filter set for 11 of the 12 remaining pointings, pairing with a second exposure of F115W. 
We also had to adjust times to balance the program within its allocation, due to small on-orbit alterations in the parallel observation timing model.
In GOODS-N, we accomplished this by decreasing the time in the 3 pointings with MIRI parallels; in one case we were unable to include F335M.
The detailed exposure times are presented in Table \ref{tab:ncexpGSmed} for GOODS-S and
Table \ref{tab:ncexpGNmed} for GOODS-N.
Most exposures use the DEEP8 readout mode, but the shorter ones use MEDIUM8.

Unfortunately, for the five second-year pointings in GOODS-S, the observatory had a series of guide star problems and a safing event during the observing period.  Only two of the five fields were observed.  Rather than wait a full year to return to form a mosaic at a single position angle, we opted to replan the remaining time to fill in specific holes in coverage around the flanks of the Deep Prime region.  As will be discussed in \S~\ref{sec:miri}, this resulted in a less favorable location of the MIRI parallels.

In particular, observation 24 and most of 20 were observed at the originally planned location.  Fortunately, these were the two most important to connecting to the first-year mosaic and making a wider mosaic around the Deep Prime region.  Observation 20 was interrupted by an observatory safing before the last filter pair, F200W and F277W.  We replanned to cover these two filters at a position angle 90 degrees rotated, using two pointings (220 and 222) to cover each of the two NIRCam modules from observation 20.  The second module was placed inside Deep Prime to increase the time-domain coverage of that field.

The remaining 3 pointings were replanned at PAs later in the observing window to address flaws in the coverage.  Observation 219 served to cover the module gap of observation 20 and a thin remaining wedge between Deep Prime and the year-1 Medium mosaic.  Observation 223 served to cover the module gap of observation 24 and increase time-domain coverage of the Deep Prime region.  Observation 22 was placed to cover a remaining portion of the Medium/HST spectroscopy.  All 3 of these are mildly shorter than observation 24 due to the need to balance the total time budget.

For the 7 pointings in  GOODS-N, we pack four of them tightly in V2 to cover the
intermodule gap.  The other three have no intermodule coverage, and we further have chosen these to have less exposure time, as the MIRI parallels are more flexible than the NIRSpec MOS parallels.  The
northern portion of the mosaic is therefore somewhat shallower, but able to cover more of our 
early NIRSpec Medium/HST data.  We decided that it was more important
to maintain a 6-fold dither than 
to insist on a filled footprint, expecting that this field will 
likely attract larger coverage in future Cycles.  The mosaic was
designed and observed at V3 PA of 241$^\circ$.

Regarding the dither strategy,
when operated with MIRI parallels, we use 3 dithers with the INTRAMODULEX
pattern and 2-point subdithers.  These were based on the MIRI F1800W PSF 
in GOODS-N and F2100W in GOODS-S, 
chosen to increase the dither step size for background subtraction.  This 
dither uses small steps at 45-degrees in V2 and V3, stepping over both
SW chip gaps in each.  Hence most points in the chip gaps receive 4 
SW exposures; only small overlaps from the cross in the middle generate 
only 2.  We remind that in most cases, the center of each arm is covered
by the other pointing in the pair.

When operated with NIRSpec parallels, the strategy is mildly different.
Here, we split the 6 exposures into two sets of 3.  Each triplet is
a different MSA design with largely independent targets, to be
described further in \S\ref{sec:nirspec}.  In each triplet, 
NIRSpec executes its 3-step nod along the slits of the MSA.
We note that these steps are roughly at $45^\circ$ on the NIRCam
pixel grid.  For the next triplet, we step the central pointing purely in
the V3 direction by an amount to cover the V2-parallel chip gap.
The V3-parallel gap is not covered within this pointing, but usually
is by the partner in the mosaic.

In summary, this part of JADES was an investment of 195 hours, 123 in GOODS-S
and 72 in GOODS-N, resulting in 131 open-shutter hours (with the
formally prime instrument), a utilization of 67\%.

\begin{figure*}[ht]
\includegraphics[width=\textwidth]{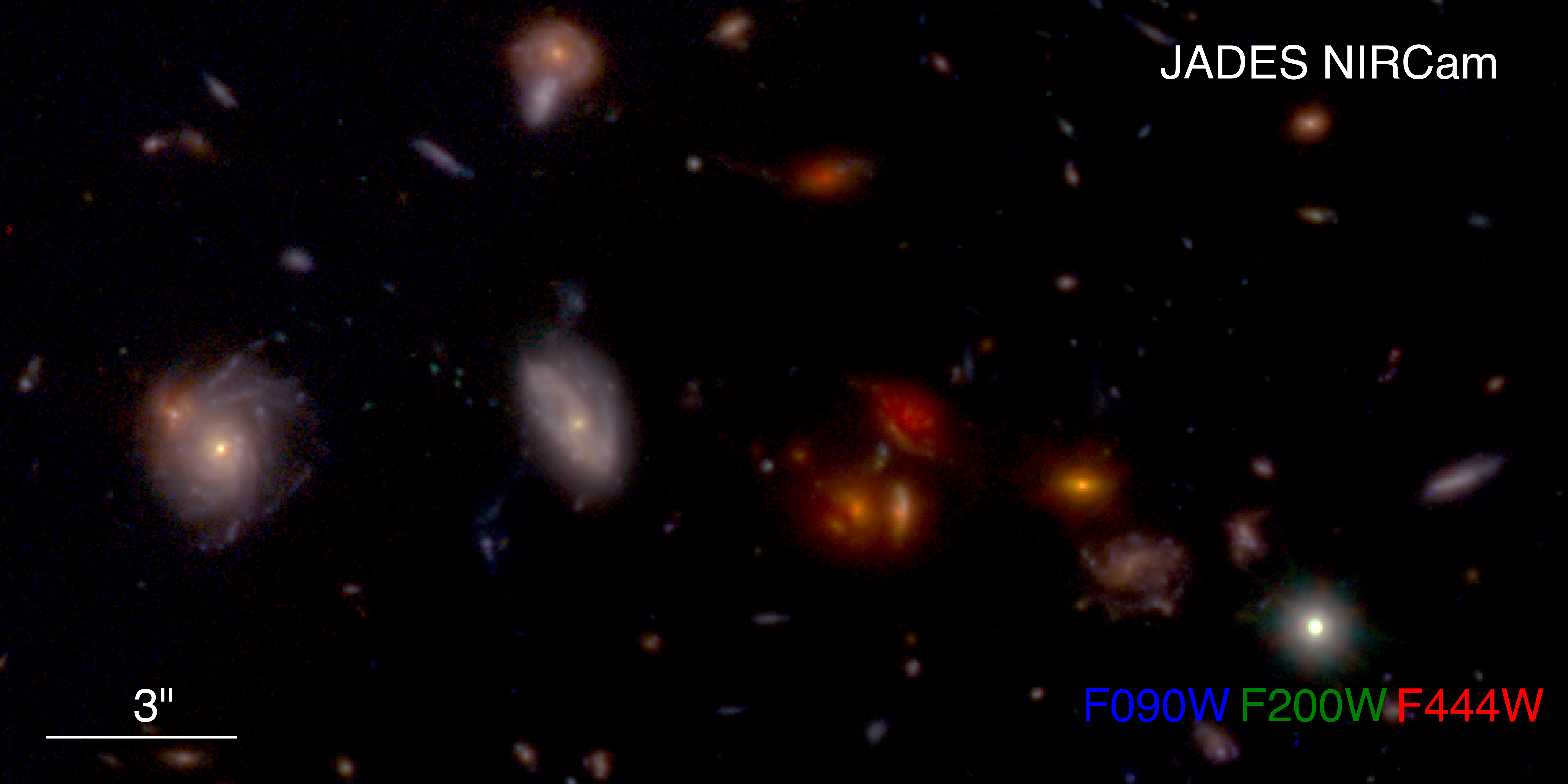} 
\caption{A small portion of JADES NIRCam GOODS-S imaging, combining F090W, F200W, and F444W filters in the Deep Parallel region of program 1210.  This image shows the great diversity of galaxies revealed in every JWST image, with a wide variety of colors and morphologies.  North is up.
}
\label{fig:cosmic_rose}
\end{figure*}

\subsection{NIRCam Deep Parallel}

In GOODS-S, JADES executes two long NIRSpec MOS pointings (\S\ref{sec:nirspec}),
each of 200 ks open-shutter spectroscopy.  Each of these is used to make
a long NIRCam parallel exposure, the location of which depends on the 
position angle and observing window, which in turn was subject to the
programmatic constraints of the NIRSpec targeting.

It turned out, fortuitously, that the first of these pointings, program 1210 (PI: L\"utzgendorf), was
at a position angle (V3 PA of 321$^\circ$) that caused the NIRCam
parallel to fall on top of the deep MIRI parallels produced by the
NIRCam Deep Prime program.  This very deep region then became the optimal location for the deep medium-band imaging and deep spectroscopy of the Cycle 2 program 3215 \citep[Program 3215; ][]{eisenstein23jof}.  This region has now been named the JADES Origins Field, with over a week of NIRCam observation on a single pointing.

The second pointing, program 1287 (PI: L\"utzgendorf), was required to be late in the observing window so that the
targets resulting from the analysis of the full NIRCam data set could be used.  
We were scheduled for V3 PA of 53$^\circ$, placing the NIRCam parallel northwest of the main mosaic.  The exact location was set by the location of the most interesting high-redshift candidates for the prime spectroscopy.

The first deep parallel use the same nine filters as the Deep Prime
program.  For the second deep parallel, we added three additional medium-band filters, based on the success of this mode in program 3215; this will be discussed in \S~\ref{sec:ncfilters}.  Each uses the same readout pattern as Deep Prime: 7 groups yielding
1375 second integrations.  Both use 144 such integrations 
for a total
of 55 hours of open-shutter imaging, indeed mildly deeper than a single
Deep Prime pointing (but without the overlaps of the mosaic).  The
exposure times per filter are shown in Table \ref{tab:ncexp}.

One limitation of these data is that NIRSpec only employs 9 dither locations, 
3 nod locations in each of 3 different MSA configurations.  We were careful
to arrange that each filter is observed at least twice at every location.
In detail, the NIRSpec exposures are just over twice as long as the 
NIRCam integration, so each nod position results in a pair of back-to-back
otherwise identical NIRCam integrations, 72 in all.  We note that the dither 
pattern is set by the geometry of the NIRSpec MSA and therefore not
tuned to the pixel scale of NIRCam; that said, the steps are not
commensurate with the NIRCam pixel scale and so the intra-pixel 
behavior is sampled. 

Another limitation is that the 3 MSA configurations in NIRSpec are
stepped only a short distance $\sim$0.8$''$, 
to limit the effects of
distortions across the MSA and thereby increase the ability to keep high-priority targets well centered in shutters.  This means that the data set does not
fill the SW chip gaps.  On the flip side, the data maximize depth
in the region that is covered. An example of the superb quality obtained in this program is shown in Figure \ref{fig:cosmic_rose}.

\subsection{NIRCam Medium Parallel}

JADES executes twelve medium-depth NIRSpec MOS pointings, 4 in GOODS-N
and 8 in GOODS-S, to be described in \S\ref{sec:nirspec}.  Each of these 
produces a NIRCam coordinated parallel observation, with a total of 
45 exposures.  The exposure times are given in Tables \ref{tab:ncexpGSmed} 
and \ref{tab:ncexpGNmed}.

As these parallel fields were at risk to fall off of the HST GOODS
and CANDELS imaging, depending on the final position angles, we opted
to include the F070W filter in addition to the nine filters used
in the Deep imaging.  This improves isolation of $z\sim5.5$ Lyman $\alpha$
dropouts.

As with the NIRSpec Deep program, these observations also use 9
closely spaced dither pointings, via 3 nod locations in each of 3
slit configurations.  The SW chip gaps are not covered.  We ensure
that each NIRCam filter is observed in at least 2 of the 3 slits
and hence at 6 dither locations.  2 of the 3 slits have all ten
filters observed; the remaining one is missing F070W, F200W, F335M,
and F356W.  Therefore, the area covered by all ten filters is
reduced by a tiny amount.

As the prime spectroscopy are located on the NIRCam Prime mosaic,
these Medium parallels in both GOODS-S and GOODS-N fall almost entirely
outside of the NIRCam Prime footprint, thereby providing a substantial
amount of additional area at medium depth.  In GOODS-N, 
we observed at V3 PA of 150.48$^\circ$ and 132.92$^\circ$, placing the new imaging
on the northeastern portion of the HST GOODS-N field.  
Because of the location of high-priority spectroscopic targets, 
these do not form a regular grid, but they are close enough to
map a sizable near-contiguous region.
By coincidence, the orientation of NIRCam in this Parallel imaging is almost
90$^\circ$ rotated from the Prime imaging, leaving an obvious pattern for
future observations to fill in the gap between the two.  Unfortunately, a guide-star acquisition failure caused Observation 8 to be skipped, requiring a replan (Obs 98) that was observed a few weeks later with 18$^\circ$ of rotation relative to the other three pointings.  We mention that although the NIRCam imaging in Observation 98 was successful, this observation suffered from problems in NIRSpec that resulted in its being reobserved as Observation 198 a year later.  The pure parallel opportunity for that visit was used by program 2514 \citep{williams25pan} to add additional medium-band filters to this pointing.

In GOODS-S, we spread the spectroscopic pointings over the full Medium
mosaic.  In the original plan, these would all have been late in the first
observing window, but given the scheduling delay, most were moved to the second year.
The first field (observation 1) was observed on January 12--13, 2023, at a 
V3 PA of 56.17$^\circ$.  
The second field (observation 5) was observed on October 19, 2023, at a V3 PA of 319.87$^\circ$.
The remaining six were observed December 12--19, 2023, at a V3 PA of 30.23$^\circ$.
We chose to use observation 5 on a NIRSpec region that placed the NIRCam parallel next to the JADES Origins Field in order to make this a wider footprint for future spectroscopy.

\subsection{Comments on Filter Selection}
\label{sec:ncfilters}

Early on in the design of JADES, we recognized that the strong
increase in zodiacal emission longward of 4.5~\mic\ would increase the background in F444W, such that F410M could be competitive despite its narrower bandpass.
We therefore opted
to include both filters to increase the resolution of the 
spectral energy distribution.  This is particularly important
because of the strong rest-optical emission lines expected in 
some high-redshift galaxies, as indicated by Spitzer imaging and
now confirmed with early JWST spectroscopy.  The spacing of 
the H$\alpha$+[NII] and H$\beta$+[OIII] complexes is such that 
only one can fall into F444W at any given redshift, and including
F410M means that the comparison can separate the line from 
continuum emission at most redshifts (with ambiguity if the 
lines fall on the shoulder of the filter curves).  This separation
is important for stellar population modeling, as we want to 
measure the rest-optical continuum color relative to the ultraviolet.

As we studied this, we concluded that F335M and F356W offered a 
similar opportunity and that the likelihood of strong H$\alpha$ or 
H$\beta$+[OIII]
emission recommended splitting this exposure time as well.
This has been borne out in practice: we have found the abilty to isolate
strong emission lines at 3--5~$\mu$m to be very useful and interesting.
In addition to the lines themselves, the extra spectral resolution helps
to isolate the Balmer jump at high redshift and to measure the rest-optical
continuum.  

Where possible, we observe F277W somewhat longer than F356W because
of the extra coverage from F335M.
We keep F410M and F444W longer to compensate for the larger zodiacal 
background.
Finally, we chose to slant the exposure time toward making F115W deeper, 
emphasizing the selection of $z>9$ candidates.

In program 3215, we found that increasing the number of medium bands 
continued to provide very desirable access to emission lines and improved
photometric redshifts.  Therefore we opted to echo much of this in our 
final deep parallel field, program 1287.  By adding F300M, F182M, and F162M,
we provide a single medium-band filter in each of F277W, F200W, and F150W.
As before, this allows one to isolate a strong line from the continuum in 
each of these filters with a minimum of extra exposure time.  We chose F182M and F300W
as the most sensitive in each of these wide bands.  We preferred F162M to F140M to keep consistent with program 3215 and because F162M and F182M provide improved Ly$\alpha$ dropout localization at redshifts above 12.

In summary, we have found the NIRCam coverage in the nine base bands
to be highly effective for photometric redshifts.  For example, at
$z\approx 7$, one observes the Lyman $\alpha$ drop in F090W and
H$\beta$+[OIII] in the longest bands.  We further have found that the 
combination of medium and wide bands, particularly at the redder filters, 
is extremely helpful, as many of the higher redshift galaxies have strong emission lines.

\subsection{Data Quality Caveats}
\label{sec:ncliens}

A detailed description of the data reduction and performance
is presented elsewhere \citep{rieke23r,deugenio2025}.  Here we describe some issues that 
we have already seen and that might be of interest to other users.

Our observing was split into many separate visits rather than long
campaigns, and we encountered substantial persistence at the start
of some visits, left over from the immediately preceding program.
These signals last for several hours and will require detailed
modeling and/or masking.  The effect is more severe on the SW chips A3, B3, and B4;
the other SW and both LW chips are much milder. 

Most dramatically, in the Deep program, observations 7 and 10, 
were observed immediately after observations of the bright Trapezium
nebula, leaving substantial diffuse emission over portions of SW
chips A3, B3, and A4.  This is most severe in the F090W filter (the
first used), but there are faint traces in the next filter (F115W),
3.5 hours later.

We stress that the persistence is not simply coming from bright stars
in the previous images, but the change in the diffuse background 
illumination level from the previous program.  This greatly increases
the affected area, particularly in A3 and B4.  The decay time for A3 
is particularly long.  Observation 4 in 1181 shows a similar morphology
of persistence incurred by a change in the background level relative 
to the previous program.

We also see persistence in A3 in visits following wavefront sensing
operations, creating a moderate-size ($\sim$100 pixel across) hexagonal image.  While such sensing is common, the reference star is planned
to fall on a consistent part of the chip.  We suspect that the frequent
re-occurence of this signal could allow that area of the chip to be
particularly well characterized, but we have not attempted this.  1180 observations 15 and 27, 1181 observation 9, 1286 observation 4, and 1287 visit 2 are 
affected by this.
The chip used for wave-front sensing has since been changed, avoiding this issue entirely.

We have also seen a case (1180 observation 18) in which a bright
star happened to fall on the NIRCam field
during the preceding MIRI observation, even though NIRCam was not
being used in parallel.  NIRCam is typically left open to the sky 
when not used.  One can see the imprint of the whole MIRI
dither pattern, as well as the trail when the telescope slewed away
from the field.

Of course, these regions do also incur persistence from brighter sources
in our own observations.  For instance, in 1181 observation 2, there
is a bright star in the bad portion of A3 that creates a recurrent glow
in all 6 dither locations.  And 1286 observation 8 contains the persistent glow of a bright star in observation 6.

Mitigating these persistent signals is particularly vexing when one is
using a small-angle dither, as a pixel may never find a blank-sky location
away from a larger galaxy.  We therefore increased our sub-pixel
dither selection in the later portion of our observations.  However, this is
not always possible when observing jointly with NIRSpec.  We caution that
we have not yet considered the effect of persistence on our photometry,
but in the worst regions of these chips, we believe this should be studied.
By construction, the dither pattern is repeated between filters, so the 
persistence from the first filter will affect the next.

Like many NIRCam observations, we occasionally have noticeable illumination
of the SW detector through an off-axis stray-light path, producing the 
so-called ``wisps''.  These affect F200W and F150W most strongly.  While
these signals are known to modulate in amplitude due to the brightness of
stars in the source region on the sky, we have found that the exact morphology
of the pattern also varies within our program.  We are still analyzing this,
but hope that a low-dimensional set of templates will suffice to remove them.

We have found it very helpful to visualize the calibrated exposures 
of a dither sequence in animations, fixed in pixel coordinates so that
the true objects move and the detector artifacts stay still.

We have found it easier to disentagle persistence from wisps when F090W or 
F115W are observed first in a visit, so that the persistence has decayed 
away before the wisp-affected F150W and F200W bands are observed.

We now turn to rarer problems. 
The second half of one pointing (observation 30) of the Medium Prime mosaic in GOODS-S had
to be skipped due to an on-board issue unrelated to our program.
There was not enough time in the observing window (constrained by
the spectroscopic coordinated parallel) to try again.  Fortunately,
this pointing is at
the edge of the mosaic.  The first half of this same pointing had its
LW data badly contaminated by a glowing short circuit in the NIRSpec MSA.  
We repeated this entire pointing in October 2023 to complete the 6-dither coverage; this means that the SW data in this pointing is 50\% longer than normal.

One half of another pointing (observation 25, first 4 dither sequences) was also affected by the NIRSpec glowing short \citep{rawle22}.  In this case,  because the location of the field was favorable to access in a different way in the year 2 observing, we opted not to repeat this imaging location but instead combine the NIRSpec re-do with another pointing.

In both of these cases, we found that the LW data were badly affected.
The background was roughly doubled, but further there are many
patterns of concentric rings, with spacing depending on wavelength,
likely due to some diffractive pattern from where the light from
NIRSpec has bounced off of the tertiary mirror.  As the short circuit 
is apparently not particularly hot, the effect on SW is much less and we
believe this data is usable.  There are a handful of faintly detected rings
in F200W, chips A1, A2, A4, and B4.  

Next, in the 3 deep visits of 1210, we find an enigmatic set of
arcsecond-scale blobs near an edge of B3.  These are bright in the
second visit, but detectable in the other two as well as in the 3215 imaging
in year 2.  They are therefore
not due to persistence, and we are confident they are not astrophysical
as they do not appear in 1180 images of the same region.  They also appear in other nearby pointings at this PA (observation 1 of program 5997) despite an arcminute change in pointing.  The morphology
is very different from a wisp, and importantly they appear in both SW and LW images. 

Some of the 1210 exposures have a plume of what appears to be scattered
light in a corner of B4.  We hypothesize that this may be due to a
bright star striking the chip mask, as the signal changes slightly
between the three visits of 1210, which move only at the arcsecond
level, suggesting a well-focused source.
However, we also see it in observations 27, 28, 29, and 30 of 1180, 
so it seems that the cause is not particularly rare.

Observation 219 of program 1180 suffers from a spray of light in A4 that we believe to be a bright star glinting off something just above the focal plane, as the amplitude is modulated by small dithers.  There is also weaker stray light in A2 that we think is connected to this.

Observation 98 of program 1181 displays stray light on multiple chips, 
probably from a bright star on the B4 chip.   This stray light appears 
in both modules and both SW and LW, and it is not morphologically connected to 
the bright star.  Clearly there is some complicated reflection path.  Fortunately, much of the image area is usable.  We see a weaker version of this in observation 11 of program 1181.
  
\input{nircam_depth}

%% file: nircam_depth.tex
\subsection{NIRCam Imaging Depth}

\begin{table*}[t]
    \noindent\begin{center}\hspace*{-1.5in}
    \begin{tabular}{|c|ccc|ccc|ccc|}
    \hline
    & \multicolumn{3}{|c|}{Deepest (9 arcmin$^2$)} 
    & \multicolumn{3}{|c|}{Deep (33 arcmin$^2$)} &
    \multicolumn{3}{|c|}{Medium (167 arcmin$^2$)} \\
    Filter & Time & \multicolumn{2}{c|}{PS Depth (AB)} & Time & \multicolumn{2}{c|}{PS Depth (AB)} & Time & \multicolumn{2}{c|}{PS Depth (AB)}  \\
           & (ks) & ETC & On-sky & (ks) & ETC & On-sky & (ks) & ETC & On-sky \\
    \hline
F070W$^a$ &\nodata&\nodata & \nodata&\nodata  &\nodata &  \nodata&    7.1 &  28.14 & 28.20 \\
F090W &  71.0 &  29.59 & 29.70 &   44.1  &   29.33 & 29.42 &   11.2 &  28.58 & 28.70 \\
F115W & 122.9 &  29.99 & 29.98 &   69.2  &   29.68 & 29.67 &   14.4 &  28.83 & 28.87 \\
F150W &  72.5 &  29.96 & 29.90 &   43.7  &   29.69 & 29.61 &    9.8 &  28.87 & 28.85 \\
F162M$^b$ &\nodata&\nodata & \nodata&   24.7  &   28.87 & 28.90 &\nodata &\nodata & \nodata\\
F182M$^b$ &\nodata&\nodata & \nodata&   24.7  &   29.10 & 29.12 &\nodata &\nodata & \nodata\\
F200W &  52.6 &  29.93 & 29.86 &   31.8  &   29.66 & 29.60 &    8.0 &  28.91 & 28.87 \\
\hline
F277W &  75.4 &  30.23 & 30.21 &   41.0  &   29.90 & 29.90 &    9.5 &  29.10 & 29.17 \\
F300M$^b$ &\nodata&\nodata & \nodata&   24.7  &   29.13 & 29.10 &\nodata &\nodata & \nodata\\
F335M$^c$ &  51.7 &  29.58 & 29.58 &   28.6  &   29.26 & 29.24 &    7.1 &  28.51 & 28.61 \\
F356W &  51.7 &  30.02 & 30.04 &   34.2  &   29.80 & 29.80 &    8.0 &  29.01 & 29.08 \\
F410M &  71.7 &  29.40 & 29.34 &   47.5  &   29.18 & 29.17 &   11.1 &  28.39 & 28.50 \\
F444W &  73.7 &  29.61 & 29.69 &   49.7  &   29.40 & 29.44 &   11.7 &  28.61 & 28.72 \\
    \hline
    \end{tabular}
    \end{center}
    \caption{A Summary of Average Exposure Times and Depths in the NIRCam Deep and Medium Surveys.  The 6 Deep pointings cover a total footprint of $\sim$42 arcmin$^2$, with $\sim$9 of these being double-covered and marked as Deepest.
The 30 Medium pointings cover a total additional footprint of $\sim$167 arcmin$^2$ exclusive of the Deep footprint.  Within each, we compute the average exposure time per filter, including the contribution of Medium to the deeper regions.  
We then present the 10-$\sigma$ $0.2''$ diameter aperture depth, including point source aperture correction, for these average exposure times, based on the JWST Exposure Time Calculator and separately from the variance of random apertures measured on the mosaics.
    $^a$The F070W filter is used in only a subset of Medium pointings, covering an area of $\sim$82 arcmin$^2$.  
    $^b$The F162M, F182M, and F300M filter is used only in the single 1287 Deep pointing, covering an area of $\sim$9 arcmin$^2$.  
    $^c$The F335M filter is not used in some Medium pointings, so that this filter covers a Medium footprint of $\sim$140 arcmin$^2$. In all 3 cases, we quote the average exposure time and depth in the smaller region.
    }
    \label{tab:ncsum}
\end{table*}

As shown in Figures \ref{fig:south} and \ref{fig:north}, the NIRCam pointings often overlap, so that the survey
is deeper than what appears in Table \ref{tab:ncexp}--\ref{tab:ncexpGNmed}.  Further, the depth
varies because of the geometry of these overlaps.  To provide a useful summary of the NIRCam program, in Table \ref{tab:ncsum}
we divide the footprint into 3 disjoint regions: Deepest, Deep, and
Medium.  Deepest refers to the area where two or more of the Deep Prime
pointings overlap.  Deep refers to the remainder of the area of the Deep
Prime and Parallel fields.  Medium refers to the rest of the area, including
both GOODS-S and GOODS-N.  In each case, 
the quoted areas are based on exposure time maps where we have required both F090W and F356W data and have required at least 2400 sec in each filter to remove the outermost trim with only 2-3 dithered exposures. 
Only exposures from the Cycle 1 JADES program (PIDs 1180, 1181, 1210, 1286, 1287) are included.
Masked pixels are included in the area estimate.
We then compute the mean
exposure time in each filter in each footprint.
As this method uses the total exposure time maps, Medium pointings overlapping
Deep regions are included in the Deep tally.
We remind that the exposure times do vary within the 
regions; for example, the northern portion of GOODS-N is shallower by a factor of 2-3 in exposure time than the bulk of the Medium region, while in some places multiple Medium Parallel pointings overlap to approach Deep depth.
Nevertheless, these summaries are reasonable averages for forecasts and contextual 
comparisons.  

We then convert the representative exposure times to anticipated 10-$\sigma$
depths for background-limited $0.2''$ diameter apertures with point source
aperture corrections.  We do this in two ways.  The first is to use the JWST Exposure Time Calculator.  Here we assume the integration length of the Deep Prime program and observations in October.  Spotchecking for other observations in JADES suggests that variations from this would be at the few percent level in depth.  The second method is to measure the variance in many apertures in blank regions of
the mosaic that have similar exposure time to the mean of the tier, scaling those apertures by the square root of the ratio of each exposure time to the reference value.   
We find that these two estimates are very similar, which is both a successful validation of the JWST Exposure Time Calculator and a finding that JADES is achieving the hoped-for scaling of depth with exposure time, i.e., there is not a floor from systematic imaging errors.

We remind that the Deep area is all in GOODS-S, while the medium area is split approximately evenly between GOODS-S and GOODS-N.  The average
depths in GOODS-N is mildly shallower than in GOODS-S, but 
the two fields are sufficiently similar that we do not separate them for
this summary.  

\begin{figure}
    \centering
    \includegraphics[width=0.45\textwidth]{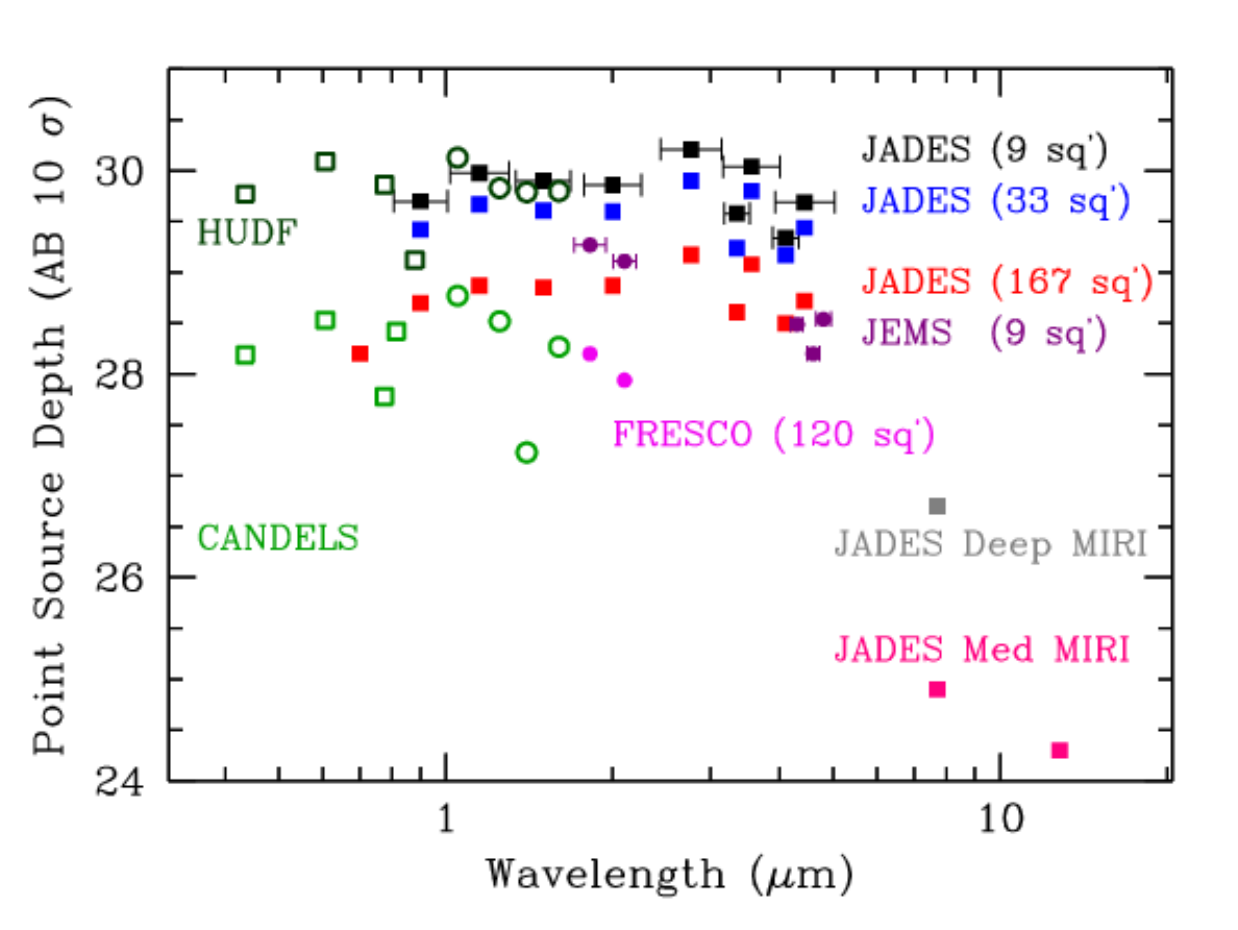}
    \caption{Depth versus Wavelength for JADES and other data sets.  Black, blue, and red solid squares show the 10-$\sigma$ point source depth for JADES Deepest, Deep, and Medium NIRCam data, using the variation in 0.2$''$ blank-sky apertures.  Purple and magenta circles show the depth of the JEMS and FRESCO medium-band imaging.  Horizontal ranges show the filter widths.  The JADES Deep and Medium MIRI depths are shown with grey and pink points on the right.  The 0.2$''$ aperture is mildly too small for the HST WFC3 image quality, causing these estimates of depth to be too optimistic by 0.1--0.2 mag compared to larger apertures.  Comparison is shown (dark and light green) to the depths measured in the same method from the Hubble Legacy Field \protect\citep{whitaker19} mosaics, separating the single HUDF pointing from the broader CANDELS region.  The HUDF ACS (WFC3) footprint is 11 (4.7) square arcminutes, comparable to the area of the Deepest JADES data.  The CANDELS area exceeds the JADES Medium area.  
    JADES improves over even the HUDF in area and spatial resolution, and at wavelengths longward of 1.6~$\mu$m, the gains in depth and resolution are immense.  
    }
    \label{fig:depth}
\end{figure}

In Figure \ref{fig:depth}, we show a 
visualization of the estimated depth of the JADES, JEMS, and FRESCO NIRCam and 
JADES MIRI imaging, along with the estimates for HST ACS and WFC3 imaging \citep{whitaker19} in the HUDF and CANDELS fields that JADES overlaps. 
We have measured the JEMS and FRESCO depths in the same manner as that of JADES, using our own reductions of these data.
One sees that the JADES imaging is comparable in depth in the deepest HUDF data, which covered only a single HST pointing.
One also sees that the ratio of optical ACS to infrared JADES depth is 
less favorable in the broader CANDELS region compared to that of the HUDF.

%% file: nirspec.tex
As introduced in \S \ref{sec:survey}, the JADES NIRSpec MOS data fall into 
3 tiers: Deep, Medium/JWST, and Medium/HST.
All 3 tiers use the low-resolution prism as well as several gratings, with
grating spectra available for most of the prism targets.  
Table \ref{tab:ns} provides a summary of disperser configurations and exposure times for each tier. This section presents some of the features common to all tiers before describing each tier in turn.

Each of the NIRSpec gratings are used with a matching long-pass filter to prevent overlap of first order spectra by  higher orders. For the band 1 G140M grating there are two available filters; F070LP and F100LP. F100LP blocks all light below 1\,$\um$, thereby ensuring no second order overlap within the nominal spectral range up to 1.8\,$\um$. F070LP allows through light at $>0.7\,\um$, so it has the advantage of enabling observation in the range 0.7 to 1.0\,$\um$. This corresponds to the wavelength of the Lyman-$\alpha$ transition at redshifts of 4.8 to 7.2, bridging the epoch of the end of cosmic reionization \citep{robertson22a}. Therefore we chose to use the F070LP filter for all the JADES G140M spectroscopy. We accept that there will be second order overlap from the sky, increasing the sky background at $>1.4\,\um$, and from galaxies at redshifts below 7 that have flux in this wavelength region. However, much like the overlapping grating spectra allowed by our MSA configurations described in the following subsection, this increased continuum flux has little impact on our emission line measurements from the G140M spectra.

\subsection{MSA Configuration Design}
{\label{sec:msa}}

JADES designed its NIRSpec multi-object observations using the
tool eMPT \citep{bonaventura23a} that provided key features beyond what was
found in the baseline tools for MSA design.  In particular,
eMPT allowed us to:
\begin{enumerate}
\item Constrain the NIRSpec pointings to a rigid mosaic of NIRCam fields, once the position angle is specified.
\item Impose a detailed prioritization system for our targets and have complete control over the order in which each class of targets is attempted placed on the MSA at a given pointing.
\item Optimize repeated observations across multiple overlapping MSA designs to maximize exposure time on the highest priority targets.
\item Identify and eliminate beforehand targets having contaminating objects falling within their (nodded) slitlets.
\item Avoid the use of shutters leading to prism spectra truncated by the NIRSpec detector gap or contaminated by the spectra of failed open shutters.
\item Enable overlap of the grating spectra (except for some high priority targets whose grating spectra are protected from overlap) to maximize the gratings multiplexing, while keeping those of the prism distinct.
\item Open additional blank-sky shutters that disperse onto unused detector real estate to support master background subtraction. 
\end{enumerate}
To do this, eMPT contains the full NIRSpec model of the astrometric
distortions and multi-shutter geometry and constraints, from which
it can accurately predict how given astrometric positions will fall
onto shutters, whether those shutters are available to use, and where the
resulting spectra will fall on the detectors (and thereby whether
they will overlap).  The code thereby revealed the detailed outcome
of each target, with which one can proceed to accept targets in
complex priority orders.  After this process determines which
shutters were to be opened, the final optimal pointings and matching MSA masks were imported into the standard APT/MPT workflow for further execution.

The MSA shutters are on a rigid grid, and the opaque regions between the
shutters block enough light from compact sources that one typically
chooses to retain only the fraction of targets that are sufficiently
well centered in their shutters. Not all shutters function properly with 22 failed open that always disperse light onto the detector and 17.5\% of the unvignetted shutters that are permanently closed \citep{boker23}. Our MSA masks require a 3-shutter-high slitlet for nodding and background subtraction. Locations where one can open a 3-shutter-high slitlet are therefore limited by this MSA operability leading to the concept of a `viable slitlet' map. In all the JADES tiers we perform MSA reconfigurations with small (always $<$10 and mostly $<$1 arcsec) offsets where we attempt to obtain spectra of at least the highest priority targets in multiple configurations. Simulations have shown that these offsets need to be kept small to maximize the overlap of the viable slitlet maps and to avoid the astrometric distortion at the NIRSpec MSA plane that cause some objects to become insufficiently centered. For most tiers of JADES, the main constraint on these offsets is to ensure maximal coverage of the highest priority targets in multiple MSA configurations.

Because of these constraints, one needs a very high target density
in order to achieve a high multiplex of assigned targets.  JADES
typically supplies at least 200 targets per square arcminute, 
yielding about 150 assigned targets on the prism designs, 
corresponding to an average assignment rate of only 9\%.
An obvious consequence of this low average rate is that one does not
want to serve the rarer higher-value targets in this limited way.
We therefore developed a detailed prioritization of the targets, largely by
redshift and flux (see \S~\ref{sec:targets}). The eMPT allows us to 
assign slits in a greedy order, assigning higher priorities first.
While this slightly decreases the total multiplex, it yields much
higher assignment rates on high value targets.

NIRSpec prism spectra have relatively short traces, allowing multiple columns
of non-overlapping spectra. For a given prism MSA
design, there is a matching grating MSA design that is nearly identical.
Importantly, we keep nearly all of the same shutters open for the grating configuration so that most of our galaxies have information at multiple spectral resolutions.  The
longer traces of the gratings may overlap, as may the zero-order emission, but the emission lines are sparse
in these spectra. Emission lines can be associated to their parent object
in multiple ways: the location along the three shutter tall $1.5''$ slit, the prism
spectrum, where the lines appear unoverlapped, and the wavelength
ratio of multiple detections. The dispersed continua of the
typical faint targets is below the detector noise in the grating spectra and hence the continua of the overlapping
spectra do not substantially increase the noise.  
For the highest priority
targets and for the infrequent brighter targets, we do close some
shutters to avoid overlap; so a small fraction of objects
are observed only with the prism.  

\begin{table*}[tb!]
\begin{center}
\begin{tabular}{|l|l|ccc|ccccc|}
\hline
& & & & & \multicolumn{5}{c|}{Exposure Times (ksec)} \\
Subsurvey & Program & \# Fields & Subpointings & \# Targets & Prism & G140M & G235M & G395M & G395H \\
\hline
GOODS-S Deep/HST    & 1210 & 1 & 3 & 253 & 100 & 25  & 25  & 25  & 25  \\
GOODS-S Deep/JWST   & 1287 & 1 & 3 & 235 & 100  & 25  & 25  & 25  & 25  \\
\hline
GOODS-S Medium/HST  & 1180 & 5$^{ab}$ & 1 & 677 & 3.8 & 3.1 & 3.1 & 3.1 & --- \\
GOODS-S Redo Obs 134    & 1180 & 1$^c$ & 2 & 185 & 7.5 & 6.2 & 6.2 & 6.2 & --- \\
GOODS-S Redo Obs 135    & 1180 & 1$^c$ & 3 & 185 & 11.3 & 9.3 & 9.3 & 9.3 & --- \\
GOODS-S Redo Obs 136    & 1180 & 1$^c$ & 2 & 169 & 7.5 & 6.2 & 6.2 & 6.2 & --- \\
GOODS-S Medium/JWST & 1286 & 8 & 3 & 1490 & 8.0 & 8.0 & 9.3 & 9.3 & 8.0 \\
\hline
GOODS-N Medium/HST  & 1181 & 8$^a$ & 1 & 853 & 6.2 & 3.1 & 3.1 & 3.1 & --- \\
GOODS-N Medium/JWST & 1181 & 4 & 3 & 950  & 9.3 & 9.3 & 9.3 & 9.3 & 9.3 \\
\hline
\end{tabular}
\end{center}
\caption{\label{tab:ns}Summary of the NIRSpec MOS Observations.
For each program, we list the number of separate MSA fields, as well as the exposure time per disperser in
kiloseconds.  Each field consists of 1 to 3 sub-pointings, each with two nearly identical MSA designs: one for the 
prism and a second for the grating; the latter closes a few shutters to protect certain high-priority spectra from overlap.
The quoted times are summed over the sub-pointings, but not all targets can be placed on all sub-pointings. The number of unique targets in each subsurvey is listed; a small number of targets are repeated between rows.
The long-pass filter choices for the gratings are F070LP, F170LP, F290LP, and F290LP, respectively.
$^a$Each Medium/HST pointing is split into two distinct MSA locations, separated by the
primary dither step across the NIRCam SW chip gaps.  As the larger dither causes us to typically observe distinct
galaxies, we account each pointing as two fields with one sub-pointing.  When the slit registrations are favorable, we do reobserve high-priority  
targets to double the exposure times on these.
$^b$Twelve MSA locations were planned, but only 5 were completed due to instrument problems.  
$^c$These 7 re-dos were organized to be similar to three Medium/JWST locations, one with 3 sub-positions and two with 2.
}
\end{table*}

\subsection{Target Prioritization}
\label{sec:targets}

Here we summarize the design of the JADES spectroscopic target selection process.
The main criteria for placing galaxies into priority classes are their redshifts and fluxes. Redshifts are estimated via photometric redshift algorithms and/or Lyman-break color selection. The highest redshifts are prioritized both because they are rare and because one of the main scientific goals of JADES is to understand the earliest phase of galaxy evolution. Galaxies with higher fluxes (in continuum or predicted emission lines, depending on the category) are prioritized since higher S/N spectra allow a wider range of science investigations.

The details of the priority classes depend on the tier (Deep, Medium/JWST or Medium/HST) and whether the targeting is based only on HST imaging or on the JADES imaging. Deeper spectroscopic observations have lower flux limits for similar classes so that the achieved S/N will be similar for the different tiers. 

The highest priority class contains relatively bright galaxies at the highest redshifts; $z>8.5$ for Deep and Medium/JWST and $z>5.7$ for Medium/HST program.  After these, we prioritize fainter galaxies at the same redshift and then progressively lower redshift bins, favoring the brighter galaxies.  Through this, we aim to build up a statistical sample between redshift $1.5<z<5.7$ in the lower priority classes over the tiers, with the shallower tiers contributing to the bright end and deeper tiers contributing to the faint end. In addition to these classes we also include a small fraction of galaxies identified as special in other data, e.g. with ALMA, Chandra, Lyman-$\alpha$ emitters selected with the ESO VLT/MUSE instrument. 

We also prioritize a few bright ($H_{AB}<23.5$) moderate-redshift ($z>1.5$) galaxies, enabling the
collection of exquisite infrared spectra from objects around cosmic noon.
After these, we prioritize in photometric redshift bins, favoring
the rarer brighter examples.  Galaxies with photometric redshifts
below 1.5 are used as a low priority filler sample; nevertheless,
the large number of these targets yields a substantial observed set.
Further details on the target prioritization, including variations per tier, are provided in the JADES data release papers \citep{bunker23r,deugenio2025}. 

\begin{figure}[t]
\includegraphics[width=0.45\textwidth]{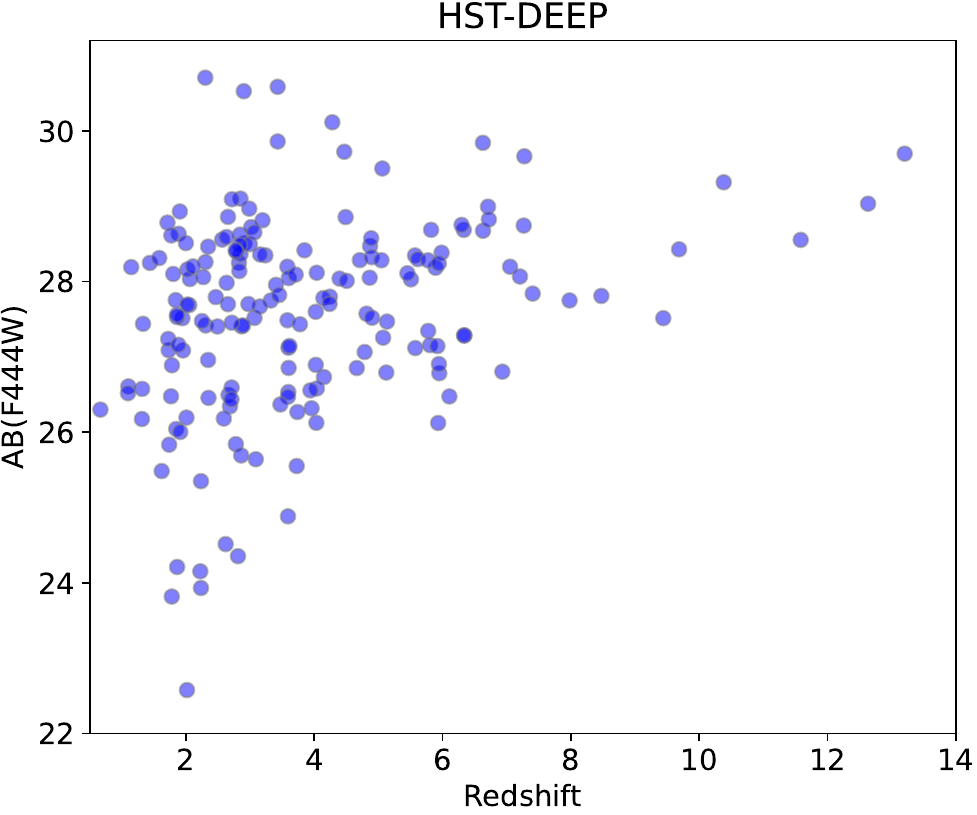}
\includegraphics[width=0.45\textwidth]{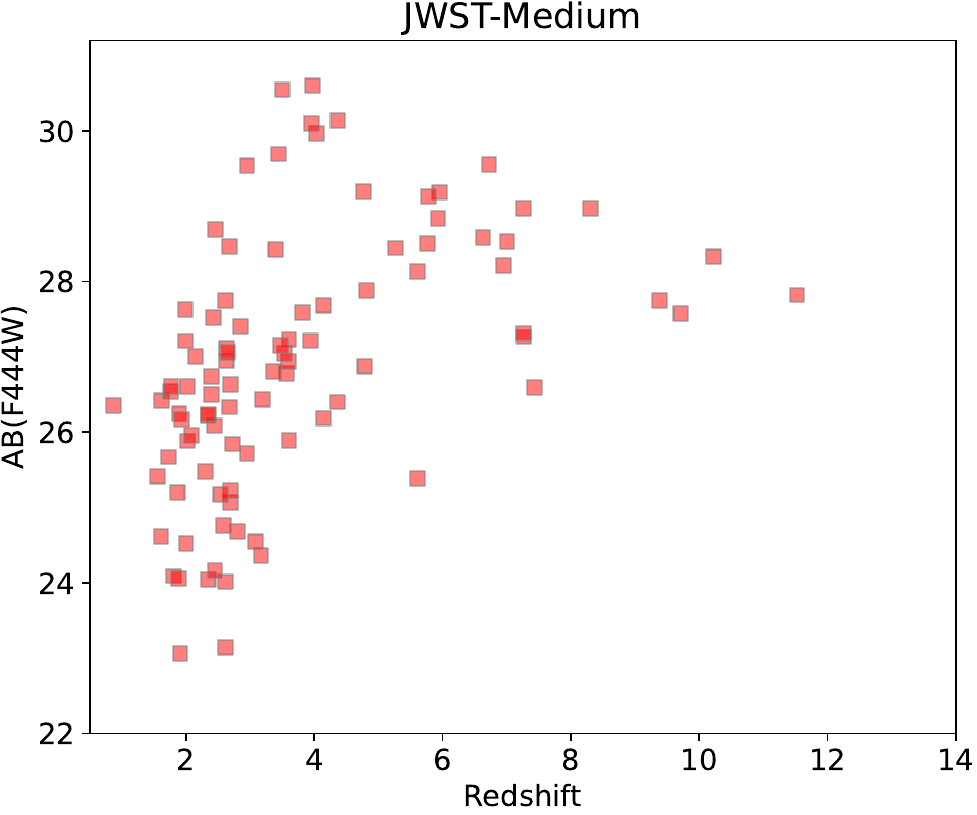}
\caption{The observed distribution of F444W AB magnitude versus spectroscopic redshift for the Deep/HST pointing (top) and the first Medium/JWST pointing (bottom).  The target selection has successfully weighted toward higher redshift galaxies, resulting in a more even redshift distribution.
}
\label{fig:ns_z_mag}
\end{figure}

The distribution of spectroscopic redshift and F444W magnitudes are shown in Figure \ref{fig:ns_z_mag}.  One sees that NIRSpec is recovering redshifts to extremely faint flux levels, particularly in the Deep pointing.

\subsection{MSA Target Acquisition}

Successful use of NIRSpec MOS depends critically on high-quality astrometry.
Astrometric distortions in the target coordinates tend to perturb targets away from the
centers of their shutters, lowering performance.  More insidiously,
NIRSpec MSA Target Acquisition relies on a few bright compact sources
to align the MSA to the desired location on the sky, so astrometric
errors on those few objects can cause the entire MSA to be misaligned.

JADES was fortunate to be able to utilize recent reductions of HST
imaging in the GOODS fields that had been aligned to the Gaia DR2 reference
frame (G. Brammer priv. comm., \citealt{GaiaMission, GaiaDR2}).  This alleviated concerns of distortions or mismatches between faint targets and bright acquisition sources.  

However, the GOODS-S and GOODS-N fields, being intentionally placed
in regions of very low stellar density, do present a severe deficit
of stars suitable for target acquisition.  Instead, we had to use
the HST imaging to identify compact galaxies, using the longest
exposure time for the acquisition image to reach down to 24-27 mag. 
All the JADES target acquisitions so far were successful, although the increased centroiding error for such extended targets may somewhat have reduced
the alignment accuracy of the NIRSpec observations.  
In later spectroscopy, when NIRCam data was available, we used its imaging
to select roughly circular compact galaxies and stars, using the multi-band near-IR photometry to more confidently estimate the fluxes in the NIRSpec CLEAR and F140X target acquisition filters and to enforce isolation criteria.  All candidates were visually inspected to reject a portion of objects that were lopsided, flattened, or insufficiently isolated; we aimed to be conservative in providing compact sources.

\subsection{NIRSpec Deep Spectroscopy}

JADES features two long NIRSpec multi-object observations, each of
200 ksec total exposure time and both located in GOODS-S on the UDF and mostly inside the
NIRCam Deep Prime footprint.  The first pointing was designed to
be observed early and be targeted using HST GOODS and CANDELS
imaging, supplemented with other pre-JADES data.  The second deep
pointing was observed at the end of the program, using
targets from the full JADES imaging.  We refer to these two pointings
as Deep/HST and Deep/JWST, respectively.

The NIRSpec Deep/HST observations occurred over three visits between 21 and 25 October 2022 at a V3PA of 321$^\circ$. 
As described above, these observations were intended to be targeted solely on pre-JWST data. Shortly before the scheduled visits, NIRSpec suffered some shorts on the MSA that required us to replan the MSA configurations for these observations. The NIRCam Deep Prime data were taken at the start of October, and an early reduction of the multi-band images and photometry catalogs were available. This enabled us to include some JWST-selected targets, at the expense of lower priority HST-selected targets, in the NIRSpec observations. We also re-prioritised the catalog using the NIRCam data to improve high-redshift photometric redshift accuracy, and to homogenize the selection in the lower priority classes with respect to the Deep/JWST to be based on the F444W filter. We note that the pointings of the observations were not changed, only the choice of which shutters to open. Two of the additional target galaxies with photometric redshifts from the NIRCam imaging \citep{robertson23} were spectroscopically confirmed in Deep/HST to lie at the highest redshifts known of 12.6 and 13.2 \citep{curtis-lake23}.

The NIRSpec Deep/JWST observations occurred over three visits on January 10--12, 2024 at a V3PA of 53$^\circ$.  While we has originally expected for this pointing to be in the Deep Prime region, we found that the most attractive high-redshift candidates in JADES imaging were in and around the JADES Origins Field, with the very deep program 1210 and 3215 imaging.  We therefore shifted the pointing.  Unfortunately, the second of the three visits was skipped due to a telescope guide star acquisition failure; this visit was observed a year later in January 12--13, 2025 with no changes from the original design.
 
The NIRSpec Deep spectroscopy utilizes five dispersers:
the prism and four gratings: G140M/F070LP, G235M/F170LP, G395M/F290LP,
and G395H/F290LP.  The prism is observed for 100 ksec, and the gratings
for 25 ksec each.  

Each pointing uses slitlets of 3 shutters, with the 3-point nod.  To
provide additional pixel diversity and some dithering in the
spectral direction, we design 3 sub-pointings for each pointing, typically separated by 3-5 shutters in the dispersion direction and 1-2 shutters in the spatial direction such that the optimal common coverage of the highest priority targets in all three dithered pointings is achieved.
This results in the target light from a given wavelength for the majority of sources appearing
in up to 9 pixel locations.   
As described earlier, there are actually 2 MSA designs for each sub-pointing (6 in total) because we use a separate configuration for the prism relative to that of the gratings.  
Each integration uses NRSIRS2 readout with 19
groups, yielding 1400 second apiece.  We conduct two integrations
per exposure.  For the gratings, this means that each nod location
is visited only once for 2 consecutive integrations.  For the prism,
the telescope repeats the nodding 4 times, with 2 integrations per time.

As discussed in \S~\ref{sec:msa}, it is inevitable that not all targets
can be placed on all three sub-pointings, even though this is our preference for high-priority targets.  
We accept this and fill in some targets with only 1 or 2 sub-pointings.  
Using a smaller dither step increases the ability to repeat targets.  For lower priority targets, we prefer to change targets between sub-pointings, so as to maximize the number of distinct objects, as described in \citet{bunker23r}.

This part of JADES is an investment of 145 hours and results in
111 prime open-shutter hours, a utilization of 76\%.  The exceptional line-flux
sensitivity achieved as a function of wavelength is shown in 
Figure \ref{fig:ns_depth}.

\subsection{NIRSpec Medium/JWST Spectroscopy}

JADES includes 12 medium-depth pointings that are targeted
from JWST imaging but otherwise scaled down versions of the
Deep pointings.  Relative to Deep, the prism is
scaled down more than the gratings, reflecting the goal of studying
the galaxy spectra in more detail.
Exposure times are listed in Table \ref{tab:ns}.
The original plan was 8665 sec in each of the 5 dispersion modes.
However, small on-orbit changes in the timing model for parallel observations 
caused us to make small adjustments in the exposure times to be more efficient
with the NIRCam parallels and fit into the program allocation.

As with Deep, the observing uses 3-point nods with 3-shutter slits
at each of 3 sub-pointing locations, for a total of 9 pixel dither
locations.  Each of these exposures is a single integration, using NRSIRS2 readout 
with 12 or 14 groups.
As for the Deep pointings, some targets can only be placed on one
or two sub-pointings, resulting in proportionally lower exposure time.

Four of the twelve Medium/JWST pointings are placed in GOODS-N and eight are in GOODS-S that has wider JADES NIRCam imaging. Three of the GOODS-N pointings were observed between April 30 and May 5, 2023 at V3 PA 150.48$^\circ$, covering a large fraction of the NIRCam Medium Prime mosaic.  The fourth was delayed by an observatory failure to acquire guide stars and was observed at V3 PA 132.93$^\circ$ on May 27, 2023.

Unfortunately, this final JWST/Medium observation in GOODS-N was affected by a short circuit in the NIRSpec MSA for some configurations, which when triggered produced a glow of light that flooded the detectors, ruining the NIRSpec data \citep{rawle22}. Half of the configurations were affected by these shorts (two of three prism and one of three grating). The other three configurations did not address the susceptible column of the MSA and therefore did not have the glow.  We re-observed the missing configurations in May 2024.  The NIRCam coordinated parallel imaging data show no impact of the short.

For GOODS-S, four pointings are in the NIRCam Deep Prime mosaic, with the other four on the Medium Prime mosaic. Most of the GOODS-S Medium/JWST was delayed until Cycle 2, save one pointing (observation 1) observed on January 12 and 13, 2023.  Observation 5 was observed October 19, 2023, and the other six in December 2023.  Pointing locations were designed to maximize coverage of bright high-redshift galaxy candidates from the NIRCam data.

\begin{figure}[t]
\includegraphics[width=0.45\textwidth]{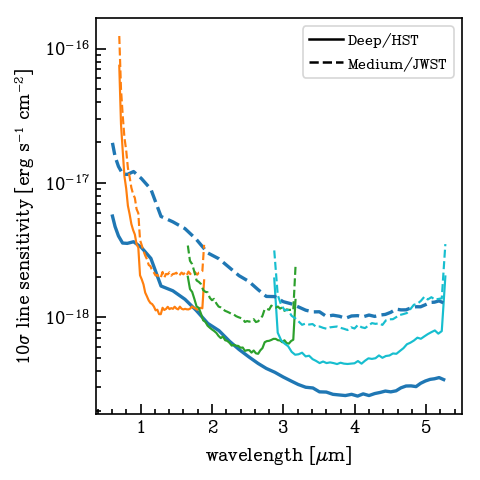}
\caption{The unresolved line detection limit as a function of wavelength for two JADES NIRSpec tiers.  We assume a well-centered point source and use a $10\sigma$ detection threshold.
The blue long lines are the prism, while the orange, green, and cyan lines are the three $R=1000$ gratings.  The solid lines show the depth in the Deep/HST pointing, while the dashed lines are a representative Medium/JWST pointing.
For an unresolved line, the G395H grating has a similar line detection limit to G395M plotted here.
}
\label{fig:ns_depth}
\end{figure}

This part of JADES is an investment of 222 hours, 77 in GOODS-N and 145 in GOODS-S, resulting in
146 prime open-shutter hours, a utilization of 66\%.
As before, the line detection sensitivity as a function of wavelength is shown in 
Figure \ref{fig:ns_depth}.

\subsection{NIRSpec Medium/HST Spectroscopy}

The final tier is the shallowest and results from ``parallel''
observations during ten of the NIRCam Medium Prime fields, four in GOODS-N and six in GOODS-S.  We
remind that our naming convention is following the instrument that
is driving our science design; in all coordinated parallels including NIRSpec MOS, NIRSpec is formally the
prime instrument.
The HST name refers to the fact that in the survey design, the JWST
imaging was not yet available and hence the targeting was from HST data.
We describe explicitly the few cases we could use JWST-based targets,
due to interruptions in the program.

As the imaging program requires a large $7''$ offset to step over the V2-parallel SW chip gap,
we opt to split each parallel opportunity into two largely distinct sets of
targets.  Offsets of this size when combined with the NIRSpec astrometric
distortion at the MSA plane would otherwise cause many targets to become
poorly registered within their slits. Each pointing in the mosaic was given 
a small $1''$ freedom of motion to optimize the slit centration of a few highest priority targets. 

For each of these 20 target sets, we use 4 dispersers, omitting the G395H
higher resolution grating.  Exposure times are listed in Table \ref{tab:ns}.
GOODS-S was observed first, and here we used NRSIRS2 readout with 17 groups
for the prism and 14 groups for the gratings, each observed once at each of
3 nod locations. Based on the first tranche of data in GOODS-S, we replanned 
GOODS-N so that the prism was observed twice with 14 groups at each pointing, 
to increase the S/N.

As this program results from the parallels of a reasonably tightly
packed imaging mosaic, these NIRSpec MOS fields overlap substantially.
However, any given object in the footprint of a single MSA pointing
will often be poorly centered in its possible shutter, leading to
many targets being rejected due to the low expected throughput.
Collisions of prism spectral traces block many other targets.
Therefore, several returns to a given area can be supported without
much duplication.  We do allow our targets to be observed twice (and our highest priority targets up to four times), if the slit registration is favorable. 

Our observations of the GOODS-S portion of this
mosaic (observations 25--30 of program 1180) were affected by two short circuits in 
the NIRSpec MSA, similar to those described in the previous section. Some configurations did not use these
columns of the MSA and therefore did not have the glow.
The NIRCam coordinated parallel imaging data was unaffected in most cases, but the first half of observations 25 and 30 suffer from a short so bright that the illumination even reached NIRCam, as described in \S~\ref{sec:ncliens}.  

Only 4 of the 12 target sets were successfully observed without a bright short glow, and even in these cases
there is some persistence that affects the prism exposure in a small portion of 
the image.  These are MSA configurations 2, 4, 6, and 10, which are the second 
half of observations 25, 26, 27, and 29, respectively.
For a 5th target set (configuration 5, the first half of observation 26), the
grating exposures are unaffected but the prism exposures were flooded by the 
short.  These were re-observed in October 2023, completing this configuration.

One of the target sets was not observed due to an unrelated telescope issue
(configuration 12, second half of observation 30).  As the first half of observation 30
was badly affected by shorts, we performed a complete repeat of this pointing as observation 136
in October 2023 at the original position angle, but with the same target selection as Medium/JWST.  We allowed a small translation from the original mosaic to optimize
the MSA for high-priority targets.

The other 5 target sets were re-observed on January 27 \& 28, 2023, without new NIRCam parallels.  Because the new position angle was already going to require a complete re-plan, we opted to collect the 5 single-nod designs into two pointings with smaller dithers, akin to the Medium/JWST program.  Observation 134 has two sub-pointings; observation 135 has three.  For these, we use the same NIRCam-based target selection that was used for the observation 1 of Medium/JWST in GOODS-S.

\subsection{Data Quality Caveats}

The NIRSpec MOS observations have been processed by using a pipeline developed by the ESA NIRSpec Science Operations Team and the NIRSpec GTO Team. The major steps of the data processing are described in \citet{bunker23r} and \citet{deugenio2025}, while a detailed description of the pipeline and its performance will be reported in a forthcoming paper (Scholtz, in prep.). Here we discuss the main issues encountered during the data processing and analysis phase.  

As mentioned in the previous sections, program 1180 year 1 was affected by short circuits in the NIRSpec MSA that produced bright glows of artificial light, ruining most of the exposures. In particular, 76 out of 132 exposures are ruined by such a bright glow. The other exposures did not suffer from this problem, but we found some persistence for the prism exposures that contaminate the spectra of $\sim10\%$ targets in the MSA design.  In some cases, the signal of the persistence is as high as the sky background emission.  

By inspecting the count rate maps before background subtraction, we noticed that some  shutters dedicated to the targets did not open \citep{rawle22}. We excluded these temporarily failed shutters from the data processing workflow. In each pointing, we found, on average, that $\sim$1\% of the targets in the MSA masks are affected by this issue. In most cases, only one of the 3-slitlet shutters was unexpectedly closed, but there are the same targets in which two or even all three shutters did not open during the observations. In these cases, the noise of the final products increases as the total exposure time is reduced.

Although the target selection was optimized to adopt the 3-point nod strategy for the background subtraction process, some background shutters were contaminated by either background or foreground source. We have thus exploited either HST or NIRCam images to identify automatically contaminated shutters and excluded them in the background subtraction steps of the pipeline. This increases the noise of the background subtraction image but avoids a possible over-subtraction of background emission that could alter the final spectra of the targets.

%% file: miri.tex
\subsection{MIRI Deep Parallel}

\begin{figure}[t]
\includegraphics[width=0.45\textwidth]{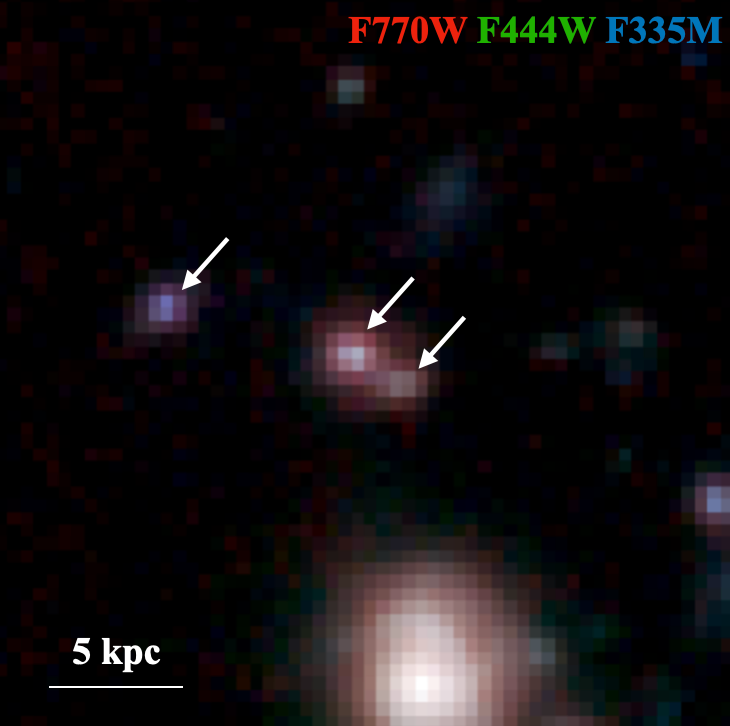}
\caption{RGB (F770W, F444W, F335M) image of a trio of $z\sim6$ photometric redshift candidates detected in the GOODS-S JADES Deep/MIRI parallel (24-25.6 AB with $\mathrm{SNR}\sim20-80$ in F770W).  All three are likely emission line galaxies, with H$\alpha$ in F444W. The leftmost candidate additionally has significant [OIII]5007 emission in the F335M medium band filter, resulting in a more purple color.  The F770W probes the rest-frame $1\mu$m emission, a powerful constraint on the properties of older stellar populations.}
\label{fig:miri}
\end{figure}

\begin{table*}[t]
\begin{center}
\hspace*{-60pt}
\begin{tabular}{|l|cc|ccc|ccc|}
\hline
Subsurvey & Number & Area & \multicolumn{3}{c|}{Exposure Times (Ksec)} & \multicolumn{3}{c|}{10$\sigma$ AB Mag Limit} \\
& Pointings & $\square'$ &  F770W & F1280W & F1500W &  F770W & F1280W & F1500W \\
\hline
GOODS-S Deep                    &  4 & 10 & 155 &  \nodata &\nodata & 27.5 &  \nodata &  \nodata \\
GOODS-S Medium Obs 24           &  1 & 3 &  5.6     & 10.9    &  13.5 &  25.6     & 24.6   &  24.2 \\
GOODS-S Medium Obs 20           &  1 & 3 &  \nodata & 13.5    &  10.9 &  \nodata  & 24.8   &  24.2 \\
GOODS-S Medium Obs 220, 222     &  2 & 6 &  \nodata & \nodata &  5.0  &  \nodata  & \nodata &  23.9 \\
GOODS-S Medium Obs 22, 219, 223 &  3 & 9 &  5.0     & 10.9    &  10.9 &  25.6     & 24.7   &  24.3 \\
GOODS-N Medium            &  3$^a$   &  9 &  6.0    &  9.0  &  \nodata &  25.7    & 24.6   &  \nodata\\
\hline
\end{tabular}
\end{center}
\caption{\label{tab:miri} Summary of MIRI Observations.
The number of pointings, area in arcmin$^2$, exposure times per
filter, and the final AB magnitude detection threshold per filter for a 10-$\sigma$ point source. 
The aperture-corrected depths are estimated from the variance in blank-sky apertures of 0.42$^{\prime\prime}$, 0.42$^{\prime\prime}$, and 0.49$^{\prime\prime}$ diameter in F770W, F1280W, and F1500W, respectively, chosen to include 65\% of the encircled energy.
These are about 0.2 mag deeper than the JWST Exposure Time Calculator in F770W and 0.3 mag deeper in F1280W and F1500W.
Only Obs 20 of the Medium/MIRI program overlaps the Deep/MIRI pointings; this observation omitted F770W to avoid redundancy.  The mild overlap of Observation 222 and 223 as well as the mild overlap of two of the GOODS-N pointings has been ignored in the area computation.
$^a$One of the GOODS-N pointings is mildly shallower, with 
6.6 ks of exposure in 
F1280W.
}
\end{table*}

The NIRCam Deep Prime program creates very deep MIRI parallels in
GOODS-S, totaling 43.1 hrs of open-shutter time in each of the four
fields.  The fields overlap only slightly, so that the deep area
is about 10 arcmin$^2$. 

The fields were designed to overlap the NIRCam Medium
Parallel mosaic, but it turned out that the NIRCam 
Deep/HST Parallel field substantially overlaps as well.  This 
allows these MIRI images to provide a very deep look at the
high-redshift universe.  We note that while the MIRI fields
are not on the UDF itself, they do fall near the center
of the Chandra Deep Field South.

We decided to focus this time to the study of rest-frame
1--2~\mic\ imaging of galaxies at redshifts above 3.  For
this, we selected the F770W filter as the most promising
compromise between the rising background to the red and the
lever arm relative to the deeper F410M and F444W data.  An example of robust F770W detections of a trio of $z\sim6$ photometric redshift candidates in the GOODS-S Deep/MIRI footprint can be seen in Figure~\ref{fig:miri}. 
F770W is somewhat less sensitive in AB magnitude than F560W, but it more than
doubles the logarithmic wavelength gap relative to 4.4~\mic,
allowing it to better detect power-law deviations in the 
slope of the near-IR SED.  The longer band is also less likely
to suffer from rest-optical emission line contamination of
the continuum light measurement; H$\alpha$ falls in F560W
at $6.6<z<8.4$, whereas F770W probes the rest-frame $I$-band at these redshifts 
\citep{2025arXiv250602099H}.

The MIRI data must of course follow the NIRCam exposure times and
dither pattern.  To keep the data volume within the allowed rate, we use SLOWR1 readout mode with 57 groups to yield
a single integration of 1361 seconds per exposure.  Each pointing
then has 114 of these exposures, taken at 22 different dither
points, for a total of 155.2 ksec.  Within each visit, the MIRI data is taken
cycling through 9 or 4 subpixel dither locations.  The resulting 10$\sigma$ point source sensitivity is 27.5 AB (Table \ref{tab:miri}).

\subsection{MIRI Medium Parallel}

In addition to the deep data, we conduct MIRI parallels with eight
of the NIRCam Medium Prime pointings.  Five of these are in GOODS-S
and three in GOODS-N.  

As these data are considerably shallower and yet only mildly more 
area, we opt to focus on the science of intermediate-redshift
galaxies ($z\sim3-5$), where we can place strong constraints on the stellar emission SED, such as the regime of the contribution from TP-AGB stars, robustly identify the rising continuum associated with AGN, and look for unusual SEDs.  To accomplish these goals, we take moderately deep exposures in F770W and then use most of the time in the F1280W and F1500W filters, which gain in sensitivity over the WISE W3 band by a factor of $\sim1000$.  These filters present a notable increase in wavelength reach relative to F770W, while getting to sufficient depth to study intermediate redshift galaxies.  We could not use the reddest filters, as these would require the high-data-rate FAST readout. 
In GOODS-N, 2/3 of the MIRI pointings fall off of the planned NIRCam coverage, but are covered by CANDELS.

The layout in GOODS-S was originally designed to place the MIRI parallels largely on the Medium Prime NIRCam imaging.  However, interruptions of the observing forced a replan, with only observations 20 and 24 falling at their original location.  Observations 22, 219, and 223 are therefore rotated in PA, placing their MIRI parallels north-west of the JADES Prime mosaic.  The MIRI data from observations 219 and 223 substantially overlap NIRCam imaging from programs 1286, 1287, 2514, and 3990; those from observation 22 currently have no NIRCam coverage.  In addition, because observation 20 was cut short, the two re-observation pointings (220 and 222) taken to replace the missing NIRCam filter have short F1500W MIRI imaging, again with a PA that places the MIRI field to the northwest.  Only the data of 222 overlap current NIRCam data.  The failed observation 23 also produces a small amount of F1500W imaging, not otherwise mentioned; future users of these exposures should be careful that the telescope may have been drifting without a guide star.

The exposure times are listed in Table \ref{tab:miri}.  In all pointings,
the data set uses 6 dither locations.  The dither locations have 3 close
pairs with relatively long strides between them, and hence there
is a relatively large boundary region that has only 2 or 4 exposures.
However, since MIRI is Nyquist sampled, this was considered acceptable.
We always use the SLOWR1 readout, so as to reduce the data rate.
F770W uses 1 integration per readout; F1280W and F1500W use 2 or 3 to avoid saturation on the background.
In GOODS-S, where the available exposure times are longer, we observe F770W once per dither location and return to F1280W and F1500W two times each.  In GOODS-N, our exposures are shorter and we do not have any coverage from the Deep MIRI parallels; we therefore opt to include two exposures of F770W and 2-3 of F1280W per dither location.  

\subsection{Data Quality Caveats}

The reduction of the MIRI data is described in \citet{Alberts24MIRI}.
We have not encountered any substantial concerns with the MIRI data acquired, but note three minors issues.  
First, optimal background subtraction through the combination of multiple, spatially independent pointings requires the observations be taken in a short time interval due to the time-varying background.  Some of the GS medium observations due to guide star failures were re-observed outside of this time interval.  This had a minor impact on the quality of the background subtraction.   
Second, we did find that the subpixel dither pattern used in the first Deep data, based on the F770W PSF size, was smaller than we would have preferred to use for the generation of sky flats.  We find this can be mitigated by generating sky flats from roughly contemporaneous exposures over multiple pointings. Nevertheless, we have increased the dither steps in later observations.  
Last, astrometry corrections for MIRI pointings without overlapping NIRCam or HST coverage from CHARGE were done using matches to Gaia and have larger astrometric uncertainties due to the reliance on a small number of stars.

%% file: preparations.tex
Like many JWST observing programs, the JADES team engaged in substantial
preparations for the data set.  We were particularly driven by the tight
time scale, likely at most 6 weeks, to provide targets for multi-object
spectroscopic follow-up from the NIRCam and MIRI imaging.  This central
goal of the program requires image reduction, mosaicing, source detection,
source photometry, photometric redshift generation, target selection,
and MSA design to be ready to run in quick order.  We also sought
to prepare for analysis of the spectroscopy, most obviously for the 
data reduction, extraction, and spectral analysis, as 
clearly the spectroscopic results would be desired to feed into the
processing of the second field to be observed.  

Part of this preparation was inherent in the needs of the instrument
teams to support a wider range of commissioning and early science 
observations.  Most obviously, we rely on the effort to develop 
exposure-level reduction software such as NCDhas written by K.~Misselt and the pre-processing pipeline developed by the ESA NIRSpec Science Operations Team \citep{Birkmann2022}, which subsequently became the basis for the STScI stage 1 and 2 pipelines.
To build and validate these tools pre-launch required creation of detailed codes to simulate instrument data: {Guitarra\footnote{{\url{https://github.com/cnaw/guitarra}}} for NIRCam and the NIRSpec Instrument Performance Simulator (IPS; \citealt{dorner16}).

For NIRCam, we used Montage\footnote{\url{http://www.ascl.net/1010.036}}  to combine the individual exposures into a mosaic. In order to identify outlier pixels (such as cosmic rays that have not been picked up by NCDhas), we first construct a median-based mosaic, which we then projected this median-based mosaic back to the individual exposures. We identified and masked outlier pixels that are 3$\sigma$ outliers. In a final step, we constructed the mean-based mosaic and fully propagated the errors. 

An important additional aspect of JADES preparation included optimizing our instrument configurations, integration times, and area to maximize our key science goals (e.g. bottom panel of Figure 4). To this end, we developed JAGUAR, a novel phenomenological model of galaxy evolution out to $z\sim15$ \citep[JAdes extraGalactic Ultradeep Artificial Realizations;][]{Williams2018}\footnote{\url{https://fenrir.as.arizona.edu/jwstmock}}, incorporating known galaxy abundances, flux, color, and morphology relations across redshift. A key utility of JAGUAR includes both mock SEDs and full-resolution spectra, which we generated using BEAGLE \citep{chevallard16} based on self-consistent models of stellar radiation and its transfer through the interstellar and intergalactic medium. Beyond survey design, JAGUAR also enabled simulation of realistic galaxy fields using mock imaging tools like Guitarra, mock NIRSpec spectra and optimizing our MSA design procedures.

With these tools, JADES then performed data challenges, simulating mock
galaxy fields down to individual ramps and then reducing the data to 
make high-level products.  For NIRCam, this included mosaicing, 
object detection, photometry, and photometric redshifting.  For
NIRSpec, mock target lists were used to build MSA designs with 
the eMPT code \citep{bonaventura23a}, and simulated spectra \citep{chevallard18} were run through
reduction and extraction to develop tools for redshift and line
flux estimation \citep{giardino19}.
A key advantage of these data challenges was to define data models
for the interfaces between segments of the analysis.  They also
allowed us to generate test suites to validate each segment.

JADES conducted Data Challenge 1 (DC1) to initiate this process.
The NIRCam component of DC1 concentrated on a single visit of Proposal 1180 (Observation 7, Visit 2), which is part of the NIRCam Deep Prime pointings. The DC1 simulations used 9 ramps
with 7 DEEP8 groups for all filters used in the original JADES deep survey design
(F090W, F115W, F150W, F200W, F277W, F356W, F410M and F444W). 
The source catalog used for DC1 was derived solely from the JAGUAR mock catalog, 
where we assigned random positions in R.A and declination for the selected galaxies but in such a way to maintain the same surface density of galaxies as the JAGUAR parent sample.

The NIRSpec component of DC1 was based on the NIRSpec Deep/JWST program. 200 point source galaxies with realistic JAGUAR spectra were simulated in the prism and medium resolution grating dispersers in each of the three Deep/JWST pointings. The data were processed with the NIRSpec IPS Pipeline Software} (NIPS; \citealt{dorner16}). The resulting output was simulated, flux-calibrated, combined spectra for 370 galaxies, used to validate our choice of integration times for these faint targets \citep{giardino19}.

Following improvements in the code base, we 
then conducted Data Challenge 2 (DC2) as a set of more comprehensive exercises.
DC2 covered about 2/3 of the area of the JADES survey in GOODS-S ($\sim$80 arcmin$^2$) and included Deep and Medium NIRCam pointings, using the same setup as the planned observations (e.g., exposure time, read out mode, dither positions) as recovered from the APT file. In contrast to DC1, the DC2 simulations included the
field-of-view distortions, particularly important to test the ability to make astrometrically correct mosaics in sky coordinates and accounting for pixel sub-sampling when constructing these mosaics. Cosmic-ray hits were also added to the individual ramps.
The DC2 sample used a combination of the CANDELS \citep{grogin11a,koekemoer11a} catalog with S\'ersic parameters estimated by
\citet{vanderWel12} and objects from JAGUAR, which also
provides S\'ersic parameters. The latter enabled including
objects beyond the apparent magnitude limit and redshift cutoff of the
HST data.  In this process, we used the positions and shapes of all observed galaxies and supplemented these with mock galaxies where a fraction of the latter were included until the counts in apparent magnitudes and redshifts were as close as possible to the JAGUAR magnitude-redshift distribution.
In addition to the galaxy catalog, we also created a separate set of images with stellar sources, that were used to verify
the photometric calibration procedure.
In both Data
Challenges a few objects with abnormal colors and Population III galaxies \citep{zackrisson11} were added to test the
efficacy of algorithms being tailored to detect outliers. For DC2, HST
fluxes were also calculated (though no HST images created) which were used
to estimate the number of low-redshift contaminants in the photometric
redshift calculation. For the JAGUAR galaxies in DC2, noise was added to the mock HST fluxes and uncertainties were estimated according to the 
depth of the available ancillary HST imaging. 

For NIRSpec, these data challenges not only served to verify and practice ingesting NIRCam-generated and -formated target catalogs and images into the NIRSpec target selection and MSA mask design work flow, but also provided a critical opportunity to augment the eMPT with needed features.
In particular, we modified eMPT to be able to point the `prime' NIRSpec instrument such that the `secondary' NIRCam instrument achieved its intended elaborate mosaicking of the NIRCam Medium Prime fields described in Sections 4.2 and 5.6, while simultaneously exploiting the $\simeq1$~arcsec level permissible deviations from the nominal NIRCam pointing pattern to optimize the parallel NIRSpec exposures such that the largest possible number of the highest priority HST targets were captured by the MSA. 
This task was further complicated by the peculiar manner in which the roll orientation of the NIRSpec MSA assigned to an observation by STScI does not refer to the center of the MSA, but rather to a reference point defined by median location of all targets contained in the NIRSpec input catalog entered into the APT \citep{bonaventura23a}. 
Limiting the impact of this complication over the $6.4^\prime$ lever arm between the field centers of the two instruments required an iterative approach in which the NIRSpec input catalog was gradually trimmed down to match the outer envelop of the final NIRSpec footprint.

The NIRSpec component of DC2 simulated an approximation of the GOODS-South Medium/HST tier. The NIRCam source scene described above was used to assign spectra and morphologies to the known HST prioritized target catalog. The eMPT was exercised to determine the optimum set of six pairs of pointing locations, within the small tolerance allowed given that in the real Medium/HST a NIRCam mosaic would be made in parallel, that maximized the number of highest priority targets assigned shutters. The eMPT was then run to assign targets to shutters in order of priority class. Spectra were simulated and processed in a similar manner to DC1. One difference is that contaminants that would fall within the target or background shutter were included in the simulation to assess the effects of contamination.

To manipulate these Data Challenges and to prepare for the real
data, JADES also built visualization tools.  To browse the sky, we
developed FitsMap \citep{hausen22}, inspired in part by the 
Legacy Survey viewer led by D.\ Lang \citep{dey19}. 
FitsMap allows us to zoom and pan the
sky, easily changing between image layers, with overlays from various
catalogs that provide pop-up access to the database information.
To study the SEDs and photometric redshift outputs, we developed
JADESview\footnote{\url{https://github.com/kevinhainline/JADESView}}, which shows image thumbnails, photometric
SEDs with best-fit template overlays, and photometric redshift
likelihoods versus redshift.

We have found these preparations to be invaluable in handling the
in-flight data.  That said, unsurprisingly the real data have presented
additional challenges to which the team (and the community more
broadly) must adjust.  We have described some of these challenges here; others are described in our data release papers \citep{rieke23r,bunker23r,eisenstein23jof}.

%% file: conclusion.tex
The JWST Advanced Deep Extragalactic Survey is bringing an ambitious
deep imaging and spectroscopic infrared view of the GOODS-S and
GOODS-N fields in the first cycle of JWST observing.  With JADES,
we use 545 hours of open-shutter dual-band NIRCam imaging and 240
open-shutter hours of MIRI imaging to cover about 210 arcmin$^2$
to very faint flux levels in 12 distinct bands.  We then conduct
extensive multi-object infrared spectroscopy using 339 open-shutter
hours of NIRSpec MOS, observing over 5000 faint targets with both
prism and grating dispersers.

The resulting JADES imaging and spectra provide an exquisite
sample for the study of galaxy evolution.  The data set has
yielded many candidates at redshifts above 8 
\citep{hainline23r} and
provided early spectroscopic confirmation galaxies at $z>10$
\citep{robertson23,curtis-lake23,tacchella23gz,bunker23gz,carniani24,witstok25}.  The amount of detail
in both imaging and spectroscopy is very impressive and is revealing
high-redshift galaxies to be a diverse set, with clear variations
in morphology, emission-line ratios, and star-formation histories
\citep[e.g.][]{dressler23r,endsley23r,looser23r}.  The spectra reveal the imprint of reionization
through variations in Lyman $\alpha$ emission \citep{saxena23,witstok23r} and
signatures of the Gunn-Peterson damping wing \citep{curtis-lake23}.

JADES also provides a useful design example for deep surveys, which we have
documented in this paper.  We have found great value in the medium-band
F335M and F410M imaging and provide examples to achieve high pixel-diversity
in both imaging and spectroscopy.  We have demonstrated how 
the multiplex of grating spectroscopy can be increased by allowing these 
spectra to overlap and using the shorter
prism spectra to disambiguate emission lines.
We are also confronting a number of
challenges in carrying out the survey, such as recovering from lost data
in a survey with substantial geometrical constraints and concerns with
NIRCam persistence.  We expect these will be useful learning experiences
as the JWST mission matures.

As listed in \S~\ref{sec:external}, JADES is one of several extragalactic surveys being carried out in Cycle 1
of the JWST mission.  These span a range of depth, areas, filter sets, and
fields, and there is a productive complementarity in these choices.  JADES 
is important because of its deep and reasonably wide coverage of the 
GOODS-S/HUDF and GOODS-N/HDF fields, where there is an awesome amount 
of multi-wavelength imaging and spectroscopy, and because of its close 
coordination of JWST imaging and spectroscopy.  

The first release of JADES data, focusing on year 1 Deep NIRCam imaging and 
NIRSpec multi-object spectroscopy on the HUDF, is presented in \citet{bunker23r} and \citet{rieke23r}.
The second release, containing the initial data in the JADES Origins Field and the year one GOODS-S Medium Prime mosaic, is presented in \citet{eisenstein23jof}.
The third release \citep{deugenio2025} contains NIRSpec data taken before November 2023, as well as the GOODS-N Prime imaging.
All are 
available at \url{https://archive.stsci.edu/hlsp/jades}, \dataset{DOI: 10.17909/8tdj-8n28}, and the latest data can be viewed at \url{http://jades.idies.jhu.edu/}.
Additional releases will follow in the coming year, and we post science updates from the survey at the JADES Collaboration website,  \url{https://jades-survey.github.io}.
JADES will provide the foundation for JWST's study of these two premier
deep fields, and we look forward to many years of utilization and 
extension of this data set.